\newif\ifdraft

\documentclass[11pt]{article}

\usepackage{xspace}
\usepackage[small,bf]{titlesec} %
\usepackage{fancyhdr}
\usepackage{nth}

\usepackage[tight]{subfigure}

\usepackage{amsmath}
\usepackage{amsthm}

\usepackage{enumitem}
\usepackage[sort,numbers]{natbib}	%
\usepackage{listings}
\usepackage{graphicx}

\ifdraft
\usepackage{soul} %
\fi

\ifdraft
\usepackage{lmodern}
\usepackage{everysel}
\EverySelectfont{%
\fontdimen2\font=0.4em%
\fontdimen3\font=0.2em%
\fontdimen4\font=0.1em%
\fontdimen7\font=0.1em%
\hyphenchar\font=`\-%
}
\else
\usepackage[charter]{mathdesign}
\fi
\usepackage[T1]{fontenc}

\usepackage{setspace}

\usepackage[a4paper,margin=2.3cm,top=2.6cm,bottom=2.6cm]{geometry} 
\usepackage{microtype}
\usepackage{textcomp}

\usepackage{footmisc}

\usepackage[hidelinks]{hyperref} %

\title{\textbf{Multiprocessor Real-Time Locking Protocols} \\ \emph{A Systematic Review}}

\author{
Bj\"orn B. Brandenburg \\[0.15cm]
\emph{Max Planck Institute for Software Systems (MPI-SWS)} \\
\emph{Kaiserslautern, Germany}
} 

\date{}

\newcommand{\eg}{{\it e.g.}\xspace}
\newcommand{\ie}{{\it i.e.}\xspace}
\newcommand{\etc}{{\it etc.}\xspace}

\newcommand{\apriori}{{\it a priori}\xspace}

\newcommand{\xth}{\ensuremath{^{\text{th}}}\xspace}

\newcommand{\vs}{{vs\mbox{.}}\xspace}
\newcommand{\resp}{{respectively}\xspace}
\newcommand{\wrt}{{w.r.t\mbox{.}}\xspace}

\newcommand{\mus}[0]{\,\ensuremath{\mu}s\xspace}

\newcommand{\ms}[0]{\,ms\xspace}

\newcommand{\res}{\ell}
\newcommand{\lmax}{L^{max}}

\newcommand{\efont}[1]{\textbf{#1}}
\newcommand{\eone}{\efont{(i)}\xspace}
\newcommand{\etwo}{\efont{(ii)}\xspace}

\newcommand{\litmus}[0]{LITMUS\textsuperscript{RT}\xspace}

\newcommand{\fmlpp}[0]{FMLP\textsuperscript{+}\xspace}
\newcommand{\rrdglp}[0]{R\textsuperscript{2}DGLP\xspace}

\newcommand{\secref}[1]{Section~\ref{sec:#1}\xspace}
\newcommand{\secrefs}[2]{Sections~\ref{sec:#1} and~\ref{sec:#2}\xspace}
\newcommand{\secrefx}[3]{Sections~\ref{sec:#1}, \ref{sec:#2}, and~\ref{sec:#3}\xspace}
\newcommand{\secrefxx}[4]{Sections~\ref{sec:#1}, \ref{sec:#2}, \ref{sec:#3}, and~\ref{sec:#4}\xspace}
\newcommand{\secrefr}[2]{Sections~\ref{sec:#1}--\ref{sec:#2}\xspace}
\newcommand{\figref}[1]{Figure~\ref{fig:#1}\xspace}
\newcommand{\figrefs}[2]{Figures~\ref{fig:#1} and~\ref{fig:#2}\xspace}

\newcommand{\defref}[1]{Definition~\ref{def:#1}\xspace}
\newcommand{\defrefs}[2]{Definitions~\ref{def:#1} and~\ref{def:#2}\xspace}

\newcommand{\incfig}[2]{
\begin{figure}[t]
  \centering
  \includegraphics[width=\textwidth]{figures/#1}
  \caption{#2}
  \label{fig:#1}
\end{figure}
}
\titlespacing{\paragraph}{0in}{0.04in}{0.08in}
\ifdraft
\titleformat*{\paragraph}{\it}
\fi

\ifdraft
\renewcommand{\emph}[1]{\ul{#1}}
\fi

\ifdraft
\titlespacing*{\subsection}{0in}{1.5em}{0.5em}
\titlespacing*{\section}{0in}{2em}{1em}
\else
\titlespacing*{\section}{0in}{1.75em}{0.5em}
\titlespacing*{\subsection}{0in}{1.25em}{0.25em}
\titlespacing*{\subsubsection}{0in}{0.75em}{0em}
\fi

\fancypagestyle{titlepage}{
   \lhead{ \ifdraft \it  Incomplete  Draft --- Do Not Distribute  \else Revision 1   \fi } %
   \chead{}
   \rhead{\ifdraft \today \else September 2019 \fi}
   \cfoot{}
   \rfoot{}
   \lfoot{\small \ifdraft \texttt \fi{\copyright~2017--2019 B. Brandenburg. All rights reserved.}}
}

\fancypagestyle{main}{
   \lhead{\small{}Multiprocessor Real-Time Locking Protocols --- A Systematic Review \ifdraft (DRAFT) \fi}
   \chead{}
   \rhead{\small B. Brandenburg}
   \cfoot{\footnotesize \thepage}
   \rfoot{}
   \lfoot{}
}

\newtheoremstyle{mydef}%
	{3pt}%
	{3pt}%
	{\normalfont}%
	{}%
	{\bfseries}%
	{.}%
	{.5em}%
	{\thmname{#1}\thmnumber{#2}\thmnote{#3}}%

\newtheoremstyle{mylemthm}
	{6pt}
	{3pt}
	{\slshape}
	{}
	{\bfseries}
	{.}
	{.5em}
	{}

\theoremstyle{mylemthm}

\theoremstyle{mydef}
\newtheorem{definition}{Def.\ }

\begin{document}

\maketitle
\doublespacing

\thispagestyle{titlepage}
\pagestyle{main}

\section*{Abstract}

We systematically survey the literature on analytically sound multiprocessor real-time locking protocols from 1988 until 2018, covering the following topics: 
\begin{itemize}
	\item progress mechanisms that prevent the lock-holder preemption problem (\secref{progress}), 
	\item spin-lock protocols (\secref{spin}), 
	\item binary semaphore protocols (\secrefs{susp}{rpc}), 
	\item independence-preserving (or fully preemptive) locking protocols (\secref{ip}), 
	\item reader-writer and $k$-exclusion synchronization (\secref{relx}),  
	\item support for nested critical sections (\secref{nested}), and
	\item implementation and system-integration aspects (\secref{impl}).
\end{itemize}
A special focus is placed on the suspension-oblivious and suspension-aware analysis approaches for semaphore protocols, their respective notions of priority inversion, optimality criteria,  lower bounds on maximum priority-inversion blocking, and matching asymptotically optimal locking protocols.

\section{Introduction}\label{sec:intro}

In contrast to the thoroughly explored and well understood  uniprocessor real-time synchronization problem, the multiprocessor case considered herein is still the subject of much ongoing work. 
In particular, uniprocessor real-time locking protocols that are both \emph{optimal} and \emph{practical} are readily available since the late 1980s and early 1990s~\cite{B:91,SRL:90,R:91}, can flexibly support mutual exclusion, reader-writer synchronization, multi-unit resources, and have been widely adopted and deployed in industry (in POSIX, OSEK/AUTOSAR, \etc). Not so in the multiprocessor case: there is no widely agreed-upon standard protocol or approach,  most proposals have focused exclusively on mutual exclusion to date (with works on reader-writer, $k$-exclusion, and multi-unit synchronization starting to appear only in the past decade), and questions of optimality, practicality, and industrial adoption are still the subject of ongoing exploration and debate. 
Another major difference to the uniprocessor case is the extent to which \emph{nested critical sections} are supported: a large fraction of the work on multiprocessor real-time locking protocols to date has simply disregarded (or defined away) \emph{fine-grained nesting} (\ie, cases where a task holding a resource may dynamically acquire a second resource), though notable exceptions exist (discussed in \secref{nested}). No such limitations exist in the state of the art \wrt uniprocessor real-time synchronization. 

We can thus expect the field to continue to evolve rapidly: it is simply not yet possible to provide a ``final'' and comprehensive survey on multiprocessor real-time locking, as many open problems remain to be explored. Nonetheless, a considerable number of results has accumulated since the multiprocessor real-time locking problem was first studied more than three decades ago. The purpose of this survey is to provide a \emph{systematic} review of the current snapshot of this body of work, covering most papers in this area published until the end of 2018. 

We restrict the scope of this survey to real-time locking protocols for \emph{shared-memory} multiprocessors (although some of the techniques discussed in \secref{rpc} could also find applications in distributed systems), and do not consider alternative synchronization strategies such as lock- and wait-free algorithms, transactional memory techniques, or  middleware (or database) approaches that provide a higher-level transactional interface or multi-versioned datastore abstraction (as any of these techniques warrants a full survey on its own). We further restrict the focus to \emph{runtime} mechanisms as commonly provided by real-time operating systems (RTOSs) or programming language runtimes for use in \emph{dynamically scheduled} systems and exclude fully static planning approaches (\eg, as studied by \citet{X:93}) that resolve (or avoid) all potential resource conflicts statically during the construction of the system's scheduling table so that no runtime resource arbitration mechanisms are needed.

Our goal is to structure the existing body of knowledge on multiprocessor real-time locking protocols to aid the interested reader in understanding key problems, established solutions, and recurring themes and techniques.  We therefore focus primarily on ideas, algorithms, and provable guarantees, and place less emphasis on empirical performance comparisons or the chronological order of developments. 
\section{The Multiprocessor Real-Time Locking Problem}\label{sec:problem}

We begin by defining the core problems and objectives, summarizing common assumptions, and surveying key design parameters. Consider a shared-memory multiprocessor platform consisting of $m$  identical processors (or \emph{cores}) hosting $n$ \emph{sequential tasks} (or \emph{threads}) denoted as $\tau_1,\ldots,\tau_n$. Each \emph{activation} of a task is called a \emph{job}. In this document, we use the terms `processor` and `core` interchangeably, and do not precisely distinguish between jobs and tasks when the meaning is clear from context.

The tasks share a number of software-managed \emph{shared resources} $\res_1, \res_2, \res_3,\ldots$ that are explicitly acquired and released with \texttt{lock()} and \texttt{unlock()} calls.\footnote{Multicore platforms also often feature hardware-managed, \emph{implicitly} shared resources such as last-level caches (LLCs), memory controllers, DRAM banks, \etc; techniques for managing such hardware resources are not the subject of this survey.} Common examples of such resources include shared data structures, OS-internal structures such as the scheduler's ready queue(s), I/O ports, memory-mapped device registers, ring buffers, \etc that must be accessed in a mutually exclusive fashion. (We consider weaker exclusion requirements later in \secref{relx}.)

The primary objective of the algorithms considered herein is to serialize all \emph{resource requests}---that is, all \emph{critical sections}, which are code segments surrounded by matching \texttt{lock()} and \texttt{unlock()} calls---such that the timing constraints of all tasks are met, despite the \emph{blocking delays} that tasks incur when waiting to lock a contested resource. In particular, in order to give nontrivial response-time guarantees, one must bound the \emph{maximum} (\ie, \emph{worst-case}) \emph{blocking delay} incurred by any task due to contention for shared resources. To this end, a \emph{multiprocessor real-time locking protocol} determines which type of locks are used, the rules that tasks must follow to request a lock on a shared resource, and how locks interact with the scheduler. We discuss these design questions in more detail below. 

\subsection{Common Assumptions}\label{sec:model}
Although there exists much diversity in system models and general context, most surveyed works share the following basic assumptions (significant deviations will be noted where relevant). Tasks are typically considered to follow a \emph{periodic or sporadic activation pattern}, where the two models can be used largely interchangeably since hardly any work on multiprocessor real-time locking protocols to date has exploited the knowledge of future periodic arrivals.\footnote{Notable exceptions include proposals by \citeauthor{CTB:94}~\cite{CTB:94} and \citeauthor{SUBC:19}~\cite{SUBC:19}.} The timing requirements of the tasks are usually expressed as \emph{implicit or constrained deadlines} (rather than arbitrary deadlines), since arbitrary deadlines that exceed a task's period (or minimum inter-arrival time) allow for increased contention and cause  additional analytical complications.%

For the purpose of schedulability analysis, but \emph{not} for the operation of a locking protocol at runtime, it is required to know a \emph{worst-case execution time} (WCET) bound for each task, usually including the cost of all critical sections (but \emph{not} including any blocking delays), and also individually for each critical section (\ie, a \emph{maximum critical section length} must be known for each critical section). Furthermore, to enable a meaningful blocking analysis, the maximum number of critical sections in each job (\ie, the maximum number of \texttt{lock()} calls per activation of each task) must be known on a per-resource basis. 

While it is generally impossible to make strong \apriori response-time \emph{guarantees} without this information being available at analysis time (for at least some of the tasks), to be practical, it is generally desirable for a locking protocol to work as intended even if this information is unknown at runtime. For example, RTOSs are typically used for many purposes, and not all workloads will be subject to static analysis---the implemented locking protocol should function correctly and predictably nonetheless. Similarly, during early prototyping phases, sound bounds are usually not yet available, but the RTOS is expected to behave just as it will in the final version of the system.

With regard to scheduling, most of the covered papers assume either \emph{partitioned} or \emph{global} multiprocessor scheduling.
Under partitioned scheduling, each task is statically assigned to exactly one of the $m$ processors (\ie, its \emph{partition}), and each processor is scheduled using a uniprocessor policy such as \emph{fixed-priority} (FP) scheduling or \emph{earliest-deadline first} (EDF) scheduling. The two most prominent partitioned policies are
 \emph{partitioned FP} (P-FP) and \emph{partitioned EDF} (P-EDF) scheduling.

Under global scheduling,  all tasks are dynamically dispatched at runtime, may execute on any of the $m$ processors, and migrate freely among all  processors as needed. Widely studied examples include \emph{global FP} (G-FP) and \emph{global EDF} (G-EDF) scheduling, as well as optimal policies such as \emph{Pfair} scheduling~\cite{BCPV:96,SA:06}.  

A third, more general notion is \emph{clustered scheduling}, which generalizes both global and partitioned scheduling. Under clustered scheduling,  the set of $m$ cores is split into a number of disjoint \emph{clusters} (\ie, disjoint subsets of cores), tasks are statically assigned to clusters, and each cluster is scheduled locally and independently using a ``global'' scheduling policy (\wrt the cores that form the cluster). We let $c$ denote the number of cores in a cluster. Global scheduling is a special case of clustered scheduling with a single ``cluster'' and $m=c$, and partitioned scheduling is the other extreme with $m$ clusters and $c=1$.

While clustered scheduling is a more general assumption (\ie, any locking protocol designed for clustered scheduling also works for global and partitioned scheduling), it also comes with the combined challenges of both global and partitioned scheduling, and is hence generally much more difficult to deal with. Historically, most authors have thus focused on either global or partitioned scheduling. 

Under partitioned and clustered scheduling, it is useful to distinguish between \emph{global} and \emph{local} resources. Under partitioned (\resp, clustered) scheduling, a shared resource is considered local if it is accessed only by tasks that are all assigned to the same core (\resp, cluster).  In contrast, a resource is global if it is accessed by at least two tasks assigned to different partitions (\resp, clusters). The advantage of this distinction is that local resources can be managed with existing, simpler, and often more efficient
protocols: 
\begin{itemize}
	\item under partitioned scheduling, local resources can be managed using one of the known, optimal uniprocessor protocols (\eg, the PCP~\cite{SRL:90} or the SRP~\cite{B:91}); and
	\item  under clustered scheduling, local resources can be managed  using a (usually simpler) protocol for global scheduling (instantiated within each cluster). 
\end{itemize}
In this survey, we therefore consider only global resources, which are the more challenging kind of shared resources to manage.

\subsection{Key Design Choices}

Given the common setting outlined so far, there are a number of key design questions that any multiprocessor real-time locking protocol must address.  We next provide a high-level overview of these issues.

\subsubsection{Request Order}
The first key design parameter is the serialization order for conflicting requests. 
Whenever two or more requests for a resource are simultaneously blocked, the protocol must specify a policy for sequencing the waiting requests. 

The two most common choices are \emph{FIFO queuing}, which ensures basic fairness (\ie, non-starvation), is easy to implement, and analysis friendly, and \emph{priority queuing}, which allows control over the amount of blocking incurred by different tasks. As a third choice, there also exist some situations (discussed in \secrefs{sob}{ip}) in which the use of \emph{hybrid queues} consisting of both FIFO- and priority-ordered segments can be advantageous.

When using priority queues, each request must be associated with a priority to determine its position in the wait queue. The common choice is to use a job's scheduling priority, but it is also possible to use a separate request priority, which can be selected on a per-resource or even on a per-critical-section basis. The latter obviously provides a considerable degree of flexibility, but is not commonly considered since it introduces a nontrivial configuration problem. FIFO and priority queuing can be generalized in a straightforward way by requiring that equal-priority requests are satisfied in FIFO order.

Finally, implementation concerns may sometimes necessitate the use of \emph{unordered locks} (\eg, primitive test-and-set spin locks), which do not provide any guarantees regarding the order in which conflicting critical sections will execute. 
Unordered locks are decidedly not analysis-friendly, but sometimes unavoidable.

\subsubsection{Spinning \vs Suspending} 
The second main question is how tasks should wait in case of contention.  The two principal choices are \emph{busy-waiting} (\ie, \emph{spinning}) and  \emph{suspending}. 
In the case of busy-waiting, a blocked task continues to occupy its processor and simply executes a tight delay (or \emph{spin}) loop, continuously checking whether it has been granted the lock, until it gains access to the shared resource.  Alternatively, if tasks wait by suspending, a blocked task yields the processor and is taken out of the scheduler's ready queue until it is granted the requested resource. 

Determining how blocked tasks should wait is not an easy choice, as there are many advantages and disadvantages associated with either approach. On the one hand, suspension-based waiting is conceptually more efficient: busy-waiting obviously wastes processor cycles, whereas suspension-based waiting allows the wait times of one task to be overlaid with useful computation of another task. On the other hand, busy-waiting is easier to implement, easier to analyze, requires less OS support, and the cost of suspending and resuming a task can easily dwarf typical critical section lengths. Spin locks also provide predictability advantages that can aid in the static analysis of the system (\eg, a busy-waiting task ``protects'' its processor and cache state, whereas it is virtually impossible to predict the cache contents encountered by a resuming task).

Whether spinning or suspending is more efficient ultimately depends on workload and system characteristics, such as the cost of suspending and resuming tasks relative to critical section lengths, and it is impossible to categorically declare one or the other to be the ``best'' choice.  Generally speaking, ``short'' critical sections favor busy-waiting, whereas ``long'' critical sections necessitate suspensions, but the threshold between ``short'' and ``long'' is highly system- and application-specific. We discuss spin-based locking protocols in \secrefx{spin}{ip}{relx} and suspension-based locking protocols in \secrefr{susp}{relx}.

\subsubsection{Progress Mechanism} 
A third major choice is the question of how to deal with the \emph{lock-holder preemption} problem, which is tightly coupled to the choice of scheduler, the employed analysis approach, and the constraints of the target workload. If a lock-holding task is preempted by a higher-priority task, then any other task waiting for the  resource held by the preempted task is \emph{transitively} delayed as well. This can give rise to (potentially) \emph{unbounded priority inversions} (\ie, excessive  delays that are difficult to bound), which must be avoided in a real-time system. To this end, it is at times necessary to (selectively) \emph{force} the execution of lock-holding tasks by means of a \emph{progress mechanism}.  As this is a crucial aspect of multiprocessor real-time locking protocols, we dedicate \secref{progress} to the lock-holder preemption problem and common solutions.

\subsubsection{Support for Fine-Grained Nesting} 
Fine-grained locking, where a task concurrently acquires \emph{multiple} locks in a nested, incremental fashion, is a major source of complications in both the analysis and the implementation of multiprocessor real-time locking protocols. In particular, it comes with the risk of deadlock, and even if application programmers are careful to avoid deadlocks, nesting still introduces intricate transitive blocking effects that are extremely challenging to analyze accurately and efficiently. Furthermore, in many cases, support for fine-grained nesting leads to substantially more involved protocol rules and more heavy-weight OS support. 

As already mentioned in \secref{intro}, consequently many works on multiprocessor real-time locking protocols simply disallow the nesting of critical sections altogether, or under partitioned scheduling, restrict nesting to local resources only, where it can be resolved easily with classic uniprocessor solutions~\cite{B:91,SRL:90,R:91}.

Another common approach is to sidestep the issue by relying on \emph{two-phase locking schemes} with all-or-nothing semantics, where a task either atomically acquires all requested locks, or holds none,  or simple \emph{group lock} approaches that automatically aggregate fine-grained, nested critical sections into coarse-grained, non-nested lock requests. From an analysis point of view, the two-phase and group locking schemes are conveniently similar to protocols that disallow nesting altogether, but from a developer's point of view, they impose limitations that may be difficult to accommodate in practice. 

Only in recent years have there been renewed efforts towards full, unrestricted support for fine-grained nesting (\eg,~\cite{WA:12,BBW:16}), and there remains ample opportunity for future work. We discuss the issues surrounding fine-grained nesting and the state of the art in \secref{nested}, and until then focus exclusively on non-nested critical sections.

\subsubsection{In-Place \vs Centralized Critical Sections} \label{sec:prob:loc}
The final choice is where to execute a critical section once the lock has been acquired. In shared-memory systems, the typical choice is to execute critical sections \emph{in place}, meaning that a task executes its critical sections as part of its regular execution, on the processor that it is (currently) assigned to by the scheduler. 

However, that is not the only choice. It is also possible to \apriori designate a \emph{synchronization processor} for a particular resource, to the effect that all critical sections (pertaining to that resource) must be executed on the designated synchronization processor. This can yield analytical benefits (\ie, less blocking in the worst case~\cite{B:13a}), a reduction in worst-case overheads~\cite{CVB:14}, and throughput benefits due to improved cache affinity~\cite{LDTL:12}. Furthermore, for specialized hardware resources, such as certain I/O devices, it might simply be unavoidable on some platforms (\eg, in a heterogenous multiprocessor platform, one processor might be a designated I/O processor). We discuss protocols that rely on designated synchronization processors in \secref{rpc} and for now focus exclusively on in-place execution.

\subsection{Analysis and Optimization Problems}

In addition to the just-discussed design choices, which determine the \emph{runtime} behavior of the protocol, there are also a number of  challenging \emph{design-time} problems related to \apriori timing and schedulability analysis of the system, fundamental optimality questions, and system optimization and design-space exploration problems. 

Most prominently, as a prerequisite to schedulability analysis, the \emph{blocking analysis problem} asks to bound the worst-case delay due to resource conflicts. That is, given a workload, a multiprocessor platform, and a specific multiprocessor real-time locking protocol, the objective is to compute a safe (and as accurate as possible) upper bound on the maximum additional delay due to resource contention encountered by a given task in any possible execution of the system.
Once such a bound is known for each task, hard timing guarantees can be made with a  blocking-aware schedulability or response-time analysis. In some instances, it can be more accurate to carry out both the blocking analysis and the blocking-aware schedulability analysis jointly in a single step (\eg, \cite{YWB:15}). 

A large number of ad-hoc, protocol-specific blocking analyses have been developed over the years. Additionally, a more general \emph{holistic blocking analysis} framework~\cite{B:11} and a blocking analysis approach based on \emph{linear programming} (LP) and \emph{mixed-integer linear programming} (MILP) have been introduced to systematically reduce analysis pessimism~\cite{B:13a,WB:13a,YWB:15,BB:16,BBW:16}. These more recent	 analysis frameworks represent a general approach that can be (and has been) applied to many different locking protocols; we will note their use and discuss advantages in the context of specific locking protocols in \secrefxx{spin:part}{saw:global}{saw:part}{nested}.

Clearly, it is generally desirable for blocking bounds to be as low as possible. However, if locks are used to resolve contention at runtime, it is also obvious that, in the worst case, some delays are inevitable. This observation naturally leads to the question of \emph{asymptotic blocking optimality}: generally speaking, what is the \emph{least} bound on maximum blocking  that \emph{any} protocol can achieve? This question has been studied primarily in the context of suspension-based locking protocols (since the spin-based case is relatively straightforward), and a number of protocols with  provably optimal asymptotic blocking bounds have been found. We will discuss this notion of optimality and the corresponding protocols in \secref{susp}.

Another notion of optimality that has been used to characterize multiprocessor real-time locking protocols is resource augmentation and processor speedup bounds, which relate a protocol's timing guarantees to that of a hypothetical optimal one in terms of the additional resources (or processor speed increases) needed to overcome the protocol's non-optimality. While this is a stronger notion of optimality than asymptotic blocking optimality---speedup and resource-augmentation results consider both blocking and scheduling, whereas asymptotic blocking optimality is concerned solely with the magnitude of blocking bounds---existing speedup and resource-augmentation results have been obtained under very restrictive assumptions (\eg, only one critical section per task) and yield quite large augmentation and speedup factors. We briefly mention some relevant works in \secref{misc}.

Last but not least, a large number of challenging system optimization and design-space exploration problems can be formalized under consideration of synchronization constraints. Prominent examples include:
\begin{itemize}
	\item \emph{task mapping problems}---given a processor platform, a scheduling policy, and a locking protocol, find an assignment of tasks to processors (or clusters) that renders the system schedulable, potentially while optimizing some other criteria (\eg, average response times, memory needs, energy or thermal budgets, \etc);
	\item \emph{resource mapping problems}---select for each shared resource a designated synchronization processor such that the system becomes schedulable;
	\item \emph{platform minimization problems}---given a workload, scheduling policy, and locking protocol, minimize the number of required cores;
	\item \emph{policy selection problems}---given a workload and a platform, identify (potentially on a per-resource basis) a locking protocol (or an alternative synchronization approach) that renders the system schedulable, again potentially while simultaneously optimizing for other criteria;  and
	\item many variations and combinations of these and similar problems. 
\end{itemize}
Not surprisingly, virtually all interesting problems of these kinds are NP-hard since they typically involve solving one or more bin-packing-like problems.
While a detailed consideration of optimization techniques is beyond the scope of this survey, we briefly mention some representative results that exemplify  these types of system integration and optimization problems in \secref{placement}.

\subsection{Historical Perspective} \label{sec:hist}

Historically, the field traces its roots to the 1980s. While a discussion of the challenges surrounding multiprocessor synchronization in real-time systems, including a discussion of the respective merits of spin- and suspensions-based primitives, can be already found in an early critique of ADA~\cite{REMC:81} published in 1981, the first multiprocessor real-time locking protocol backed by a sound schedulability analysis taking worst-case blocking delays into account is \citeauthor{RSL:88}'s \emph{Distributed Priority Ceiling Protocol} (DPCP)~\cite{RSL:88}, which appeared in 1988. This result was followed in 1990 by the \emph{Multiprocessor Priority Ceiling Protocol} (MPCP)~\cite{R:90}, the second foundational protocol that had (and continues to have) a large impact on the field, and which in many ways still represents the prototypical suspension-based protocol for partitioned scheduling (which however should not obscure the fact that many other well-performing alternatives have been proposed since). 

Throughout the 1990s, a number of protocols and lock implementations appeared; however, as multiprocessor real-time systems were still somewhat of a rarity at the time, multiprocessor synchronization was not yet a major concern in the real-time community. This fundamentally changed with the advent of multicore processors and \emph{multiprocessor system-on-a-chip} (MPSoC) platforms in the early to mid 2000s. Motivated by these trends, and the desire to minimize the use of stack memory in such systems, \citeauthor{GLD:01} published a highly influential paper in 2001 proposing the \emph{Multiprocessor Stack Resource Policy} (MSRP)~\cite{GLD:01}. While \citeauthor{GLD:01}'s schedulability and blocking analysis has since been superseded by later, more accurate analyses~\cite{B:11,WB:13a,BB:16,BBW:16}, the MSRP remains the prototypical spin-based multiprocessor real-time locking protocol for partitioned scheduling.  Another influential paper motivated by the widespread emergence of multicore processors as the standard computing platform was published in 2007 by \citeauthor{BLBA:07}, who introduced the \emph{Flexible Multiprocessor Locking Protocol} (FMLP)~\cite{BLBA:07}, which combined many of the advantages of the MPCP and the MSRP, and which was the first to provide full support for both global and partitioned scheduling. 

The FMLP paper marked the beginning of the recent surge in interest in multiprocessor real-time locking: since 2007, every year has seen more than a dozen publications in this area---about 140 in total in the past decade, which is almost three times as many as published in the 20 years prior. We dedicate the rest of this survey to a systematic review (rather than a chronological one) of this vibrant field. 
\section{Progress Mechanisms}\label{sec:progress}

At the heart of every effective multiprocessor real-time locking protocol is a \emph{progress mechanism} to expedite the completion of critical sections that otherwise might cause excessive blocking to higher-priority or remote tasks. More specifically, a progress mechanism
 forces the scheduling of lock-holding tasks (either selectively or unconditionally), thereby temporarily overriding the normal scheduling policy. In this section, we review the major progress mechanisms developed to date and provide example schedules that illustrate key ideas. 

As a convention, unless noted otherwise, we use \emph{fixed-priority} (FP) scheduling in our examples and assume that tasks are indexed in order of strictly decreasing priority (\ie, $\tau_1$ is always the highest-priority task). All examples in this section further assume the use of basic suspension-  or spin-based locks (\ie, raw locks without additional protocol rules); by design the specifics are irrelevant.

\subsection{Priority Inversion on Uniprocessors}
To understand the importance of progress mechanisms, and why multiprocessor-specific mechanisms are needed, it is helpful to first briefly review the threat of ``unbounded priority inversions'' on uniprocessors and how it is mitigated in classic uniprocessor real-time locking protocols. 

\begin{figure}[t]
  \centering
  \includegraphics[width=\textwidth]{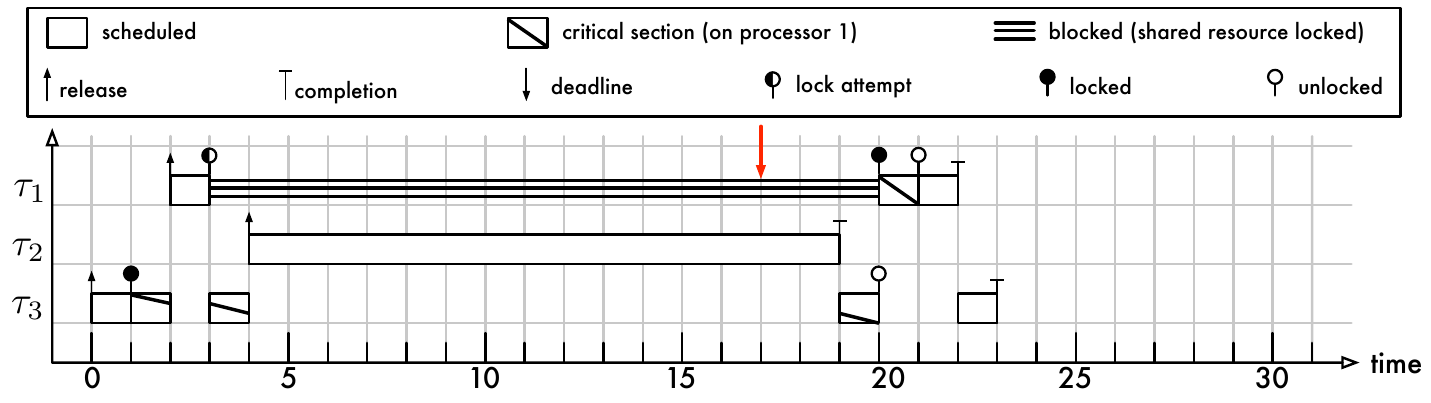}
  \caption{Example FP schedule of three tasks demonstrating an ``unbounded'' priority inversion on a uniprocessor. Irrelevant deadlines are omitted from this and all following figures to reduce clutter.}
  \label{fig:uni-unbounded-pi}
\end{figure}

\figref{uni-unbounded-pi} shows the classic ``unbounded priority inversion'' example of three tasks under FP scheduling. At time~1, the lowest-priority task $\tau_3$ locks a resource that it shares with the highest-priority task $\tau_1$. When $\tau_1$ is activated at time~2, it  also tries to lock the resource (at time~3), and thus becomes blocked by $\tau_3$'s critical section, which intuitively constitutes a \emph{priority inversion}  since $\tau_1$ is pending (\ie, it has unfinished work to complete before its deadline) and has higher priority than $\tau_3$, but $\tau_3$ is scheduled instead of $\tau_1$. When the ``middle-priority'' task $\tau_2$ is activated at time~4, it preempts the lock-holding, lower-priority task $\tau_3$, which delays the completion of $\tau_3$'s critical section, which in turn continues to block $\tau_1$, until $\tau_2$ completes and yields the processor at time~19 (at which point $\tau_1$ has already missed its deadline). 

Since $\tau_2$ has lower priority than the pending (but not scheduled) $\tau_1$, this delay also constitutes a priority inversion. And since the length of this priority inversion is determined by $\tau_2$'s WCET, which in general could be arbitrarily large, this is traditionally considered to  be an ``unbounded'' priority inversion (even though technically it is bounded by the maximum scheduling interference incurred by $\tau_3$). That is, a priority inversion is traditionally considered ``bounded'' only if its maximum-possible duration can be bounded in terms of \emph{only} the maximum critical section length and the number of concurrent critical sections, and which is independent of all tasks' WCETs, since WCETs are expected to usually be (much) larger than typical critical section lengths.

To summarize, on uniprocessors, ``unbounded'' priority inversion arises as a consequence of the lock-holder preemption problem, and it is problematic because it renders the response times of high-priority tasks (\eg, $\tau_1$'s in \figref{uni-unbounded-pi}) dependent on the WCETs of  lower-priority tasks (\eg, $\tau_2$'s in  \figref{uni-unbounded-pi}). This contradicts the purpose of priority-driven scheduling, where  higher-priority tasks (or jobs) should remain largely independent of the processor demands of lower-priority tasks (or jobs). On uniprocessors, classic progress mechanisms such as \emph{priority inheritance}~\cite{SRL:90} or \emph{priority-ceiling protocols}~\cite{SRL:90,B:91} avoid unbounded priority inversion, either by raising the priority of lock-holding tasks or by delaying the release of higher-priority tasks. 

\subsection{Priority Inversion on Multiprocessors}

\begin{figure}[t]
  \centering
  \includegraphics[width=\textwidth]{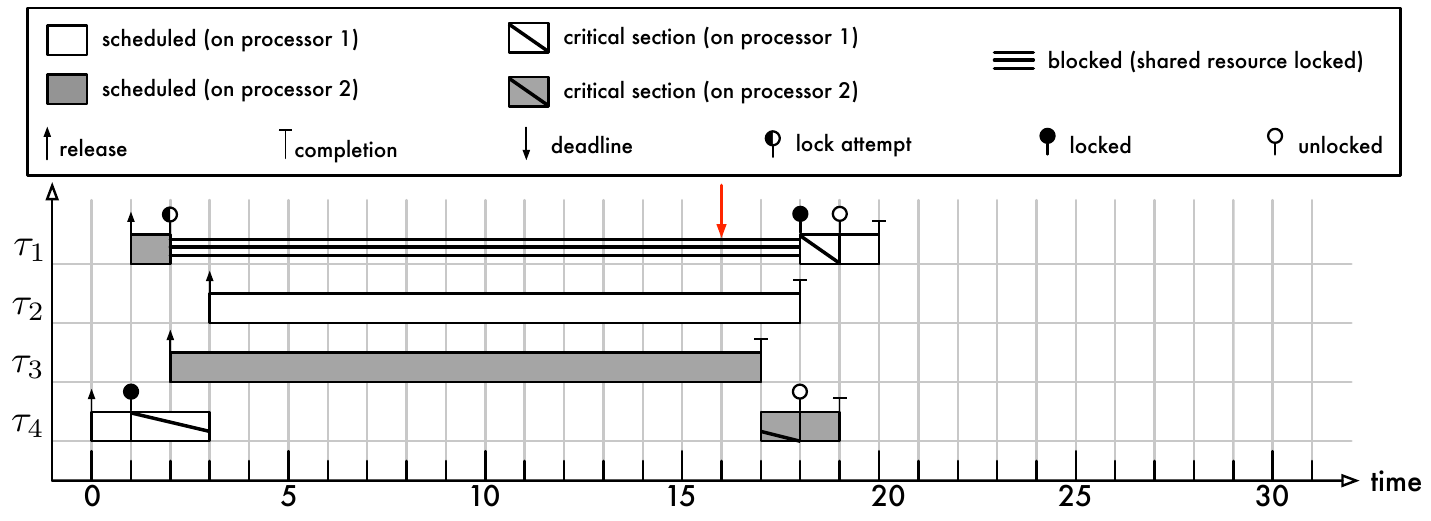}
  \caption{Example G-FP schedule of four tasks on two processors demonstrating an ``unbounded'' priority inversion on a globally scheduled multiprocessor.}
  \label{fig:mp-unbounded-pi-global}
\end{figure}

The lock-holder preemption problem of course also exists on multiprocessors. For example, \figref{mp-unbounded-pi-global}, shows a situation comparable to \figref{uni-unbounded-pi} involving four tasks on $m=2$ processors under G-FP scheduling. Analogously to the uniprocessor example, the lock-holding task $\tau_4$ is preempted due to the arrival of two higher-priority tasks at times~2 and~3, \resp, which in turn induces an ``unbounded'' priority inversion (\ie, an unwanted dependency on the WCETs of lower-priority tasks) in the highest-priority (and blocked) task $\tau_1$.

\begin{figure}[t]
  \centering
  \includegraphics[width=\textwidth]{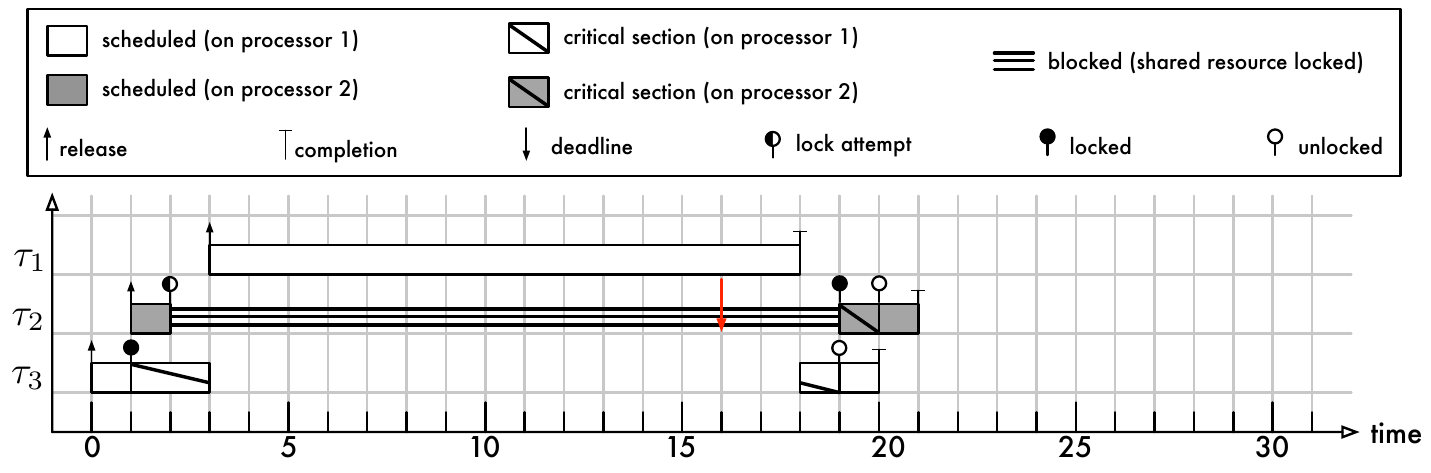}
  \caption{Example P-FP schedule of three tasks on two processors demonstrating an ``unbounded'' priority inversion due to a remote task under partitioned scheduling. }
  \label{fig:mp-unbounded-pi-part}
\end{figure}

However, in addition to creating an unwanted dependency of high- on low-priority tasks (or jobs), an untimely preemption of a lock holder  can also result in undesirable delays of \emph{remote} tasks or jobs. For instance, consider the example in \figref{mp-unbounded-pi-part}, which shows a \emph{partitioned fixed-priority} (P-FP) schedule illustrating an ``unbounded'' priority inversion due to a remote task. Compared to the uniprocessor example in \figref{uni-unbounded-pi}, the roles of $\tau_2$ and $\tau_1$ have been switched, and $\tau_2$ has been assigned (by itself) to processor~2. Again, the lock-holding task $\tau_3$ is preempted at time~3 by a higher-priority task ($\tau_1$ in this case). As a result, task $\tau_2$ transitively incurs a delay proportional to $\tau_1$'s WCET, even though $\tau_1$ is, from the point of view of $\tau_2$,  an unrelated \emph{remote} task that $\tau_2$'s response time intuitively should not depend on.

Specifically, even though $\tau_1$ may have a numerically higher priority, when analyzing each processor as a uniprocessor system (the standard approach under partitioned scheduling), the delay transitively incurred by $\tau_2$ due to $\tau_1$ must be considered an extraordinary source of interference akin to a priority inversion since there is no local higher-priority task that executes on processor~2 while $\tau_2$ is pending. 

As a result, multiprocessor systems require a more general notion of ``priority inversion.'' To capture delays due to remote critical sections, the definition of priority inversion under partitioned scheduling must include not only the classic case where a (local) lower-priority task is scheduled instead of a pending, but blocked (local) higher-priority task, but also the case where a processor idles despite the presence of a pending (but remotely blocked) higher-priority task. Analogously, under global scheduling on an $m$-processor platform, any situation in which fewer than $m$ higher-or-equal-priority tasks are scheduled while some task is waiting constitutes a priority inversion. Both cases (partitioned and global scheduling) can be captured precisely with the following definition. Recall that clustered scheduling generalizes both global and partitioned scheduling.

\begin{definition}
	\label{def:pi}
	A job $J$ of task $\tau_i$, assigned to a cluster $C$ consisting of $c$ cores, suffers \emph{priority-inversion blocking} (pi-blocking) at time $t$ if and only if 
	\begin{enumerate}
		\item $J$ is pending (\ie, released and incomplete) at time $t$,
		\item $J$ is not scheduled at time $t$, and 
		\item fewer than $c$ equal- or higher-priority jobs of tasks assigned to cluster $C$ are scheduled on processors belonging to $\tau_i$'s assigned cluster $C$.
	\end{enumerate} 
\end{definition}

Under partitioned scheduling $c=1$, and under global scheduling $c=m$. We prefer the specific term ``pi-blocking'' rather than the more common, but also somewhat vague term ``blocking'' since the latter is often used in an OS context to denote suspensions of any kind, whereas we are explicitly interested only in delays that constitute a priority inversion.

Note that \defref{pi} is defined in terms of jobs (and not tasks) to cover the full range of \emph{job-level fixed-priority} (JLFP) policies, and in particular EDF scheduling (which belongs to the class of JLFP policies). We will further refine \defref{pi} in \secref{susp} to take into account certain subtleties related to the analysis of self-suspensions. 

Applying \defref{pi} to the example in \figref{uni-unbounded-pi}, we observe that $\tau_1$ indeed incurs pi-blocking from time~3 until time~20, since  $\tau_1$ is pending, but not scheduled, and fewer than $c=1$ higher-priority jobs are scheduled, matching the intuitive notion of ``priority inversion.'' In \figref{mp-unbounded-pi-global}, $\tau_1$ suffers pi-blocking from time 2 until time~19 since fewer than $c = m=2$ higher-priority jobs are scheduled while $\tau_1$ waits to acquire the resource shared with $\tau_4$ (under global scheduling, any locking-induced suspension constitutes a priority inversion for the top $m$ highest-priority jobs).
Similarly, in \figref{mp-unbounded-pi-part}, $\tau_2$ suffers pi-blocking from time~2 until time~19 since in its cluster (\ie, its assigned core) it is pending, not scheduled, and fewer than $c=1$ higher-priority jobs are scheduled (in fact, none are scheduled at all). 

\subsection{Non-Preemptive Sections} \label{sec:prog:nps}
Several mechanisms (\ie, scheduling rules) have been proposed to ensure a bounded maximum (cumulative) duration of pi-blocking.  The most simple solution is to let tasks spin and make every lock request a \emph{non-preemptive section}: if tasks execute critical sections non-preemptively, then it is simply impossible for a lock holder to be preempted within a critical section. 

\begin{figure}[t]
  \centering
  \includegraphics[width=\textwidth]{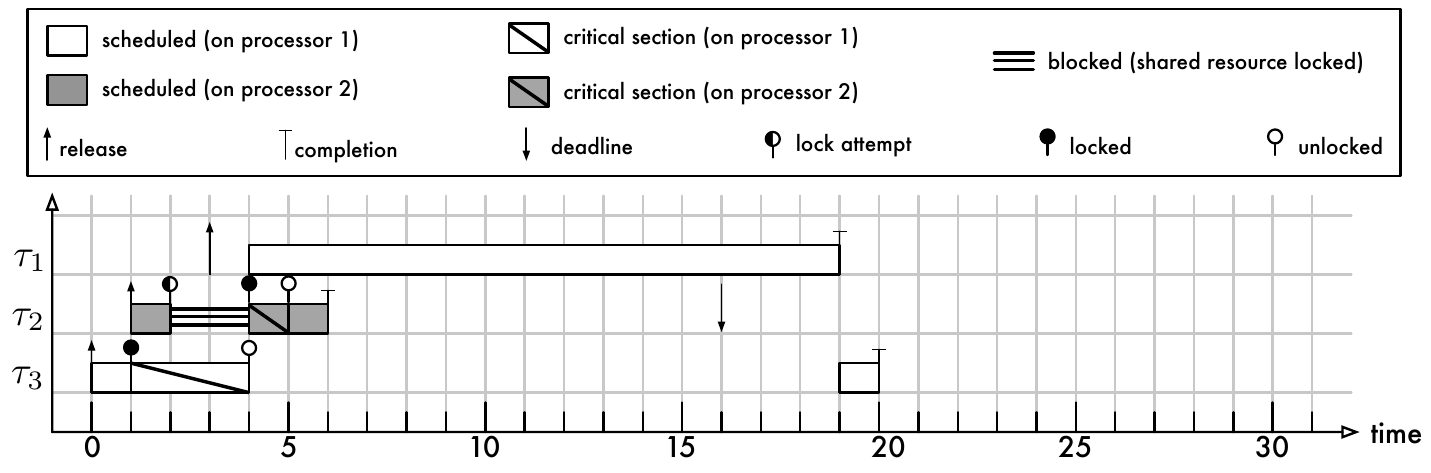}
  \caption{Example P-FP schedule of three tasks on two processors assuming non-preemptive critical sections. }
  \label{fig:mp-np-part}
\end{figure}

\figref{mp-np-part} illustrates how turning critical sections into non-preemptive sections prevents ``unbounded'' pi-blocking in the example scenario previously shown in \figref{mp-unbounded-pi-part}. In \figref{mp-np-part}, because $\tau_3$ cannot be preempted from time~1 until time~4, the preemption due to the arrival of $\tau_1$ is deferred, which ensures that the lock is released in a timely manner, and so $\tau_2$ can meet its deadline at time~16. 

However, the delay now incurred by $\tau_1$ during the interval $[3, 4)$ also constitutes pi-blocking. This highlights an important point: progress mechanisms do not come ``for free.'' Rather, they must strike a balance between the delay incurred by tasks waiting to acquire a resource (\eg, $\tau_2$ in \figref{mp-np-part}) and the delay incurred by higher-priority tasks (\eg, $\tau_1$ in \figref{mp-np-part}) when the completion of critical sections is forced (\ie, when the normal scheduling order is overridden). 

\begin{figure}[t]
  \centering
  \includegraphics[width=\textwidth]{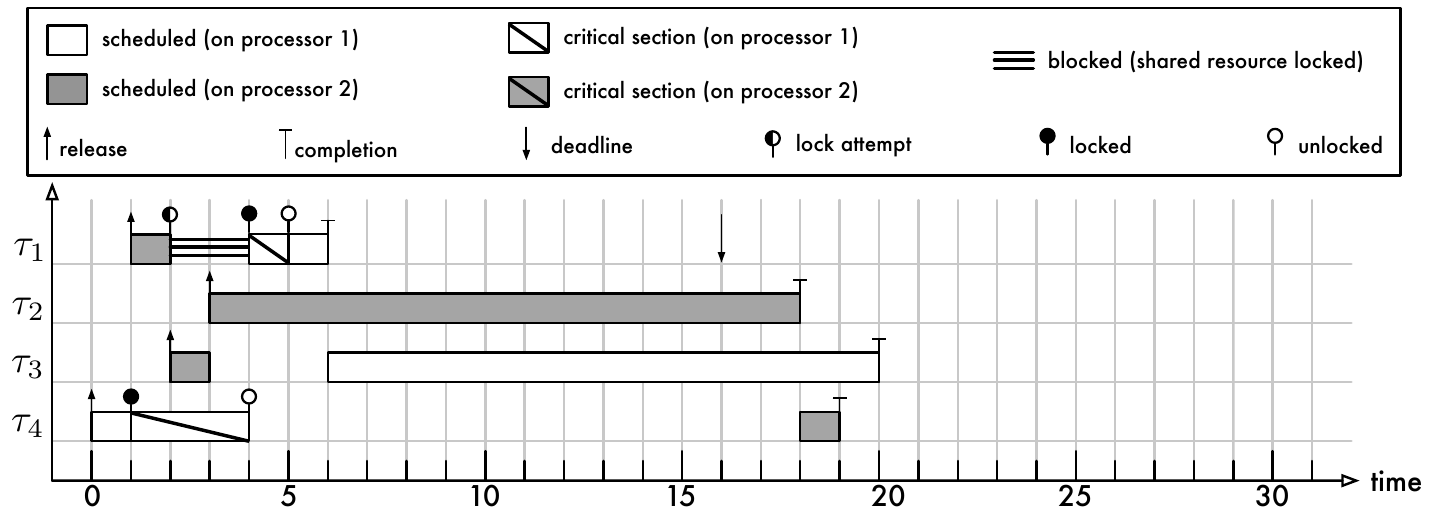}
  \caption{Example G-FP schedule of four tasks on two processors assuming non-preemptive critical sections with \emph{eager} preemptions.}
  \label{fig:mp-np-global-eager}
\end{figure}

Executing lock requests as non-preemptive sections is also effective under clustered scheduling (and hence also under global scheduling). However, there exists a subtlety \wrt how delayed preemptions are realized that does not arise on uniprocessors or under partitioned scheduling. Consider the example schedules in \figrefs{mp-np-global-eager}{mp-np-global-lazy}, which show two possible variants of the scenario previously depicted in \figref{mp-unbounded-pi-global}. In particular, since $\tau_4$ executes its critical section non-preemptively from time~1 to time~4, $\tau_2$ cannot preempt $\tau_4$---the lowest-priority scheduled task---at time~3. However, there \emph{does} exist another lower-priority task that can be preempted at the time, namely $\tau_3$. 
Should $\tau_2$ immediately preempt $\tau_3$ or should it wait until time~4, when $\tau_4$, which intuitively \emph{should} have been the preemption victim, finishes its non-preemptive section? Both interpretations of global scheduling are possible~\cite{BLBA:07,B:11}. The former approach is called \emph{eager preemptions}; the latter conversely \emph{lazy preemptions}~\cite{B:11} or \emph{link-based global scheduling}~\cite{BLBA:07}.

Whereas eager preemptions are easier to implement from an OS point of view, this approach suffers from the disadvantage that a job can  suffer  pi-blocking \emph{repeatedly} due to non-preemptive sections in \emph{any} unrelated lower-priority task  and \emph{at any point} during its execution: in the worst case, a job can be preempted and suffer pi-blocking whenever a higher-priority job is released (such as $\tau_3$ at time~3 in \figref{mp-np-global-eager}), which is difficult to predict and bound accurately. 

In contrast, on a uniprocessor (and under partitioned scheduling), in the absence of self-suspensions, a job suffers pi-blocking due to a lower-priority job's non-preemptive section at most once (\ie, immediately upon its release, or not at all), a property that greatly aids worst-case analysis. 

The lazy preemption approach, aiming to restore this convenient property,  reduces the number of situations in which a job is repeatedly preempted due to a non-preemptive section in a lower-priority job~\cite{BLBA:07,B:11}. While the lazy preemption approach cannot completely eliminate the \emph{occurrence} of repeated preemptions in all situations~\cite{BA:14}, under a common analysis approach---namely, if task execution times are inflated to account for delays due to spinning and priority inversions (discussed in \secrefs{spin}{sob})---it does ensure that pi-blocking due to a non-preemptive section in a lower-priority job has to be \emph{accounted for} only once per job (in the absence of self-suspensions)~\cite{BLBA:07,B:11,BA:14}, analogously to the reasoning in the case of uniprocessors or partitioned scheduling. In other words, lazy preemption semantics ensure analysis conditions that are favorable for inflation-based analysis.

\begin{figure}[t]
  \centering
  \includegraphics[width=\textwidth]{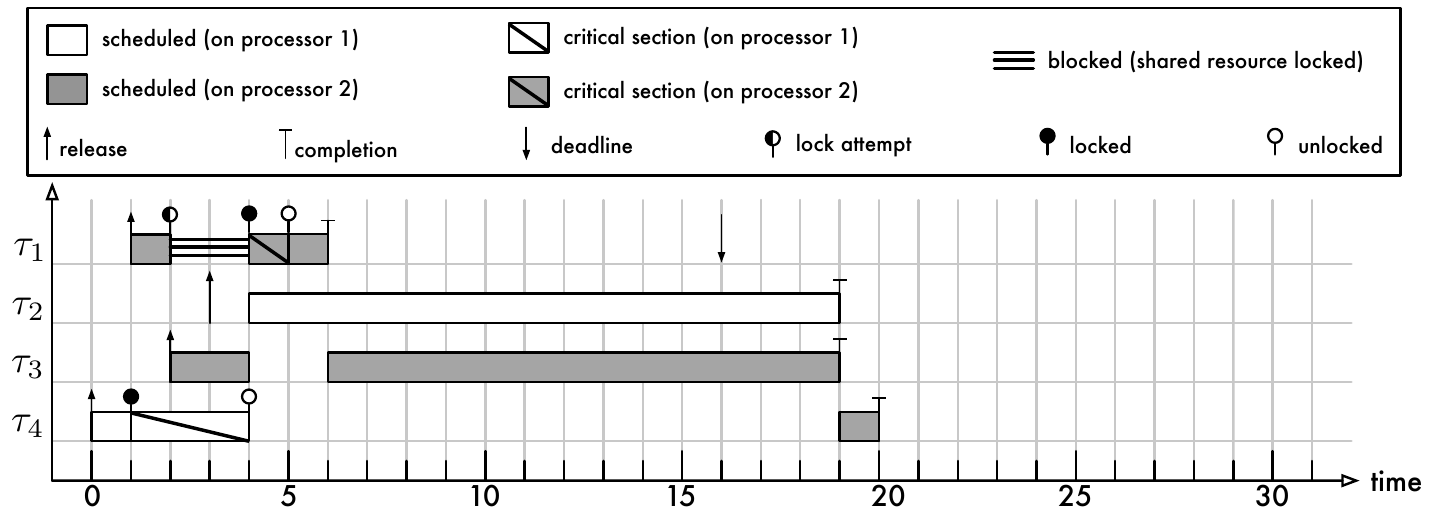}
  \caption{Example G-FP schedule of four tasks on two processors assuming non-preemptive critical sections with \emph{lazy} preemptions (\ie, assuming link-based global scheduling)}
  \label{fig:mp-np-global-lazy}
\end{figure}

Link-based global scheduling~\cite{BLBA:07,B:11}, which realizes lazy preemptions, derives its name from the fact that it establishes a ``link'' between a newly released job and the non-preemptively executing job that it \emph{should} have preempted (if any); the deferred preemption is then enacted as soon as the linked job exits its non-preemptive section, which can be implemented efficiently~\cite{B:11}. Link-based global scheduling has been implemented and evaluated in  \litmus~\cite{B:11,BCBL:08}, a real-time extension of the Linux kernel.\footnote{See \url{http://www.litmus-rt.org}.} 

Non-preemptive execution can be achieved in several ways, depending on the OS and the environment. In a micro-controller setting and within OS kernels, preemptions are typically avoided by disabling interrupts. In UNIX-class RTOSs with a user-mode / kernel-mode divide, where code running in user-mode cannot disable interrupts, non-preemptive execution can be easily emulated by reserving a priority greater than that of any ``regular'' job priority for tasks within critical sections. 

\begin{figure}[t]
  \centering
  \includegraphics[width=\textwidth]{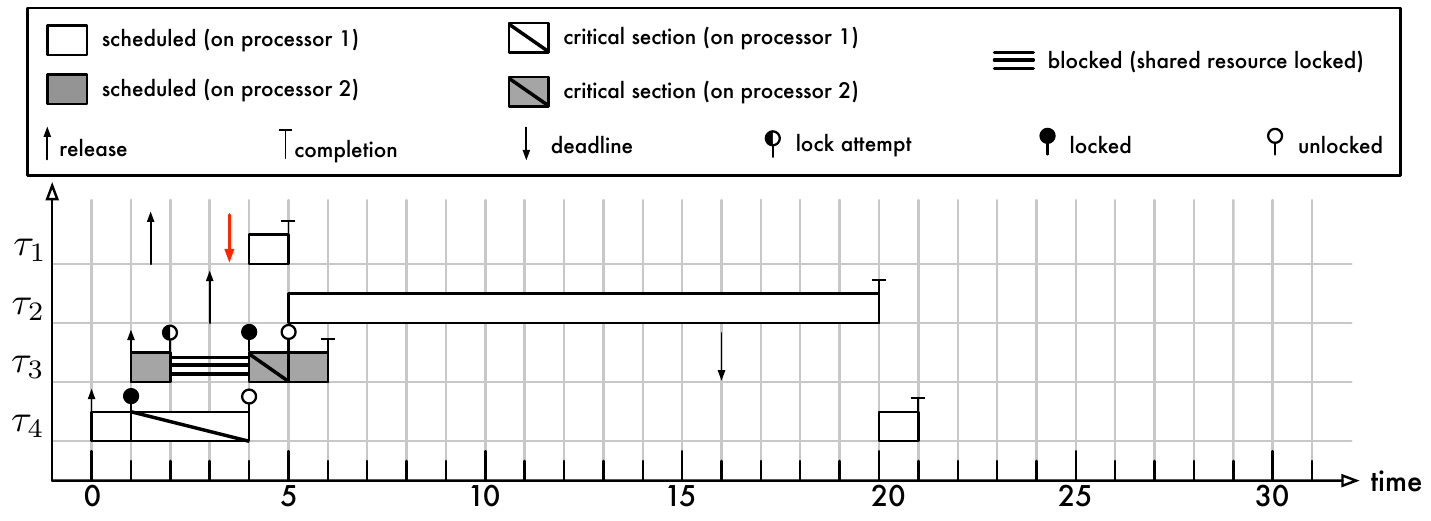}
  \caption{Example P-FP schedule of four tasks, one of which is particularly latency-sensitive, on two processors assuming non-preemptive critical sections.}
  \label{fig:mp-np-part-latency}
\end{figure}

Regardless of how non-preemptive sections are realized, the major drawback of this progress mechanism is that it can result in unacceptable \emph{latency spikes}, either if critical sections are unsuitably long or if (some of the) higher-priority tasks are particularly \emph{latency-sensitive}. For example, consider the scenario illustrated in \figref{mp-np-part-latency}, which is similar to the one depicted in \figref{mp-np-part}, with the exception that another high-priority task with a tight relative deadline of only two time units has been introduced as $\tau_1$ on processor~1. Since this task has very little tolerance for any kind of delay, it is clearly infeasible to just turn $\tau_4$'s request into non-preemptive section since it is ``too long'' relative to $\tau_1$'s latency tolerance, as shown in \figref{mp-np-part-latency}. However, not doing anything is also not a viable option since then $\tau_2$ would transitively cause $\tau_3$ to miss its deadline at time~16, similarly to the scenario shown in \figref{mp-unbounded-pi-part}. 

\subsection{Priority Inheritance} \label{sec:prog:pi}

Since it is hardly a new observation that ``long'' non-preemptive sections are problematic in the presence of tight latency constraints, better solutions have long been known in the uniprocessor case, namely the classic \emph{priority inheritance} and \emph{priority-ceiling} protocols~\cite{SRL:90}. Unfortunately, these approaches do not transfer well to the multiprocessor case, in the sense that they are not always effective \wrt bounding the maximum duration of pi-blocking. 

\begin{figure}[t]
  \centering
  \includegraphics[width=\textwidth]{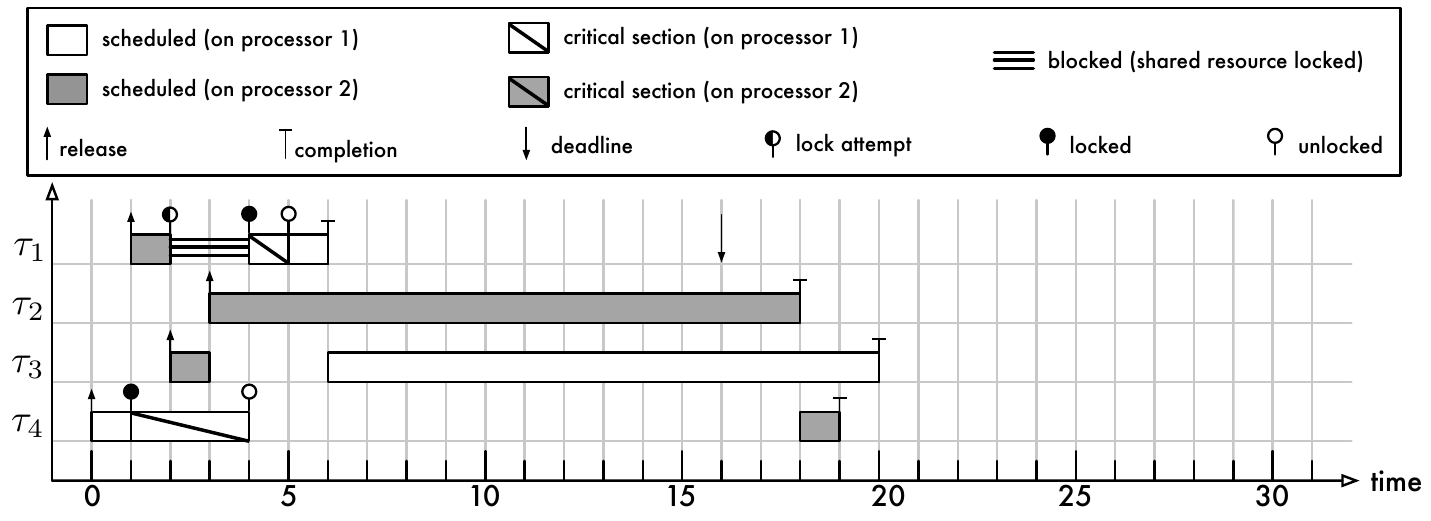}
  \caption{Example G-FP schedule of four tasks on two processors assuming priority inheritance.}
  \label{fig:mp-prio-inh-global}
\end{figure}

Priority inheritance is a good match for global scheduling, and is indeed used in multiprocessor real-time locking protocols for global scheduling (as discussed in \secrefs{sob:global}{saw:global}). Recall that, with priority inheritance, a lock-holding task $\tau_i$'s \emph{effective priority} is the maximum of its own base priority and the effective priorities of all tasks that are waiting to acquire a lock that  $\tau_i$ currently holds~\cite{SRL:90}.
\figref{mp-prio-inh-global}, which shows the same scenario as \figref{mp-unbounded-pi-global}, illustrates how this rule is effective under global scheduling: with the priority inheritance rule in place, $\tau_4$ remains scheduled at time~4 when the higher-priority task $\tau_2$ is released since $\tau_4$  inherits the priority of $\tau_1$, the maximum priority in the system, during the interval $[2, 4)$. As a result, the unrelated task $\tau_3$ is preempted instead, similar to the eager preemption policy in the case of non-preemptive sections as illustrated in \figref{mp-np-global-eager}. (To date, no preemption rule analogous  to the lazy preemption policy discussed in \secref{prog:nps} has been explored in the context of priority inheritance.) Again, this highlights that the progress mechanisms used to mitigate \emph{unbounded} priority inversions are themselves a source of \emph{bounded} priority inversions, which must be carefully taken into account during blocking analysis.

Unfortunately, priority inheritance~\cite{SRL:90} works \emph{only} under global scheduling: it is ineffective (from  the point of view of worst-case blocking analysis) when applied across cores (\resp, clusters) under partitioned (\resp, clustered) scheduling. The reason can be easily seen in \figref{mp-unbounded-pi-part}: even though the priority inheritance rule is applied across cores, $\tau_3$'s effective priority is merely raised to that of $\tau_2$, which does not prevent the preemption by $\tau_1$ at time~3 (recall that tasks are indexed in order of strictly decreasing priority). For the same reason, classic ceiling-based  protocols like the PCP~\cite{SRL:90} and the SRP~\cite{B:91} are also ineffective: given that $\tau_1$ does not access the shared resource, the ceiling priority of the shared resource is lower than $\tau_1$'s priority. Fundamentally, the root cause is that numeric priority values are, analytically speaking, \textit{incomparable} across processor (\resp, cluster) boundaries since partitions (\resp, clusters) are scheduled independently.

\subsection{Allocation Inheritance}\label{sec:prog:ai}
The solution to this problem is an idea that has appeared several times in different contexts and under various names: \emph{spinning processor executes for preempted processors} (SPEPP)~\cite{TS:97}, \emph{local helping}~\cite{HP:01,HH:01}, \emph{allocation inheritance}~\cite{HA:02a,H:04,HA:06}, \emph{multiprocessor bandwidth inheritance}~\cite{FLC:10,FLC:12}, and \emph{migratory priority inheritance}~\cite{BB:12,B:13}. The essential common insight is that a preempted task's critical section should be completed using the processing capacity of cores on which the blocked tasks would be allowed to run (if they were not blocked). 

That is, a blocking task should not only inherit a blocked task's priority, but also the ``right to execute'' on a particular core, which serves to restore analytical meaning to the inherited priority: to obtain a progress guarantee, a preempted lock holder must \emph{migrate} to the core where the pi-blocking is incurred. Intuitively, \defref{pi} implies that, if a task $\tau_i$ incurs pi-blocking, then its priority is sufficiently high to ensure that a lock holder inheriting $\tau_i$'s priority can be scheduled on $\tau_i$'s processor (or in $\tau_i$'s cluster), since the fact that $\tau_i$ incurs pi-blocking indicates the absence of runnable higher-priority jobs of tasks assigned to $\tau_i$'s processor or cluster  (recall Clause~3 in \defref{pi}).

\begin{figure}[t]
  \centering
  \includegraphics[width=\textwidth]{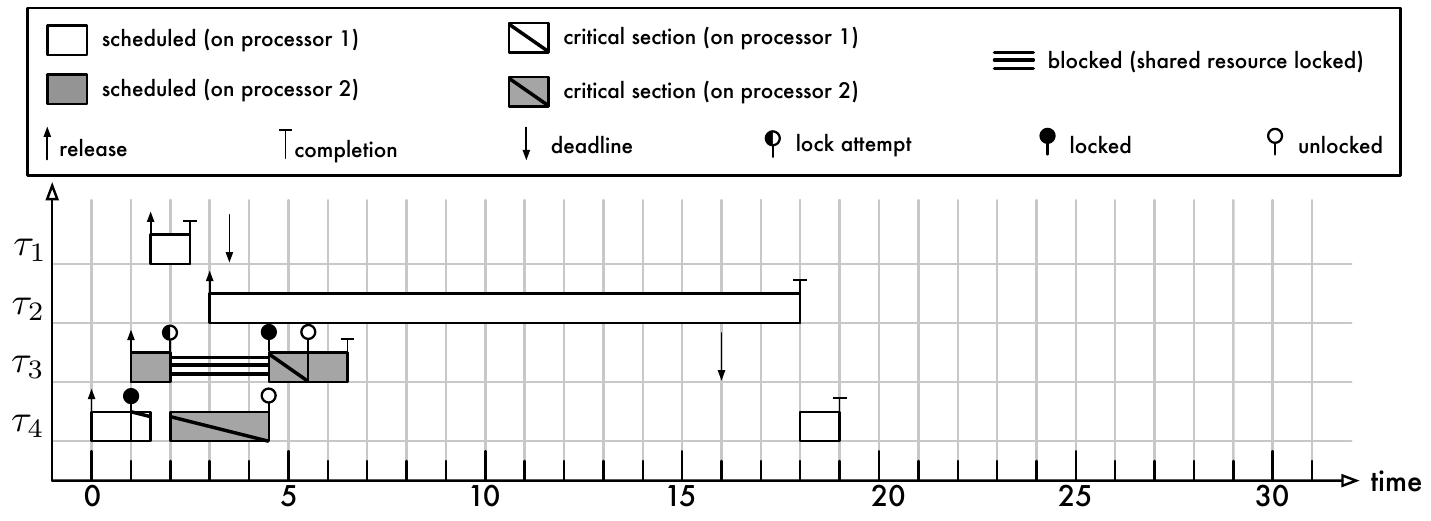}
  \caption{Example P-FP schedule of four tasks, one of which is particularly latency-sensitive, on two processors assuming migratory priority inheritance.}
  \label{fig:mp-prio-inh-part}
\end{figure}

An example of this approach is shown in \figref{mp-prio-inh-part}, which shows the same scenario involving a latency-sensitive task  previously shown in \figref{mp-np-part-latency}. In contrast to the example in \figref{mp-np-part-latency}, at time~1.5, when the latency-sensitive task $\tau_1$ is activated, the lock-holding task is preempted (just as it would be in the case of priority inheritance). However, when $\tau_3$ blocks on the resource held by $\tau_4$ at time~3, $\tau_4$ inherits the right to use the priority of $\tau_3$ \emph{on $\tau_3$'s assigned processor} (which is processor~2, whereas $\tau_4$ is assigned to processor~1). Consequently, $\tau_4$ migrates from processor~1 to processor~2  to continue its critical section. When $\tau_4$ finishes its critical section at time~4.5, it ceases to inherit the priority and right to execute on processor~2 from $\tau_3$ and thus cannot continue to execute on processor~2. Task $\tau_4$ hence migrates back to processor~1 to continue its execution at time~18 when $\tau_2$ completes. Overall, $\tau_1$ suffers no latency penalty when it is released at time~1.5, but task $\tau_3$ also suffers no undue delays while waiting for $\tau_4$ to release the shared resource.

As already mentioned, several different names have been used in the past to describe progress mechanisms based on this principle. We adopt the term ``allocation inheritance''~\cite{HA:02a,H:04,HA:06} since it clearly describes the idea that processor time originally allocated to a blocked task is used towards the completion of a blocking task's critical section.  The name also highlights the fact that this approach is a generalization of the classic priority inheritance idea. In fact, under event-driven global scheduling and on uniprocessors (the two cases where priority inheritance is effective), allocation inheritance in fact reduces to priority inheritance since all tasks are eligible to execute on all cores anyway.

From a  purely analytical point of view, allocation inheritance is elegant and highly attractive: as evident in \figref{mp-prio-inh-part}, it has no negative latency impact on unrelated higher-priority tasks, while ensuring guaranteed progress, thus restoring the strong analytical foundation offered by priority inheritance on uniprocessors. Put differently, allocation inheritance is the natural multiprocessor extension of classic priority inheritance that allows the idea to work under \emph{any} multiprocessor scheduling approach.

However, from a practical point of view, it can be difficult to support allocation inheritance efficiently: either it introduces task migrations (and the associated kernel complexities and cache overheads) into partitioned systems that otherwise would need none, or there must be some other, resource-specific way for remote processors to continue (or safely duplicate) the operation that the preempted task was trying to accomplish~\cite{TS:97,BW:13a}, which can be difficult (or even impossible) to achieve for certain kinds of resources (\eg, hardware resources such as I/O ports). 
In particular, if the latter approach is feasible (\ie, helping operations to complete without requiring a complete task migration), it is usually also possible to implement completely lock- or even wait-free solutions (\eg,~\cite{AJR:97,R:97}), which can be an overall preferable solution in such cases~\cite{BCBL:08}. 
We discuss allocation inheritance, protocols built on top of it, and practical implementations in \secrefs{ip}{impl:ai}.

\subsection{Priority Boosting} \label{sec:prog:boost}
The most commonly used progress mechanism is \emph{priority boosting}, which is conceptually quite similar to non-preemptive sections and also a much older idea than allocation inheritance. Simply put, priority boosting requires that each critical section (pertaining to a global resource) is executed at a \emph{boosted priority} that exceeds the maximum regular (\ie, non-boosted) scheduling priority of any task. As a result, newly released jobs, which do not yet hold any resources, cannot preempt critical sections, just as with non-preemptive sections. In fact, applying priority boosting to the examples shown in \figrefs{mp-unbounded-pi-global}{mp-unbounded-pi-part} would yield exactly the same schedules as shown in \figrefs{mp-np-global-eager}{mp-np-part}, \resp.  However, in contrast to non-preemptive sections, tasks remain preemptive in principle, and since other critical sections may be executed with even higher boosted priorities, it is possible that a task executing a critical section may be preempted by another task also executing a critical section (pertaining to a different resource).  In essence, priority boosting establishes a second priority band on top of regular task priorities that is reserved for lock-holding tasks.

Priority boosting is easy to support in an RTOS, and easy to emulate in user-mode frameworks and applications if not explicitly supported by the RTOS. It can also be considered the ``original'' progress mechanism, as its use (in uniprocessor contexts) was already suggested by multiple authors in 1980~\cite{L:80,LR:80}, and because the two first multiprocessor real-time locking protocols backed by analysis, the DPCP~\cite{RSL:88} and the MPCP~\cite{R:90}, rely on it.  However, while it is conveniently simple, priority boosting also comes with major latency penalties similar to non-preemptive sections, which limits its applicability in systems with tight latency constraints.

\subsection{Restricted Priority Boosting}\label{sec:prog:rb}
Priority boosting  as described so far, and as used in the DPCP~\cite{RSL:88}, MPCP~\cite{R:90}, and many other protocols, is \emph{unrestricted}, in the sense that it applies to all tasks and critical sections alike, regardless of whether or not a lock-holding task is actually causing some other task to incur pi-blocking. In contrast, both priority and allocation inheritance kick in only \emph{reactively}~\cite{SVBD:14}, when contention leading to pi-blocking is actually encountered at runtime. 

This unrestricted nature of priority boosting can be problematic from an analytical point of view since it can result in a substantial amount of unnecessary pi-blocking. To overcome this limitation, several restricted (\ie,  selectively applied) versions of priority boosting have been derived in work on locking protocols that ensure asymptotically optimal pi-blocking bounds, such as \emph{priority donation}~\cite{BA:11,B:11,BA:13}, \emph{restricted segment boosting}~\cite{B:14b}, and \emph{replica-request priority donation}~\cite{WEA:12}. We will discuss these more sophisticated progress mechanisms in the context of the protocols in which they were first used in \secrefx{sob}{saw}{kex}.

\subsection{Priority Raising} 
As a final consideration, one can pragmatically devise a rule to the effect that critical sections are executed unconditionally at an elevated priority that, in contrast to priority boosting, is \emph{not} necessarily higher than the maximum regular scheduling priority, but still higher than \emph{most} regular scheduling priorities. The intended effect is that tasks with ``large'' WCETs cannot preempt critical sections, whereas lightweight latency-sensitive tasks (\eg, critical interrupt handlers) with minuscule WCETs are still permitted to preempt critical sections without suffering any latency impact.

\begin{figure}[t]
  \centering
  \includegraphics[width=\textwidth]{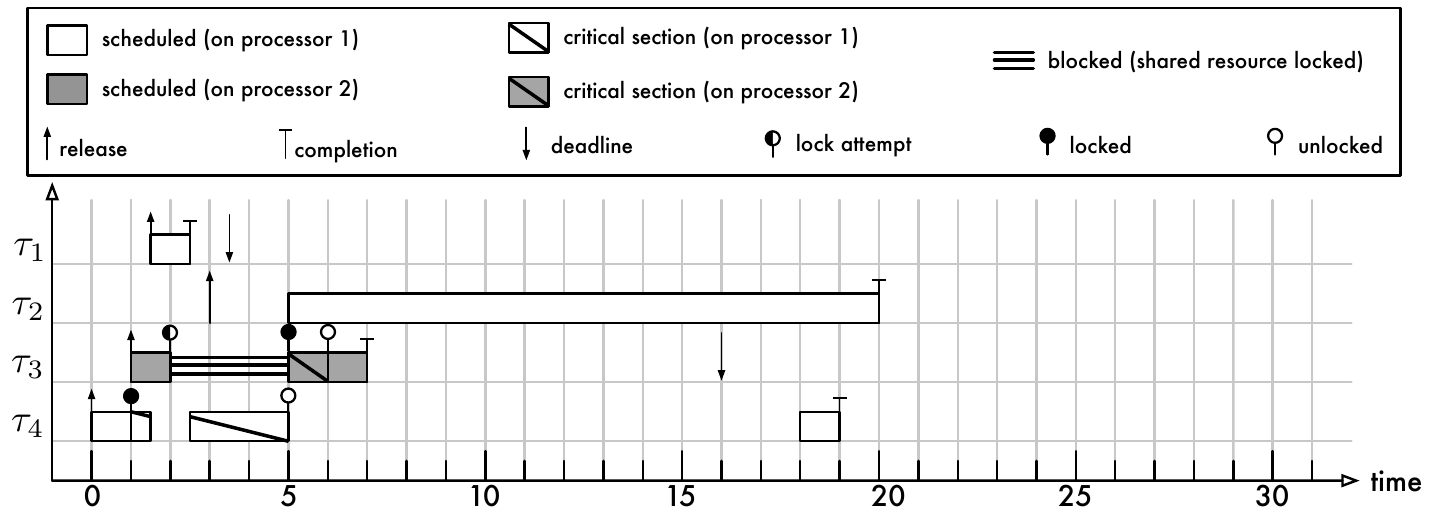}
  \caption{Example P-FP schedule of four tasks, one of which is particularly latency-sensitive, on two processors assuming the critical section priority is raised to an intermediate level in between the priorities of tasks $\tau_1$ and $\tau_2$}
  \label{fig:mp-prio-raise-part}
\end{figure}

For example, consider the schedule shown in \figref{mp-prio-raise-part}, which shows the same scenario previously depicted in \figrefs{mp-np-part-latency}{mp-prio-inh-part}. Suppose the critical section priority for the resource shared by $\tau_3$ and $\tau_4$ is chosen to be higher than the priority $\tau_2$, but below the priority of $\tau_1$. As a result, when $\tau_1$ is activated at time~1.5, it simply preempts the in-progress critical section of $\tau_4$, which transitively causes some pi-blocking for $\tau_3$. However, this extra delay is small relative to the critical section length of $\tau_4$ and the execution requirement of $\tau_3$.  Importantly, when $\tau_2$ is released at time~5, it is \emph{not} able to preempt the in-progress critical section of $\tau_4$, which ensures that $\tau_3$ does not suffer an excessive amount of pi-blocking. Again, this causes some pi-blocking to $\tau_2$, which however is minor relative to its own WCET. 

To re-iterate, this pragmatic scheme avoids both the need to migrate tasks (\ie, allocation inheritance) and the latency degradation due to non-preemptive sections and priority boosting by raising the priority of \emph{short} critical sections so that it exceeds the priorities of all tasks with ``substantial'' WCETs, while also keeping it below that of latency-sensitive tasks. Since latency-sensitive tasks must have relatively high priorities anyway, and since latency-sensitive tasks usually do not have large WCETs in practice, this scheme has the potential to work for a wide range of workloads. However, it violates a strict interpretation of the common notion of ``bounded priority inversions''  since now the pi-blocking bounds of certain remote tasks depend on the WCETs of high-priority latency-sensitive tasks (\eg, in \figref{mp-prio-raise-part}, the pi-blocking bound of $\tau_3$ on Processor~2 depends on the WCET of $\tau_1$ on Processor~1).  The ``priority raising'' approach is thus not frequently considered in the academic literature. %

This concludes our discussion of progress mechanisms. We next start our review of multiprocessor locking protocols and begin with spin-lock protocols, as they are conceptually simpler and easier to analyze than semaphore protocols. 

\section{Spin-Lock Protocols}\label{sec:spin}

The distinguishing characteristic of a spin lock is that a waiting task does not \emph{voluntarily} yield its processor to other tasks. That is, in contrast to suspension-based locks, spin locks do not cause \emph{additional} context switches.\footnote{Based on this definition, we do not consider approaches such as ``virtual spinning''~\cite{LNR:09} (discussed in \secref{saw}) to constitute proper spin locks, precisely because under ``virtual spinning'' tasks still self-suspend upon encountering contention.} However, the use of spin locks does not necessarily imply the absence of preemptions altogether; \emph{preemptable} spin locks permit regular preemptions (and thus context switches) as required by the scheduler during certain phases of the protocol. Such preemptions, however, do not constitute \emph{voluntary} context switches, and a task that is preempted while spinning does not suspend; rather, it remains ready in the scheduler's run-queue. 

All protocols considered in this section force the uninterrupted execution of critical sections, by means of either non-preemptive execution or priority boosting, which simplifies both the analysis and the implementation. 
There also exist spin-based protocols under which tasks remain preemptable at all times (\ie, that allow tasks to be preempted even within critical sections); such protocols usually require allocation inheritance and are discussed in \secref{ip}.

In the analysis of spin locks, it is important to clearly distinguish
between two different cases of ``blocking,'' as illustrated in \figref{spin-s-vs-pi}.
First, since jobs execute the protocol in part or in whole
non-preemptively, higher-priority jobs that are released while a
lower-priority job is executing a spin-lock protocol can be delayed.
As already discussed in \secref{prog:nps}, this is a
classic priority inversion due to non-preemptive execution as covered
by \defref{pi}; recall that this kind of delay is called
pi-blocking.

\defref{pi}, however, does not cover the delay that jobs incur
while spinning because it applies only to jobs that are \emph{not} scheduled,
whereas spinning jobs are being delayed despite being scheduled. To clearly distinguish
this kind of delay from pi-blocking, we refer to it as \emph{spin blocking} (\emph{s-blocking})
\cite{B:11}. Note that, if jobs spin non-preemptably, then the s-blocking
incurred by a lower-priority job can \emph{transitively} manifest as
pi-blocking experienced by a local higher-priority job. A sound blocking
analysis must account for both types of
blocking.

\incfig{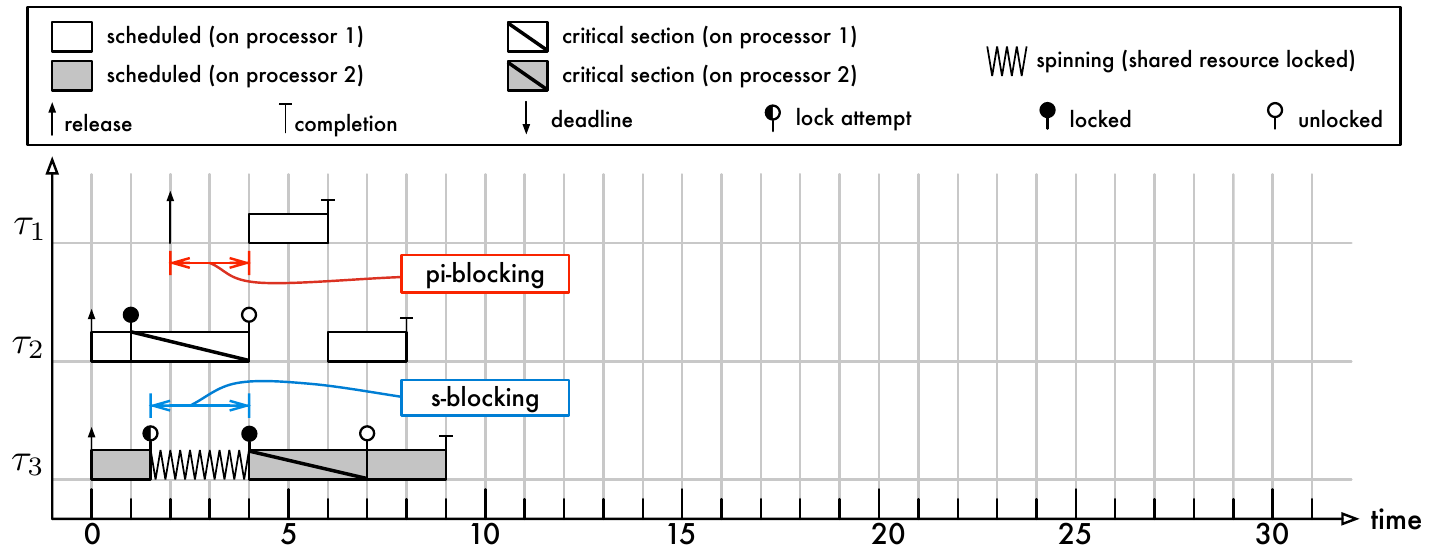}{Example P-FP schedule of three tasks on two processors illustrating the difference between s-blocking (\ie, delays due to spinning) and pi-blocking (\ie, delays due to priority inversion).}

\subsection{Spin-Lock Protocols for Partitioned Scheduling} \label{sec:spin:part}

The first spin-lock protocol backed by a sound blocking and
schedulability analysis is due to \citeauthor{GLD:01}~\cite{GLD:01}, who presented
the now-classic \emph{Multiprocessor Stack Resource Policy} (MSRP) in
2001. While a number of authors had previously explored real-time and
analysis-friendly spin-lock \emph{implementations} (which we discuss in \secref{impl:spin}),
\citeauthor{GLD:01} were the first to leverage a particular class of spin-lock implementations,
namely \emph{non-preemptive FIFO spin locks}, to arrive at an analytical
sound \emph{protocol} (\ie, a set of rules that determine how such
locks should be used in a real-time system) and corresponding blocking
and schedulability analyses.

The MSRP is a protocol for partitioned scheduling and can be used with
both FP and EDF scheduling (on each core). Like earlier suspension-based
protocols for partitioned scheduling~\cite{RSL:88,R:90,R:91}, the MSRP distinguishes
between local and global resources (recall \secref{model}). Local resources are
dealt with by means of the classic SRP \cite{B:91} and are of no concern here.

Global
resources are protected with non-preemptive FIFO spin locks.
When a task seeks to acquire a shared resource, it first becomes
non-preemptable (which resolves any local contention) and then executes
a spin-lock algorithm that ensures that conflicting requests by tasks on
different processors are served in FIFO order. The exact choice of FIFO spin-lock algorithm
is irrelevant from a predictability point of view; many
suitable choices with varying implementation tradeoffs exist (see \secref{impl:spin}).

\figref{spin-msrp} depicts an example MSRP schedule, where 
$\tau_2$'s request is satisfied after $\tau_4$'s request because $\tau_4$ 
issued its request slightly earlier. The highest-priority task $\tau_1$
incurs considerable pi-blocking upon its release at time~2 since the local lower-priority task $\tau_2$ 
is spinning non-preemptively. Here, $\tau_1$ is transitively blocked by $\tau_3$ and $\tau_4$'s critical
sections while $\tau_2$ incurs s-blocking.

The fact that tasks remain continuously non-preemptable both while
busy-waiting and when executing their critical sections has several
major implications. In conjunction with the FIFO
wait queue order, it ensures that any task gains access to a shared resource after
being blocked by at most $m-1$ critical sections, where $m$ in this
context denotes the number of processors on which tasks sharing
a given resource reside. This property immediately ensures starvation freedom and
provides a strong progress guarantee that allows for a simple
blocking analysis \cite{GLD:01}. However, as apparent in \figref{spin-msrp}, it also causes
$O(m)$ latency spikes, which can be highly undesirable in hard real-time systems
\cite{C:93,TS:94,TS:96,T:96,TS:97,B:13} and which implies that the MSRP is
suitable only if it can be guaranteed that critical sections are
\emph{short} relative to latency expectations, especially on platforms with large core counts.

From an implementation point of view, simply making the entire locking
protocol non-preemptive is attractive because it completely decouples the scheduler
implementation from locking protocol details, which reduces
implementation complexity,  helps with maintainability, and lessens overhead concerns. In
fact, of all protocols discussed in this survey, the MSRP
imposes the least implementation burden and is the easiest to integrate
into an OS.

\incfig{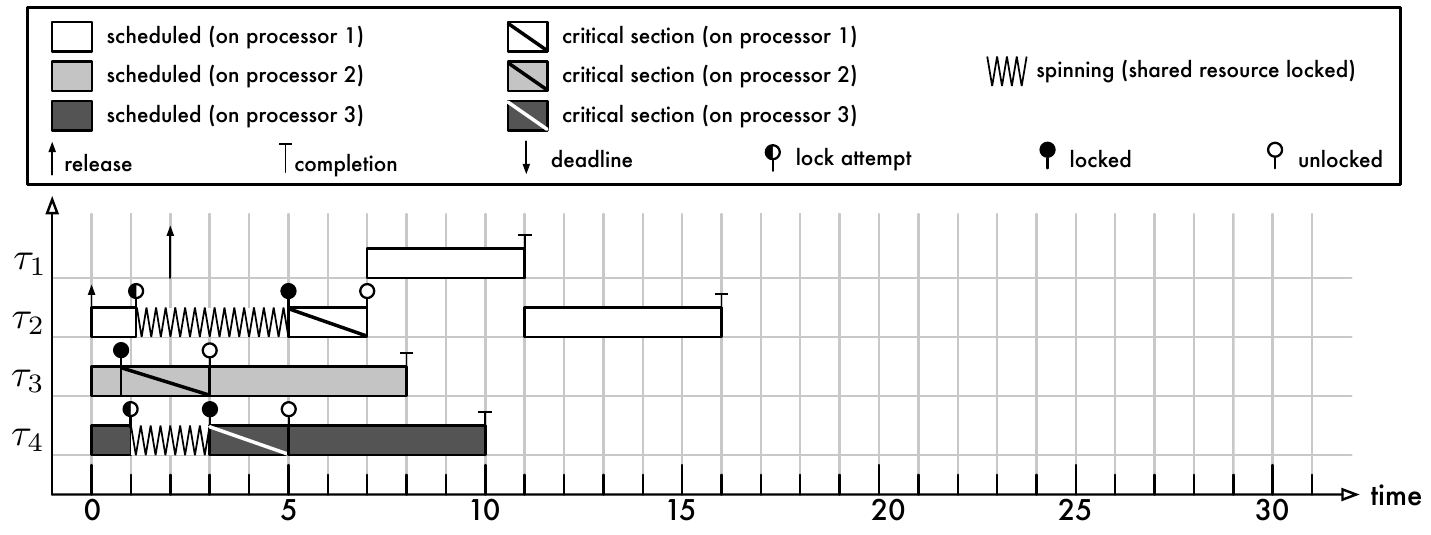}{Example P-FP schedule of four tasks on three processors sharing a global resource according to the rules of the MSRP~\cite{GLD:01}}

\citeauthor{GLD:01} presented a simple blocking analysis of the MSRP \cite{GLD:01},
which relies on \emph{execution-time inflation} and bounds pi-blocking
and s-blocking in a bottom-up fashion. First, \citeauthor{GLD:01}'s analysis
considers each critical section in isolation and determines the maximum
s-blocking that may be incurred due to the specific, single critical section,
which is determined by  the sum of the  maximum critical section lengths
(pertaining to the resource under consideration) on all other
processors. A task's cumulative s-blocking is then simply bounded by the sum of
the per-request s-blocking bounds. Finally, a task's \emph{inflated
WCET} is the sum of its cumulative s-blocking bound and its original (\ie,
non-inflated) WCET. Existing schedulability analysis can then be applied
to the inflated WCET bounds.

While \citeauthor{GLD:01}'s original analysis \cite{GLD:01} is simple and convenient,
it comes with substantial structural pessimism.
In particular, a major source of pessimism is that the maximum critical section
lengths (on remote cores) are over-represented in the final bound. Effectively,
\emph{every} blocking remote critical section is considered to exhibit the length
of the longest critical section, which is clearly pessimistic for shared
resources that support multiple operations of various cost.
For example, consider a producer-consumer scenario
where multiple tasks share a sorted list of work items ordered by urgency that may include up to 100 elements,
and suppose the list 
is accessed by two types of critical sections: \eone removal of the first element and \etwo an order-preserving insertion.
When performing a blocking analysis, it is clearly undesirable to
account for each $O(1)$ dequeue operation as an $O(n)$ enqueue
operation. 

To lessen the impact of this source of pessimism, a
\emph{holistic blocking analysis} of the MSRP was developed~\cite{B:11}.
A holistic analysis considers 
all critical sections of a task together and directly derives
an overall per-task s-blocking bound, rather than attributing s-blocking
to individual critical sections and then summing up the individual
bounds. 
Holistic blocking analysis effective in avoiding
the over-counting of long critical
sections in the analysis of a single task.
However, because it still proceeds on a task-by-task basis and relies on inflated
execution times to account for transitive s-blocking,
the holistic approach \cite{B:11} still over-estimates the impact of
long critical sections if multiple local tasks access the same global resource,
because in such a case the longest critical section is reflected in multiple inflated WCETs.
In fact, \citeauthor{WB:13a} showed
that \emph{any} blocking analysis that relies on execution-time
inflation is asymptotically sub-optimal due to this fundamental structural
pessimism~\cite{WB:13a}.

To avoid such structural pessimism altogether, \citeauthor{WB:13a} developed a novel
MILP-based blocking analysis for spin-lock protocols under P-FP scheduling that by design
prevents any critical sections from being accounted for more than
once in the final aggregate pi- and s-blocking bound \cite{WB:13a}.
Crucially, \citeauthor{WB:13a}'s analysis does not rely on execution time
inflation to implicitly account for transitive s-blocking delays (\ie,
lower-priority tasks being implicitly delayed when local higher-priority
tasks spin). Rather, it explicitly models both transitive and direct
s-blocking, as well as pi-blocking, and directly derives a joint bound on all three kinds of delay
by solving a MILP optimization problem~\cite{WB:13a}.   
\citeauthor{BB:16} \cite{BB:16} recently provided an equivalent analysis for P-EDF scheduling.

Overall, if critical sections are relatively short and contention does not
represent a significant schedulability bottleneck, then \citeauthor{GLD:01}'s
original analysis \cite{GLD:01} or the holistic approach \cite{B:11} are sufficient.
However, in systems with many critical sections, high-frequency tasks,
or highly heterogenous critical section lengths, \citeauthor{WB:13a} and \citeauthor{BB:16}'s
LP-based approaches are substantially more accurate \cite{WB:13a,BB:16}. In his thesis,
\citeauthor{W:18} discusses further optimizations, including a
significant reduction in the number of variables and how to convert the
MILP-based analysis into an LP-based approach without loss of accuracy~\cite{W:18}.

\subsubsection{Non-FIFO Spin Locks}

While FIFO-ordered wait queues offer many advantages---chiefly among them
starvation freedom, ease of analysis, and ease of implementation---there
exist situations in which FIFO ordering is not appropriate or not
available.

For instance, if space overheads are a pressing concern (\eg, if there
are thousands of locks), or in systems with restrictive legacy code constraints, it
may be necessary to resort to \emph{unordered} spin locks such as
basic \emph{test-and-set} (TAS) locks, which can be realized with a single
bit per lock. In other contexts, for example given workloads
with highly heterogenous timing requirements, it can be desirable to use
\emph{priority-ordered} spin locks, to let urgent tasks acquire
contended locks more quickly.

Implementation-wise, the MSRP design---SRP for local resources, non-preemptive spin locks for
global resources---is trivially compatible with either unordered or
priority-ordered locks (instead of FIFO-ordered spin locks). The resulting
analysis problem, however, is far from trivial and does not admit a
simple per-resource approach as followed in \citeauthor{GLD:01}'s original
analysis of the MSRP \cite{GLD:01}. This is because individual critical
sections are, in the worst case, subject to starvation effects, 
which can result in prolonged s-blocking.
Specifically, in priority-ordered locks, a continuous stream of
higher-priority critical sections can delay a lower-priority critical
section indefinitely, and in unordered locks, any request may starve indefinitely
as long as there is contention.

The lack of a strong per-request progress guarantee fundamentally necessitates a ``big
picture'' view as taken in the holistic approach \cite{B:11} to bound cumulative s-blocking
 across all of a job's critical sections. Suitable analyses for priority-ordered
spin locks were proposed by \citeauthor{NE:12} \cite{NE:12} and \citeauthor{WB:13a}~\cite{WB:13a}.

Additionally, \citeauthor{WB:13a} also proposed the first analysis applicable
to unordered spin locks~\cite{WB:13a}. While unordered locks are traditionally 
considered to be ``unanalyzable'' and thus unsuitable for real-time systems,
\citeauthor{WB:13a} made the observation that unordered
spin locks are analytically equivalent to priority-ordered spin locks if
each task is analyzed assuming that all local tasks issue requests with the lowest-possible
priority whereas all remote tasks issue high-priority requests, which maximizes the starvation potential.

\subsubsection{Preemptable Spinning}\label{sec:spin:pre}

The most severe drawback of non-preemptive FIFO spin locks (and hence
the MSRP) is their $O(m)$ latency impact
\cite{C:93,TS:94,TS:96,T:96,TS:97,B:13}. One approach to lessen this latency
penalty without giving up too much desirable
simplicity is to allow tasks to remain preemptable while busy-waiting,
as illustrated in \figref{spin-preemptible}, while still requiring critical sections to be
executed non-preemptively (to avoid the lock-holder preemption problem).
This has the benefit that preemptions are delayed by at most one critical section---the latency impact is reduced to
$O(1)$---which greatly improves worst-case scalability~\cite{C:93,TS:94,TS:96,T:96,TS:97}.

\incfig{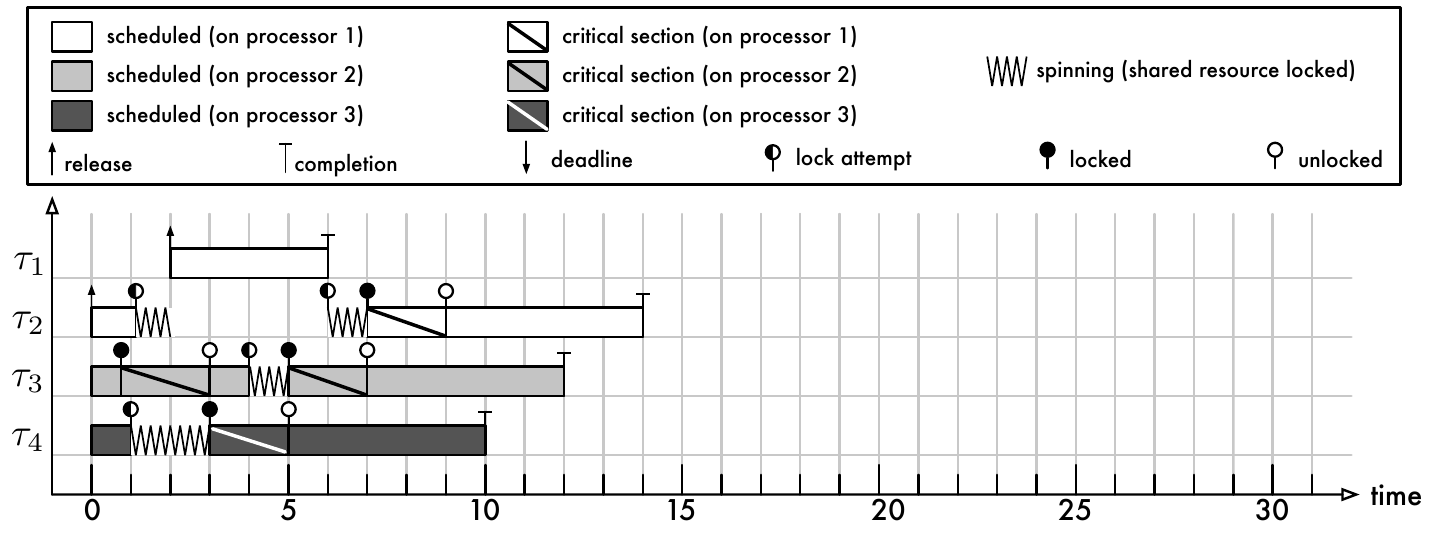}{Example P-FP schedule of four tasks on three processors illustrating preemptable spinning with request cancelation.}

Preemptable spinning poses two major challenges.
The first challenge is the
 implementation: while preemptable spinning can be easily integrated into
unordered TAS locks, it requires substantially more sophisticated
algorithms to realize FIFO- or priority-ordered spin locks that support
preemptable spinning~\cite{C:93,TS:94,TS:97,AJJ:98,HA:02a}. The reason is that a
job that is preempted while busy-waiting  must be removed from the spin queue, or marked as
preempted and skipped, to avoid granting the lock to a currently preempted job
(and hence re-introducing the lock-holder preemption problem).

As a result, preemptable spinning poses analytical problems, which
is the second major challenge. When a job continues execution after
having been preempted, it may find itself dequeued from the spin
queue---that is, after a preemption, a job may find that its lock request was
\emph{canceled} while it was preempted---to the effect that it must
\emph{re-issue} its canceled lock request, which carries the risk of
encountering additional contention. Furthermore, in the worst case, the
job may be preempted again while waiting for its re-issued lock request to be satisfied,
which will necessitate the request to be re-issued once more, and
so on. 

The additional delays that arise from the cancellation of
preempted requests hence cannot be analyzed on a per-request basis and inherently require
a holistic analysis approach that avoids execution-time inflation.
Appropriate blocking analysis for preemptable FIFO- and priority-ordered spin locks,
as well as preemptable unordered spin locks, was first proposed by \citeauthor{WB:13a}
for P-FP scheduling~\cite{WB:13a}, and recently extended to P-EDF scheduling by \citeauthor{BB:16}~\cite{BB:16}.
\citeauthor{ADJM:14}~\cite{ADJM:14} also proposed a protocol based on FIFO spin
locks and preemptable spinning.

In systems with quantum-driven schedulers, where the
scheduler preempts jobs only at well-known times, and not in reaction to arbitrarily timed events,
the cancellation penalty is reduced because no request must be re-issued more than once if critical sections are reasonably short~\cite{AJJ:98}.
As a result, it is possible~\cite{AJJ:98} to apply execution-time inflation approaches similar to the original MSRP analysis~\cite{GLD:01}.
Nonetheless, since every job release causes at most one cancellation (in the absence of self-suspensions)~\cite{WB:13a}, 
and since critical sections typically outnumber higher-priority job releases, it is conceptually 
still preferable to apply more modern, inflation-free analyses~\cite{WB:13a,BB:16} even in quantum-driven systems.

\subsubsection{Spin-Lock Protocols based on Priority Boosting}

Exploring a different direction,
\citeauthor{A:16} \cite{A:16} introduced a variation of the
preemptable spinning approach in which jobs \emph{initially} remain preemptable even when
executing critical sections, but become priority-boosted as soon as contention is encountered.
\citeauthor{A:16}'s protocol is based on the following observation:
if non-preemptive execution is applied
\emph{unconditionally} (as in the MSRP \cite{GLD:01}), then lock-holding jobs
cannot be preempted and thus cause a latency impact even if there is
actually no contention for the lock (\ie, even if there is no lock-holder
preemption problem to mitigate). Since in most systems lock contention is rare,
the latency impact of
non-preemptive execution, though unavoidable from a worst-case
perspective, may be undesirable from an average-case perspective.

As an alternative, \citeauthor{A:16} thus proposed the \emph{Forced Execution
Protocol} (FEP), which uses \emph{on-demand} priority boosting (instead
of unconditional non-preemptive execution) to reactively expedite the completion of
critical sections only when remote jobs actually request a locked
resource and start to spin. This design choice has the benefit of
reducing the \emph{average} priority inversion duration (\ie, the
average latency impact), but it comes at a high price, namely increased
worst-case pi-blocking, because high-priority jobs may be delayed by
multiple preempted critical sections. This effect is illustrated in \figref{spin-fep},
which depicts a scenario in which a single job of a higher-priority task (\eg, $\tau_1$ in
\figref{spin-fep}) may suffer pi-blocking due to \emph{multiple}
lower-priority critical sections  since  the FEP \cite{A:16} allows incomplete critical sections to be
preempted. Thus, if a remote task
forces the completion of multiple preempted critical sections  (\eg, $\tau_5$ in \figref{spin-fep}),
then a higher-priority task may be repeatedly delayed  due to priority boosting.
In contrast, the use of  non-preemptive execution in the MSRP \cite{GLD:01}
ensures that, on each processor and at any point in time, at most one
critical section is in progress.
In summary, allowing critical
sections to be preempted, only to be forced later,  makes the worst
case worse. We discuss protocols that avoid this effect by using
allocation inheritance rather than priority boosting in \secref{ip}.

\incfig{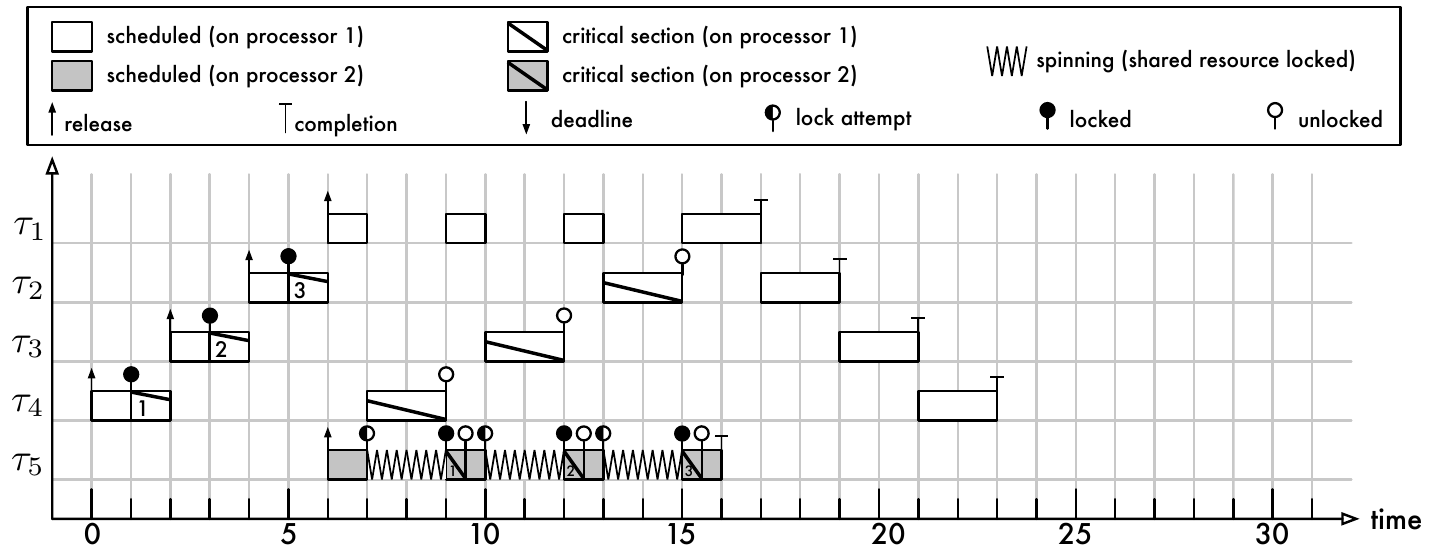}{Example P-FP schedule of five tasks on two processors illustrating repeated pi-blocking under \citeauthor{A:16}'s FEP~\cite{A:16}.}

So far we have discussed two cases at opposite ends of the preemption
spectrum: either jobs spin non-preemptively as in the MSRP, or they
remain preemptable \wrt  all higher-priority jobs while spinning. However, it is also
possible to let waiting jobs spin at some other predetermined
intermediate priority. \citeauthor{ABBN:14} \cite{ABBN:14} observed that this
flexibility allows for a generalized view on both spin locks and suspension-based protocols. 
In particular, \citeauthor{ABBN:14} noted that jobs that
spin at maximum priority are effectively non-preemptive, whereas jobs
that spin at minimum priority are---from an analytical point of
view---essentially suspended, in the sense that they neither prevent
other jobs from executing nor issue further lock requests.

\citeauthor{ABBN:14} combined this observation with (unconditional) priority
boosting and FIFO-ordered spin locks into a \emph{flexible spin-lock model} (FSLM)
\cite{ABBN:14,A:17} that can be tuned~\cite{ABBN:17} to resemble either the
MSRP \cite{GLD:01}, FIFO-ordered suspension-based protocols like the
partitioned FMLP for long resources \cite{BLBA:07} (discussed in \secrefs{sob:part}{saw:part}), or some
hybrid of the two. 
Notably, in \citeauthor{ABBN:14}'s protocol, requests
of  jobs that are preempted while spinning are not canceled, which allows for
multiple critical sections to be simultaneously outstanding on the same
core. As just discussed in the context of \citeauthor{A:16}'s FEP~\cite{A:16}, increasing
the number of incomplete requests that remain to be boosted at a later time
increases the amount of \emph{cumulative} pi-blocking that higher-priority jobs may be exposed to.

For example, \figref{spin-flex-prio} depicts a schedule illustrating \citeauthor{ABBN:14}'s protocol in which tasks
$\tau_2$, $\tau_3$, and $\tau_4$ spin at their regular priority. Jobs of all three
tasks block on a resource held by a job of task $\tau_5$, and thus when
$\tau_5$ releases the lock, the highest-priority task $\tau_1$
suffers pi-blocking due to the back-to-back execution of three
priority-boosted critical sections. This highlights that
indiscriminately lowering the priority at which tasks spin is not always
beneficial; rather, a good tradeoff must be found~\cite{ABBN:17}. For instance,
in the scenario shown in \figref{spin-flex-prio}, the back-to-back pi-blocking of $\tau_1$ could be avoided by
letting $\tau_2$, $\tau_3$, and $\tau_4$ all spin at the priority of $\tau_2$.

\citeauthor{ABBN:18}~\cite{ABBN:18}  study the spin priority assignment problem under the
FSLM approach in depth and introduce a method for
choosing elevated spin priorities that dominates unrestricted priority boosting
(\ie, it is not always best to spin at the maximum priority)~\cite{ABBN:18}. Furthermore,
\citeauthor{ABBN:18} show how to configure the FSLM such that memory requirements are not much
worse than under the MSRP (the MSRP requires only one process stack per core, the FSLM can
be configured to require at most two stacks per core)~\cite{ABBN:18}. 

\incfig{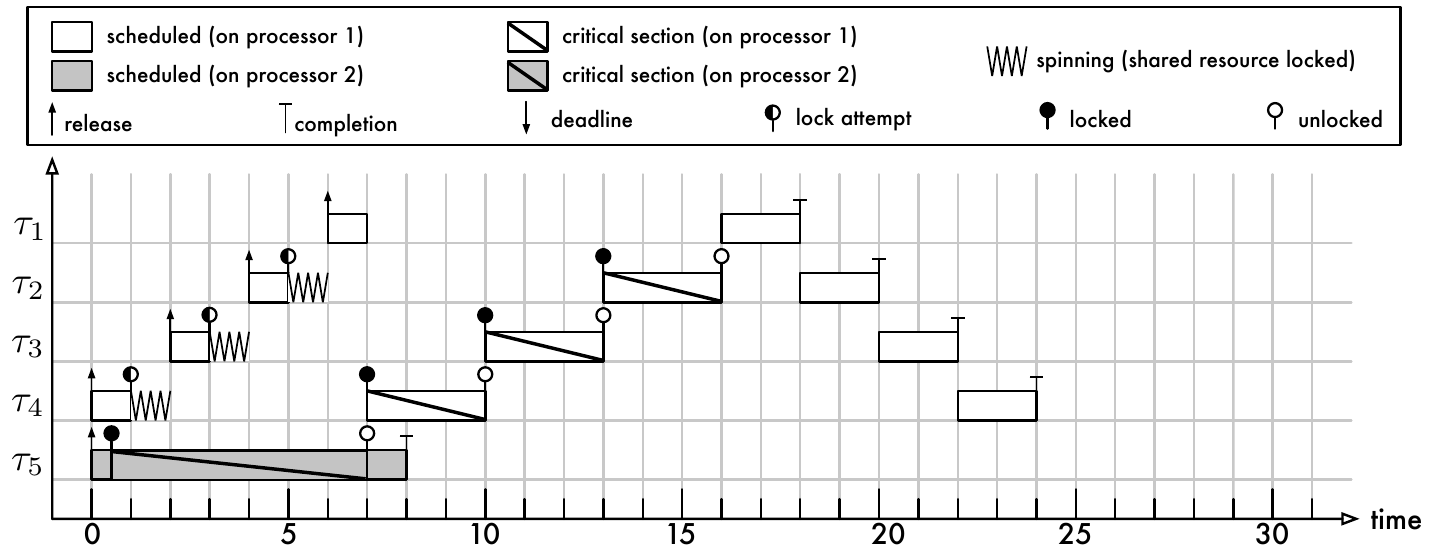}{Example P-FP schedule of five tasks on two processors illustrating preemptable spinning without request cancelation and priority boosting~\cite{ABBN:14}.}

\subsubsection{Non-Preemptive Critical Sections with Allocation Inheritance} \label{sec:spin:spepp}

It is in fact possible to achieve universally low blocking
bounds without requiring workload-specific  parameters. 
Exploring an unconventional alternative to classic spin locks,
\citeauthor{TS:97}~\cite{TS:97} proposed an elegant protocol that achieves
preemptable spinning with $O(1)$ maximum pi-blocking, FIFO-ordering of
critical sections, \emph{and} s-blocking bounds just as good
as those provided by the MSRP. \citeauthor{TS:97}'s
solution, called the \emph{Spinning Processor Executes for Preempted
Processors} (SPEPP) protocol~\cite{TS:97}, is based on the idea that the
processors of blocked jobs should work towards the completion of
blocking critical sections, rather than just ``wasting'' cycles in a spin
loop.

More specifically, in a regular FIFO-ordered spin lock, a job enqueues itself in a spin
queue, busy-waits, and then executes its \emph{own} critical section
when it finally holds the lock. \citeauthor{TS:97}'s SPEPP protocol changes
this as follows. A job first enqueues the \emph{operation} that it
intends to carry out on the shared object (as well as any associated data, \ie, in programming language terms, a \emph{closure}) in a
wait-free FIFO queue, %
and then proceeds to acquire the actual lock. As it acquires the lock,
the job becomes non-preemptable and proceeds to dequeue and execute
operations from the FIFO-ordered operations queue until its own
operation has been completed. Crucially, whenever the job finishes an
operation, it checks for deferred preemptions and interrupts, and
releases the lock and becomes preemptable if any are pending. As a
result, maximum pi-blocking is limited to the length of one operation
(\ie, one critical section length)~\cite{TS:97}, and the time that jobs spend
being s-blocked is used to complete the operations of preempted
jobs, which is an instance of the allocation inheritance principle (as discussed in \secref{prog:ai}).

An interesting corner case occurs when both a preempted and the
preempting job seek to access the same resource. In this case, simply
following the above protocol (\ie, if the preempting job just appends
its operation) could lead to the buildup of long queues, which would result in excessively pessimistic
 s-blocking bounds (\ie, in the worst case, s-blocking linear in the number of tasks $n$). \citeauthor{TS:97} devised a better solution: by
letting the preempting job \emph{steal} the preempted job's slot in the
queue, at most $m$ operations are enqueued at any time~\cite{TS:97}. As this
effectively cancels the preempted job's request, it must re-issue its
request when it resumes. However, in contrast to the preemptable spin locks discussed in \secref{spin:pre},
this does \emph{not} cause additional s-blocking---as \citeauthor{TS:97}
argue, any additional s-blocking incurred by the preempted job upon
re-issuing its request is entirely offset by a \emph{reduction} in the
s-blocking incurred by the preempting job (which benefits from stealing
a slot that has already progressed through the FIFO queue). As a result,
there is no additional net delay: \citeauthor{TS:97}'s SPEPP protocol
ensures $O(1)$ maximum pi-blocking with $O(m)$ maximum s-blocking
\cite{TS:97}.

\subsection{Spin-Lock Protocols for Global Scheduling}

\citeauthor{HA:02} \cite{HA:02,H:04,HA:06} were the first to consider spin-based real-time locking protocols
under global scheduling. In particular, \citeauthor{HA:02} studied synchronization
in systems scheduled by an optimal \emph{Pfair} scheduler \cite{BCPV:96,SA:06}, and introduced the important
distinction between \emph{short} and \emph{long} shared resources.

A shared resource is considered ``short'' if all critical sections that access it are
(relatively speaking) short, and ``long'' if some related critical sections are
(relatively) long, where the exact threshold separating ``short'' and ``long''
is necessarily application- and system-specific. To synchronize access to short resources,
\citeauthor{HA:02} proposed two protocols based on FIFO spin locks, which we discuss
next. (For long resources, \citeauthor{HA:02} proposed allocation inheritance and
semaphore protocols, as discussed in \secref{ip}.)

Without getting into too much Pfair-specific detail \cite{BCPV:96,SA:06}, it is important to
appreciate some specific challenges posed by this optimal scheduling
approach. Pfair scheduling is \emph{quantum-based}, which means that it
reschedules tasks regularly at all multiples of a \emph{system quantum}
$Q$. The magnitude of this system quantum typically ranges
from a few hundred microseconds to a few milliseconds. As a consequence,
all but the shortest jobs span multiple quanta, and thus are likely to be preempted
and rescheduled multiple times during their execution.

To avoid problematic lock-holder preemptions, which are prone to occur
if critical sections cross quantum boundaries, \citeauthor{HA:02} introduced the
notion of a \emph{frozen zone}, or \emph{blocking zone}, at the end of
each scheduling quantum: if a job attempts to commence the execution of a critical
section in this zone (\ie, if the next quantum boundary is less than a given threshold away in time), then its lock request is automatically blocked
until the beginning of its next quantum of execution, regardless of the
availability of the shared resource. If critical sections are shorter
than the system quantum length $Q$---which is arguably the case for any
reasonable threshold for ``short'' critical sections---then such a
\emph{zone-based protocol} ensures that no job is ever preempted while
holding a spin lock \cite{HA:02,H:04,HA:06}.

The remaining question is then how to deal with jobs that attempted to
lock an unavailable resource \emph{before} the start of the
automatic blocking zone, and which are still spinning at the end of the
current quantum, or which are granted the requested resource inside the
automatic blocking zone. \citeauthor{HA:02} considered two solutions to this
problem. 

First, under \citeauthor{HA:02}'s \emph{skip protocol}, such a job remains in the spin
queue and retains its position, but is marked as \emph{inactive} and can
be skipped over by later-enqueued active jobs. 
When a preempted,
inactive job receives the next quantum of processor service, it is
reactivated and becomes again eligible  to acquire the resource.
Furthermore, if a job is at the head of the FIFO spin queue, then it
immediately acquires the lock since all spin locks are available at the
beginning of each quantum in zone-based protocols \cite{HA:02,H:04,HA:06}. 
The primary
advantage of \citeauthor{HA:02}'s skip protocol is that it is starvation-free since
preempted jobs retain their position in the FIFO queue. However, this
also has the consequence that a job is blocked by potentially $n-1$
other jobs in the spin queue (\ie, every other task)---that is, unlike in the case of 
the MSRP \cite{GLD:01}, the number of blocking critical sections cannot be bounded by the number of processors $m$.
Since typically $n>m$, this leads to more
pessimistic s-blocking bounds.

\citeauthor{HA:02}'s second solution, called the \emph{rollback protocol} \cite{HA:02,H:04,HA:06},
restores $m-1$ as a bound on the maximum number of blocking critical
sections, but is applicable only under certain restrictions. Whereas the
skip protocol requires only that the maximum critical section length, denoted $\lmax$,
does not exceed the quantum length $Q$ (\ie, $\lmax \leq Q$), the
rollback protocol further requires $m \times \lmax \leq Q$. This
constraint yields the property that, if tasks on all processors attempt to execute a
critical section at the beginning of a quantum, then each task will
have finished its critical section by the end of the quantum. As a
result, there is no need to maintain a preempted job's queue position to
guarantee progress, and instead a spinning job's request is simply
canceled when it is preempted at the end of a quantum (\ie, the job is
removed from the spin queue).
Consequently, no
spin queue contains more than $m$ jobs at any time. Furthermore,
any lock attempt is guaranteed to succeed at the latest
in a job's subsequent quantum, because a preempted job immediately
re-issues its request when it continues execution, and every request issued at the beginning of a quantum is guaranteed 
to complete before the end of the quantum since  $m \times \lmax \leq Q$ \cite{HA:02,H:04,HA:06}.

\citeauthor{HA:02} \cite{HA:02,H:04,HA:06} presented blocking analyses for both the rollback and the
skip protocol. While Pfair is not widely used in practice today,
\citeauthor{HA:02}'s concept of an automatic blocking zone at the end of a job's guaranteed 
processor allocation has been reused in many locking protocols for
reservation-based (\ie, hierarchically scheduled) systems, in both
uniprocessor and multiprocessor contexts (as mentioned in \secref{placement}).

In later work on non-Pfair systems, \citeauthor{DLA:06} \cite{DLA:06}
and \citeauthor{CDW:10} \cite{CDW:10} studied FIFO-ordered spin locks under event-driven G-EDF and G-FP scheduling, respectively.
Notably, in contrast to \citeauthor{HA:02}~\cite{HA:02}, \citeauthor{DLA:06}~\cite{DLA:06} and \citeauthor{CDW:10}~\cite{CDW:10} assume non-preemptable spinning as in the MSRP~\cite{GLD:01}. 
(Non-preemptable spinning makes little sense in a Pfair
context because quantum boundaries cannot be postponed without prohibitive
schedulability penalties.)
Most recently,  \citeauthor{BMV:14} \cite{BMV:14}
proposed a spin-lock protocol for systems scheduled with the global \emph{RUN} policy~\cite{RLML:11}, another optimal multiprocessor real-time scheduler.

As already mentioned \secref{hist}, in 2007,
\citeauthor{BLBA:07} introduced the \emph{Flexible Multiprocessor Locking Protocol}~\cite{BLBA:07}, which is 
actually a consolidated family of related protocols for different schedulers and critical section lengths. In particular,
the FMLP supports both global and partitioned scheduling (with both fixed and EDF priorities),
and also adopts \citeauthor{HA:02}'s distinction between short and long resources \cite{HA:02}. For each of the resulting
four combinations, the FMLP includes a protocol variant.
For short resources, the FMLP relies on non-preemptive FIFO spin locks. Specifically,
for short resources under partitioned scheduling, the FMLP essentially integrates the MSRP \cite{GLD:01}, and 
for short resources under global scheduling, the FMLP integrate's \citeauthor{DLA:06}'s proposal and analysis \cite{DLA:06}, albeit with link-based scheduling (\ie, lazy preemptions, see \secref{prog:nps}), whereas \citeauthor{DLA:06} \cite{DLA:06} did not specify a preemption policy.
For long resources, the FMLP relies on semaphores, which we will discuss next.

\section{Semaphore Protocols for Mutual Exclusion}\label{sec:susp}

The distinguishing characteristic of \emph{suspension-based} locks, also
commonly referred to as \emph{semaphores} or \emph{mutexes}, is that
tasks that encounter contention \emph{self-suspend} to yield the
processor to other, lower-priority tasks, which allows wait times
incurred by one task to be overlaid with useful computation
by other tasks. 
Strictly speaking, the suspension-based locks considered
in this section correspond to \emph{binary} semaphores, whereas the
suspension-based \emph{$k$-exclusion protocols} discussed in \secref{kex}
correspond to \emph{counting semaphores}. We simply say ``semaphore''
when the type of protocol is clear from context.

As mentioned in \secref{prob:loc}, in the case of semaphores, there exist two
principal ways in which critical sections can be executed: either
\emph{in place} (\ie, on a processor on which a task is also executing
its non-critical sections), or on a \emph{dedicated synchronization
processor}. We focus for now on the more common in-place execution,
and consider protocols for dedicated synchronization processors later in \secref{rpc}.

In principle, semaphores are more efficient than spin locks: since wait
times of higher-priority jobs can be ``masked'' with useful computation
by lower-priority jobs, no
processor cycles are wasted and the  processor's (nearly) full capacity 
is available to the application workload.
However, there are major challenges that limit the efficiency
of semaphores in practice.

First, in practical systems, suspending and resuming tasks usually comes
with non-negligible costs due to both OS overheads (\eg, ready queue
management, invocations of the OS scheduler, \etc) and
micro-architectural overheads (\eg, loss of cache affinity, disturbance
of branch predictor state, \etc). Thus, if the expected wait time is shorter
than the cumulative overheads of suspending, then spinning can be more
efficient in practice. Whether or not runtime overheads make spinning
more attractive depends on a number of factors, including the length of
critical sections (relative to overhead magnitudes), the degree of
contention (likelihood of spinning), and the magnitude of OS and architectural overheads. As our
focus is on analytical concerns, we do not consider this aspect any
further.

The second major challenge is that semaphores are subject to more
intense worst-case contention because they allow other tasks to execute
and issue additional lock requests
while a task is waiting. That is, compared to non-preemptive spin locks,
wait queues can become much longer as the number of concurrent requests for any
resource is no longer implicitly upper-bounded by the number of
processors (as for instance in the case of the MSRP \cite{GLD:01}, recall \secref{spin:part}). Hence accurate
blocking analysis is even more important for semaphores than for spin
locks, as otherwise any practical efficiency gains are at risk of being
overshadowed by analysis pessimism.

For instance, consider the following (exaggerated) illustrative example:
suppose there are $n$ tasks sharing a single resource on $m=2$
processors, where $n \gg m$, and that each critical section is of unit length $L=1$.
With non-preemptive FIFO spin locks (\eg, the MSRP~\cite{GLD:01}), the
maximum spin time in any possible schedule is trivially upper-bounded by
$(m - 1) \times L = L$, and the maximum pi-blocking time is upper-bounded by
$m \times L = 2L$ \cite{GLD:01,GNLF:03}. If we instead change the system to use FIFO
semaphores (\eg, the FMLP~\cite{BLBA:07}), then it is easy to construct
pathological schedules in which $n-1$ tasks are simultaneously
suspended, waiting to acquire the single shared resource (\ie, the
maximum pi-blocking duration is \emph{lower-bounded} by
$(n-1)\times L \gg 2L$). This places semaphore protocols at an
analytical disadvantage. And while we have chosen FIFO queueing in this
example for simplicity, this effect is not specific to any particular
queue order; in particular, similar examples can be constructed for
protocols that employ priority queues, too \cite{BA:10}.

Another complication that suspension-based locking protocols must
address is that tasks are inherently not guaranteed to be scheduled when
they become the lock owner. That is, if a task encounters contention and
self-suspends, then it will certainly not be scheduled when it receives
ownership of the lock, and worse, it may remain unscheduled for a
prolonged time if higher-priority task(s) started executing in the
meantime. The resulting delay poses a risk of substantial
transitive pi-blocking if other blocked tasks are still waiting for the
same lock. Real-time semaphore protocols hence generally require a
progress mechanism that ensures that lock-holding tasks can (selectively)
preempt higher-priority tasks when waking up from a self-suspension. In contrast,
simple non-preemptive spin locks do not have to take resuming lock holders into account. 

Finally, and most importantly, while semaphores allow wait times to be
\emph{potentially} overlaid with useful computation, showing that this
\emph{actually} happens \emph{in the worst case} (\ie, showing that the
processor does not just idle while tasks suspend) is not always
possible. And even when it is theoretically possible, it is an
analytically difficult problem that requires identifying (or safely
approximating) the worst-case self-suspension pattern, which has proven
to be a formidable challenge \cite{CNHY:19}.

More precisely, on multiprocessors, an accurate analysis of semaphores
generally requires the use of a \emph{suspension-aware} (\emph{s-aware}) schedulability
test, that is, an
analysis that applies to a task model that incorporates an explicit
bound on a task's maximum self-suspension time. In contrast, most
schedulability analyses published to date are
\emph{suspension-oblivious} (\emph{s-oblivious}), in the sense that they make the modeling
assumption that tasks never self-suspend (\ie, jobs are either ready to
execute or complete). 

S-oblivious schedulability analyses can still be
employed if tasks (briefly) self-suspend \cite{CNHY:19}, but any
self-suspensions must be pessimistically modeled as computation
time during analysis (\ie, execution-time inflation must be applied).
For example, if a task $\tau_i$ with WCET
$C_i$ self-suspends for at most $S_i$ time units, then it would be
modeled and analyzed as having a WCET of $C_i + S_i$ when applying
s-oblivious schedulability analysis---the task's processor demand is
safely, but pessimistically, over-approximated.

Given that the primary feature of suspension-based locking protocols is
that tasks do \emph{not} occupy the processor while waiting to acquire a
lock, one may deem it intuitively undesirable to model and analyze suspension times as
processor demand. However, s-aware schedulability analyses are
unfortunately difficult to obtain and can be very pessimistic.
Case in point, prior work on s-aware schedulability analysis for
uniprocessor fixed-priority scheduling was found to be flawed in several
instances~\cite{CNHY:19}, and the best correct analyses available today
are known to be only sufficient, but not exact (in contrast
to response-time analysis for non-self-suspending tasks,
which is exact on uniprocessors). 

Another example that highlights the
challenge of finding efficient s-aware schedulability analysis
is G-EDF scheduling: an s-oblivious analysis of G-EDF was
available \cite{BCL:05} for several years before the first s-aware test for
G-EDF~\cite{LA:13} was proposed. Furthermore, the s-aware test was actually found to be
\emph{more} pessimistic than a simple s-oblivious approach when applied
in the context of a suspension-based locking protocol \cite{B:14b}.

Generally speaking, as a result of the challenges surrounding s-aware analysis and
the pessimism present in today's analyses, s-oblivious
approaches can be competitive with s-aware analyses,
and at times even yield superior performance in
empirical comparisons \cite{BA:13,B:11}. It is hence worthwhile 
to study both approaches, and to compare and contrast their properties.

Most importantly, the s-oblivious and
s-aware approaches  fundamentally differ
\wrt the best-possible bound on cumulative
pi-blocking~\cite{BA:10,B:11,BA:13}. 
More precisely, \defref{pi} can be refined for the s-oblivious case, 
and with the refined definition in place (\defref{sob} below), the
inherently pessimistic treatment of suspensions in s-oblivious
schedulability analyses allows some of this pessimism to be
``recycled,'' in the sense that \emph{less} pessimistic assumptions have
to be made when analyzing priority inversions in the
s-oblivious case, which in turn allows for lower bounds on pi-blocking.

More formally, consider \emph{maximum pi-blocking}, which for a given task set $\tau$ is the amount of
pi-blocking incurred by the task that suffers the most from
priority inversion: $ \max \{ B_i\ |\ \tau_i \in \tau \}$, where $B_i$ is the pi-blocking bound for task $\tau_i$ \cite{BA:10}.
Interestingly, the s-aware and s-oblivious analysis
assumptions yield asymptotically \emph{different} bounds on maximum pi-blocking \cite{BA:10,B:11,BA:13}. 
Specifically, there exist semaphore protocols that
ensure that \emph{any} task will incur at most $O(m)$
pi-blocking in the s-oblivious sense \cite{BA:10,BA:13},
whereas no semaphore protocol can generally guarantee pi-blocking bounds
better than $\Omega(n)$ in the s-aware case \cite{BA:10,B:14b}
(recall that $m$ denotes the number of processors and $n$ the number
of tasks, and that typically $n > m$). 
In the following, we review these bounds and the protocols that achieve them.
We first discuss locking protocols intended for
s-oblivious analysis because they are simpler and
easier to analyze, and then consider locking protocols intended for
s-aware analysis thereafter.

\subsection{Suspension-Oblivious Analysis of Semaphore Protocols}\label{sec:sob}

Under s-oblivious schedulability analysis, self-suspensions are
\emph{modeled} as execution time \emph{during analysis}. However,
\emph{at runtime}, tasks of course self-suspend; the distinction between
the s-oblivious and s-aware approaches is purely
analytical. In fact, it is possible to analyze any protocol using
either approach, and strictly speaking the protocols \emph{themselves}
are neither ``suspension-oblivious'' nor ``suspensions-aware.'' However,
certain protocols are \emph{easier} to analyze, or can be
\emph{more accurately} analyzed,  under one of the two
analysis approaches, which gives rise to the commonly used terminology
of s-oblivious and s-aware semaphore protocols.

In this subsection, we review s-oblivious locking protocols, \ie,
semaphore protocols that are primarily analyzed using
s-oblivious analysis, and that in many cases were designed
specifically with s-oblivious analysis in mind.

\subsubsection{Suspension-Oblivious Analysis and Blocking Optimality}

The key insight underlying the analysis of s-oblivious locking
protocols is that, since s-oblivious schedulability analysis
pessimistically over-approximates any self-suspension times as processor
demand, it is possible to reclaim some of this pessimism by
\emph{refining} the definition of pi-blocking (\defref{pi}) to account for this
modeling assumption.
More precisely, ``suspension-oblivious
pi-blocking'' is defined such that any times during which both
\begin{enumerate}
	\item  a job is self-suspended while waiting to acquire a
semaphore and
	\item this delay can be attributed to
higher-priority tasks (under the ``suspended tasks create processor
demand'' analysis assumption)
\end{enumerate}
are \emph{not} counted as pi-blocking, which allows tighter pi-blocking bounds to be established (without endangering  soundness of the analysis).

Intuitively, this works as follows. Consider clustered scheduling and a
task $\tau_i$ assigned to a cluster consisting of $c$ processor
cores. First, suppose a job $J$ is waiting to acquire some semaphore
and there are (at least) $c$ higher-priority \emph{ready} jobs, that is,
$J$ is self-suspended and there are $c$ higher-priority jobs
occupying the processors in $J$'s cluster. In this situation,
$J$ does not incur pi-blocking according to \defref{pi} (although it is waiting to
acquire a semaphore) since it would not be scheduled even if it were
ready, due to the presence of $c$ higher-priority ready jobs. In other words,
the self-suspension does not have to be accounted for as
\emph{additional} delay in this case because $J$ would be delayed
anyway.

Now consider the following alternative scenario:  $J$ is self-suspended and there are $c$
higher-priority jobs in $J$'s cluster, but all higher-priority jobs
are also self-suspended, each waiting to acquire some semaphore (possibly,
but not necessarily, the same that $J$ is waiting for). That is, the
higher-priority jobs in $J$'s cluster are \emph{pending}, but not
\emph{ready}. In this case, if $J$ were ready, it would be scheduled
immediately since \emph{in the real system} the suspended
higher-priority jobs do not occupy any processors, and hence intuitively
this situation represents a priority inversion for $J$ (and also according to \defref{pi}).
However,
under s-oblivious analysis, the self-suspension times of
higher-priority jobs are \emph{modeled} as execution time. Hence,
\emph{in the analyzed model of the system}, the pending higher-priority jobs are
analyzed \emph{as if} they were occupying all processors, and hence
$J$ incurs no \emph{additional} delay in the modeled situation under the
s-oblivious analysis assumption. The following definition
exploits this observation.

\begin{definition}\label{def:sob}
A job $J$ of task $\tau_i$, assigned to a cluster $C$ consisting of $c$ cores, suffers 
\emph{\textbf{s-oblivious} pi-blocking} at time $t$ if and only if
\begin{enumerate}
	\item $J$ is pending at time $t$,
	\item not scheduled at time $t$ (\ie, it is
self-suspended or preempted), 
	\item fewer than $c$ equal- or higher-priority jobs of tasks assigned to cluster $C$ are \textbf{\emph{pending}}  on processors belonging to $\tau_i$'s assigned cluster $C$.
\end{enumerate}	
\end{definition}
Note that \defrefs{pi}{sob} differ in clause (3) \wrt whether the higher-priority jobs are required to be pending (\defref{sob}) or   scheduled (\defref{pi}).

Based \defref{sob}, the suspension-oblivious schedulability
analysis approach can be summarized as
follows.\footnote{This description is somewhat simplified because it considers only self-suspensions and ignores priority inversions due to progress mechanisms to simplify the explanation. Any additional priority inversions (\eg, due to non-preemptive execution) are handled analogously by inflation.}
\begin{itemize}
	\item Suppose we are given a self-suspending task set $\tau = \{\tau_1, \ldots, \tau_n\}$,
where each task $\tau_i = (C_i, D_i, T_i, S_i)$ in $\tau$ is
characterized by its WCET $C_i$, a relative deadline $D_i$, a period
$T_i$, and a bound $S_i$ on the maximum cumulative self-suspension
duration of any of its jobs. 
	\item Further suppose that $B_i$ denotes a bound on the
\emph{maximum cumulative duration of suspension-oblivious pi-blocking}
incurred by any of $\tau_i$'s jobs (where $B_i \leq S_i$).
	\item Let $\tau' = \{\tau_1', \ldots, \tau_n'\}$ denote the corresponding \emph{inflated}, \emph{suspension-free} task set, where each
$\tau_i' = (C_i + B_i, D_i, T_i)$.
	\item Then the actual task set $\tau_i$ (with self-suspensions) does not miss any deadlines under a given preemptive JLFP policy \emph{in the presence of self-suspensions} if the inflated  task set $\tau'$ is schedulable under the same JLFP policy \emph{in the absence of any self-suspensions}.
\end{itemize}

The correctness of this approach can be shown with a simple reduction
(or \emph{schedule transformation}) argument. Suppose a job misses a
deadline in the real system, and consider the trace resulting in the
deadline miss (\ie, a concrete schedule of the given task set
$\tau$). This trace consists of a set of self-suspending jobs with
concrete release, execution, and self-suspension times (\ie, discard
any knowledge of locks; for this transformation we require only
self-suspension times). 

Now repeatedly transform this trace as
follows until no self-suspensions remain. First, for each job $J$ in the
trace, in order of decreasing priority, and for each point in time $t$ at
which $J$ is suspended, if there are fewer than $c$
higher-priority jobs in $J$'s cluster occupying a processor at time
$t$, then transform $J$'s self-suspension at time $t$ into
execution time. Otherwise, simply discard $J$'s self-suspension at
time $t$ (\ie, reduce $J$'s self-suspension length by one time
unit) since it is not scheduled at time $t$ anyway. This step does not
\emph{decrease} $J$'s response time, nor does it decrease the
response time of any other job. After this transformation, we obtain a
trace in which both \textbf{(i)} no job self-suspends and \textbf{(ii)}
a deadline is still being missed. Furthermore, let $\tau_i$ denote the task of $J$: 
since $B_i$ is a bound on the maximum amount of s-oblivious pi-blocking
as defined by \defref{sob}, \textbf{(iii)} the number of
times that an instant of self-suspension of job $J$ is converted to
an instant of execution is bounded by $B_i$.

From (i) and (iii), it follows that the transformed trace is a valid
schedule of $\tau'$, and hence from (ii) we have that $\tau$ misses
a deadline only if there exists a schedule in which $\tau'$ misses a
deadline. Conversely, if it can be shown that $\tau'$ does not miss
any deadlines, then $\tau$ also does not miss any deadlines.

Given \defref{sob}, a natural question to ask is: what is the \emph{least} upper
bound on maximum s-oblivious pi-blocking (\ie, the least $B_i$) that \emph{any} locking protocol can
guarantee in the general case? In other words, what amount of
s-oblivious pi-blocking is \emph{unavoidable}, in the sense that there
exist pathological task sets that exhibit at least this much s-oblivious
pi-blocking no matter which locking protocol is employed?

Clearly, this bound cannot be zero, as some blocking is unavoidable
under any mutual exclusion scheme. It is in fact trivial to
construct task sets in which a job exhibits $\Omega(m)$ s-oblivious
pi-blocking under any locking protocol: if a lock is requested
simultaneously by tasks on all $m$ processors, then $m$
(unit-length) critical sections must be serialized in some order, and
hence whichever task acquires the lock last is blocked by (at least)
$m-1$ critical sections (\ie, any general pi-blocking bound is
necessarily linear in the number of processors to cover this scenario). If there are exactly $c$ tasks assigned to
each cluster (\ie, if there are only $m$ tasks in total), then
according to \defref{sob} any self-suspension results in s-oblivious pi-blocking,
and the lower bound trivially follows \cite{BA:10,B:11,BA:13}.

While this \emph{lower} bound on maximum s-oblivious pi-blocking is
straightforward, finding a matching \emph{upper} bound is less obvious.
Given that up to $n$ jobs can simultaneously contend for the same
lock, one might wonder whether this is even possible. However, as we
review next, it is in fact possible to construct locking protocols that
ensure $O(m)$ s-oblivious pi-blocking for \emph{any} sporadic task set
\cite{BA:10,B:11,BA:13}, which establishes that, under s-oblivious schedulability analysis,
$\Theta(m)$ pi-blocking is fundamental \cite{BA:10,B:11,BA:13}.

\subsubsection{Global Scheduling} \label{sec:sob:global}

The earliest semaphore protocols for global scheduling are due to \citeauthor{HA:02} \cite{HA:02a,HA:02,HA:06},
who introduced support for lock-based synchronization in Pfair-scheduled systems.
However, due to the quantum-based nature of Pfair,
their analysis is not s-oblivious in the sense of \defref{sob}; we hence defer a discussion of their work until \secref{ip}.

The first multiprocessor real-time semaphore protocol explicitly studied
using the s-oblivious analysis approach is \citeauthor{BLBA:07}'s FMLP \cite{BLBA:07}.
As already discussed in
\secref{spin}, the FMLP is actually a family of related protocols for
different scheduling approaches and incorporates both spin- and
suspension-based variants. Aiming for simplicity in both implementation
and analysis, the FMLP for \emph{long} resources (\ie, the semaphore
variant) for global scheduling combines priority inheritance (recall
\secref{prog:pi}) with simple FIFO wait queues.

Priority inheritance ensures that a lock-holding job is scheduled
whenever another job that it blocks is incurring s-oblivious
pi-blocking. This is easy to see: under global scheduling (\ie, if
there is only one cluster of size $m$), a job incurs s-oblivious
pi-blocking only if it is among the $m$ highest-priority pending jobs
(\defref{sob}), and thus the lock-holding job is guaranteed to inherit a priority
that allows it to be immediately scheduled \cite{BLBA:07}.

Combined with the strong progress guarantee of FIFO wait queues, the
long FMLP for global scheduling ensures that a job incurs s-oblivious
pi-blocking for the duration of at most $n-1 = O(n)$ critical section
lengths while waiting to acquire a lock. This bound shows that, while more accurate analyses
taking actual request patterns into account are possible \cite{B:11}, in the general case the FMLP
does not ensure asymptotically optimal maximum s-oblivious pi-blocking. In
fact, no protocol relying exclusively on FIFO or priority queues can be
optimal in this regard \cite{BA:10}.

The first asymptotically optimal protocol is the
\emph{$O(m)$ Locking Protocol} (OMLP) \cite{BA:10,B:11,BA:13,BA:11}, which, as the name suggests,
ensures $O(m)$ maximum s-oblivious pi-blocking for any task set. Like
the FMLP, the OMLP is also a family of protocols for global,
partitioned, and clustered scheduling, which we review in turn.

The OMLP variant for global scheduling (\ie, the \emph{global OMLP})
\cite{BA:10} relies on priority inheritance, like the earlier global FMLP \cite{BLBA:07}. To achieve
optimality, it replaces the FMLP's simple FIFO queue with a
\emph{hybrid} wait queue, which consists of a \emph{bounded} FIFO
segment of length $m$ and a priority-ordered \emph{tail queue} that
feeds into the FIFO segment.

Jobs that request the lock enqueue directly
in the FIFO segment if there is space (\ie, if fewer than $m$ jobs
are contending for the resource), and otherwise in the tail queue, which
is ordered by job priority. The job at the head of the FIFO queue holds
the lock; when it releases the lock, it is dequeued from the FIFO queue,
ownership is passed to the new head of the FIFO queue (if any) and the
highest-priority job presently waiting in the tail queue (if any) is
transferred to the FIFO queue.

This combination of queues ensures that a job incurs s-oblivious
pi-blocking for the duration of at most $2m-1 = O(m)$ critical sections
per lock request. Clearly, once a job enters the bounded-length
FIFO queue, at most $m-1$ critical sections of jobs ahead in the FIFO
queue cause pi-blocking. Additionally, a job incurs s-oblivious
pi-blocking for the cumulative duration of at most $m$ critical
sections while it waits in the priority-ordered tail queue, which
follows from the following observation \cite{BA:10}. Suppose a job $J$
that is waiting in the tail queue is skipped over $m$ times (\ie, at
least $m$ times another job is moved to the end of the FIFO queue
while $J$ is waiting). Since the tail queue is priority-ordered,
each job that skipped ahead has a higher priority than $J$.
Furthermore, since the FIFO queue has a capacity of exactly $m$ jobs,
it follows that there are $m$ higher-priority pending jobs, which
implies that $J$ incurs no s-oblivious pi-blocking after it has been
skipped over at least $m$ times (recall clause (3) of \defref{sob}). Thus, in total, a job incurs
s-oblivious pi-blocking for a cumulative duration of at most $2m-1 = O(m)$
critical sections while moving through both queues, which is
within a factor of
$\frac{2m-1}{m-1} = \frac{2\times(m-1) + 1}{m-1} = 2 + \frac{1}{m-1} \approx 2$
of the lower bound and thus asymptotically optimal \cite{BA:10,B:11,BA:13}.
Fine-grained (\ie, non-asymptotic) analyses of the global OMLP taking
into account actual request patterns are available as well \cite{B:11,BA:13}.

\citeauthor{W:15}~\cite{W:15} studied two variants of mutual exclusion called \emph{preemptive mutual exclusion} and \emph{half-protected resources}, which are intended for synchronizing (in software) access to shared hardware resources such as memory buses and caches (which have the special property that they can be revoked without risking an inconsistent state of the shared resource, \eg, revocation of a granted cache partition carries a performance penalty but no consistency hazard). To analyze the resulting synchronization problem, \citeauthor{W:15} introduced \emph{idleness analysis}~\cite{W:15}, new blocking analysis technique that bounds the maximum amount of idleness induced on other cores by a critical section, rather than bounding the number of waiting jobs. Idleness analysis can also be applied to the analysis of regular mutual exclusion protocols such as the OMLP or the FMLP.

\subsubsection{Partitioned Scheduling} \label{sec:sob:part}

In the case of partitioned scheduling, priority inheritance is
ineffective (\secref{prog:pi}), with priority boosting being the traditional
alternative (\secref{prog:boost}). This choice was also adopted in the design of the
\emph{long FMLP for partitioned scheduling} \cite{BLBA:07}, which was the first
semaphore protocol for partitioned scheduling to be analyzed under the
s-oblivious approach.

Like all other FMLP variants, the long FMLP for partitioned scheduling
relies on FIFO queues. One additional twist that arises in conjunction
with priority boosting is that a \emph{tie-breaking policy} is required
to determine which job to schedule if there are multiple lock-holding
jobs on the same processor. In the interest of simplicity, the 
FMLP favors whichever job first \emph{acquired}
its respective lock (\ie, ``earliest-resumed job first'') \cite{BLBA:07}. This
later turned out to be a non-ideal choice in the context of
suspension-aware analysis (discussed in \secref{saw:part}), and was changed to an
\emph{earliest-\textbf{issued} request first} policy in the later \fmlpp \cite{B:11}.

However, under s-oblivious analysis, either tie-breaking rule is
problematic, as \emph{unrestricted} priority-boosting generally prevents
optimal s-oblivious pi-blocking (regardless of the order in which
blocked jobs wait), which can be inferred from the following simple example.
Consider a
job $J$ that is the highest-priority job on its processor, and
suppose for the sake of illustration that $n-1$ tasks reside on job
$J$'s processor, with the remaining task located on a second
processor. Now suppose that, just before $J$ is released, the remote
task first acquires a shared lock and then all other tasks on $J$'s
processor suspend while waiting for the same lock. If lock-holding jobs
are unconditionally priority-boosted, then $J$ will be preempted for
the duration of one critical section of \emph{each} of the $n-2$ other
tasks on $J$'s processor, which results in $\Omega(n)$ s-oblivious
pi-blocking even if $J$ itself does not acquire any locks itself.
As a result of this effect, and because FIFO queues allow for pi-blocking due to
up to $n-1$ critical sections per critical section, jobs incur s-oblivious pi-blocking of
up to $(n - 2) + (n - 1) = 2n-3$ critical section lengths under the
long FMLP for partitioned scheduling, which is not asymptotically
optimal.

The \emph{partitioned OMLP} \cite{BA:10} solves the ``too many boosted jobs'' problem with a \emph{token
mechanism} to limit the number of tasks that can simultaneously request
global resources, which implicitly restricts the maximum delay due to
priority boosting while also ensuring that global wait queues remain
short.

Under the partitioned OMLP, there is a single \emph{contention token} associated with
each processor. A processor's contention token is a local (virtual)
resource that is managed using an optimal uniprocessor protocol (such as
the PCP \cite{SRL:90} or the SRP \cite{B:91}).
Furthermore, each global resource is associated with a FIFO wait queue, and jobs
holding (global) resources are priority-boosted, just as in the earlier
FMLP. However, the key OMLP rule is that a task must hold its local
contention token before it may issue a request for a global resource. As
a result, only at most $m$ tasks compete for global resources at any
time, which in conjunction with FIFO queues and priority boosting
immediately yields a pi-blocking bound of $m-1$ critical section
lengths once a job holds its local contention token. Additionally, a job
may incur pi-blocking while it is waiting to acquire a contention token,
which however is also limited to $m$ critical section lengths
(including any priority boosting effects) \cite{BA:10}.

As a result, the partitioned OMLP guarantees a bound on s-oblivious
pi-blocking for the duration of $m-1$ critical sections per request, 
plus s-oblivious pi-blocking for the duration of up to
$m$ critical sections due to competition
for the local contention token and priority boosting, for a total of $2m-1$ critical section lengths
(assuming a task that issues only one request per job),
which is within $\frac{2m-1}{m-1} \approx 2$ of the lower bound and
thus asymptotically optimal \cite{B:11,BA:13}.
A fine-grained (\ie, non-asymptotical) analysis of the partitioned OMLP
is available \cite{BA:11,BA:13}.

\subsubsection{Clustered Scheduling} \label{sec:sob:clust}

The case of ``true'' clustered scheduling, where there are multiple
clusters (unlike the special case of global scheduling) and each cluster
contains more than one processor (unlike the special case of partitioned
scheduling), is particularly challenging because it combines the challenges
of both global and partitioned scheduling. In particular, priority
inheritance across clusters is ineffective (\secref{prog:pi}), but priority boosting,
even if restricted by a token mechanism as  in the partitioned OMLP, makes
it difficult to obtain asymptotically optimal s-oblivious pi-blocking
bounds.

For this reason, a new progress mechanism called \emph{priority
donation} was developed for the \emph{clustered OMLP} \cite{BA:11,B:11,BA:13}. As already
mentioned in \secref{prog:rb}, priority donation can be understood as a form of
restricted priority boosting. However, the key difference to the token
mechanism used in the partitioned OMLP is that there exists an explicit
relationship between the lock-holding job that is forced to be scheduled
(\ie, the \emph{priority recipient}) and the job that is \emph{not}
scheduled as a result (\ie, the \emph{priority donor}). In contrast,
priority boosting just prescribes \emph{that} a job is scheduled, but
leaves unspecified which other job is \emph{not} scheduled as a result,
which causes analytical complications if there is more than one
processor in a cluster (\ie, if there is a choice \wrt which job to
preempt).

Priority donation maintains the invariant that, in each cluster and for
each lock-holding job $J$, $J$ is either among the $c$
highest-priority jobs in its cluster, or there exists a job among the
$c$ highest-priority jobs that is the \emph{unique} and \emph{exclusive} priority donor for
$J$. As a result of this invariant, contention for global locks is
limited to at most $m$ concurrent lock requests, and lock-holding jobs
are guaranteed to make progress towards releasing the lock (\ie,
lock holders are never preempted).

A job can become priority donor only once, immediately upon its release
(\ie, before it starts executing). While it serves as priority donor,
it is suspended to make available a processor for the priority
recipient, and thus incurs s-oblivious pi-blocking. Priority donation
ceases when the critical section of the priority recipient ends. The
maximum request duration---from the time that a lock is requested until
the time that the lock is released---thus also determines the amount of
s-oblivious pi-blocking transitively incurred by the priority donor.

Under the clustered OMLP, each resource is associated with a simple FIFO
wait queue \cite{BA:11,B:11,BA:13}. Since priority donation guarantees that
lock-holding jobs are scheduled,
and since there are at most $m$ concurrent requests for global
resources in progress at any time, a job is delayed by at most $m-1$
earlier critical sections per lock request. This in turn implies that
priority donors incur s-oblivious pi-blocking for the cumulative
duration of at most $m$ critical sections \cite{BA:11}. The blocking bound for
the clustered OMLP is thus equivalent to that of the partitioned OMLP,
and it is hence also asymptotically optimal within a factor of roughly
two of the lower bound \cite{BA:11,BA:13}.

Additional protocols designed specifically for s-oblivious analysis
are discussed in \secrefs{ip}{kex}.  However, none of these protocols, nor any OMLP variant,
ensures an upper bound on maximum s-oblivious pi-blocking 
better than
within roughly a factor of  two of the known lower bound.
It is presently unknown whether it is
possible to close this gap in the general case.

\subsection{Suspension-Aware Analysis of Semaphore Protocols}\label{sec:saw}

Under s-aware analysis, any self-suspensions due to lock contention and priority
inversions due to progress mechanisms are explicitly modeled and must be 
accounted for by the schedulability test. Hence there is no opportunity
to ``recycle'' any pessimism, there is no ``analysis trick''  as in the s-oblivious case---when
targeting s-aware analysis, the
goal of the locking protocol designer is simply to bound maximum delays
as tightly as possible.

The potential upshot is that an underlying s-aware schedulability
analysis can potentially be much more accurate when characterizing the effects of contention, and substantially less pessimistic in terms of
system utilization
since execution times are not being inflated. This is an important consideration especially if self-suspensions are relatively long.
For instance, s-aware
analysis becomes essential when synchronizing access to \emph{graphics processing units} (GPUs), where
critical section lengths can easily reach dozens or even hundreds of
milliseconds \cite{EA:12}. In comparison,  when considering shared data structures, where critical
sections are typically just a few microseconds long \cite{BCBL:08,AJJ:98},
the s-oblivious utilization impact is minor.

However, while historically the first multiprocessor real-time locking
protocols \cite{RSL:88,R:90,R:91,RM:95a} have all been intended for s-aware analysis, the
understanding of self-suspensions from a schedulability point of view
has only recently begun to mature, 
and a number of
misunderstandings and misconceptions have been identified in earlier analyses of
task sets with self-suspensions \cite{CNHY:19}. Multiprocessor locking protocols
for s-aware analysis, and the required s-aware analyses themselves, are thus
presently still active areas of research. 

In the following, we provide
an overview of the current state of the art, starting with a brief
review of the definition of s-aware pi-blocking, known asymptotic
bounds, and (non-)optimality results, and then summarize major binary
semaphore protocols for global, partitioned, and clustered scheduling.

\subsubsection{Suspension-Aware Schedulability Analysis and Blocking Optimality}\label{sec:saw:opt}

In the case of s-aware schedulability analysis, any delay that does not
result from the execution of higher-priority jobs constitutes a priority
inversion that must be explicitly accounted for. This is captured by the
following definition.

\begin{definition}\label{def:saw}
A job $J$ of task $\tau_i$, assigned to a cluster $C$ consisting of $c$ cores, suffers 
\emph{\textbf{s-aware} pi-blocking} at time $t$ if and only if
\begin{enumerate}
	\item $J$ is pending at time $t$,
	\item $J$ is not scheduled at time $t$ (\ie, it is
self-suspended or preempted), and
	\item fewer than $c$ equal- or higher-priority jobs of tasks assigned to cluster $C$ are \textbf{\emph{scheduled}}  on processors belonging to $\tau_i$'s assigned cluster $C$.
\end{enumerate}	
\end{definition}
Notably, \defref{saw} is equivalent to the classic uniprocessor notion of
pi-blocking (\defref{pi}). The key difference to the s-oblivious case (\defref{sob}) is
that under \defref{saw} only the presence of $c$ \emph{scheduled}
higher-priority jobs prevents a delay from being considered a priority
inversion, whereas under \defref{sob} a priority inversion is ruled out if there
are $c$ \emph{pending} higher-priority jobs. Since any scheduled job
is also pending, \defref{saw} is weaker than
\defref{sob}, and consequently any bound on
s-aware pi-blocking is also a bound on s-oblivious pi-blocking (but the
converse does not hold) \cite{BA:10,B:11}.

The fundamental lower bound on s-aware pi-blocking is $\Omega(n)$
\cite{BA:10,B:11}, which can be easily shown with a task set in which all $n$
tasks simultaneously compete for a single shared resource \cite{BA:10}, so that
whichever task acquires the resource last is delayed by $n-1$ earlier critical sections.

While the lower bound is rather intuitive, the true challenge is again to
construct locking protocols that asymptotically match this lower bound,
that is, to find protocols that ensure $O(n)$ maximum s-aware pi-blocking for \emph{any} task set.
In fact, from a blocking optimality point of view, the s-aware
case is much more challenging than the
well-understood s-oblivious case, and required several
attempts until it was solved~\cite{BA:10,B:11,B:14b}. To date, while there exists a
protocol for clustered scheduling that achieves $O(n)$ maximum s-aware
pi-blocking---namely the generalized \fmlpp \cite{B:14b}---which suffices to establish
asymptotic tightness of the lower bound under global and partitioned
scheduling, no protocol with asymptotically optimal
s-aware pi-blocking bounds is known specifically for global scheduling~\cite{B:11,B:14b,YWB:15}.

The search for practical protocols that are also asymptotically optimal
with regard to maximum s-aware pi-blocking is complicated by several
non-optimality results. For one, any protocol relying exclusively
on priority queues is generally subject to an $\Omega(m \times n)$
lower bound on maximum s-aware pi-blocking due to starvation effects
\cite{BA:10}. Furthermore, under global scheduling, any protocol relying on
priority inheritance or (unrestricted) priority boosting is subject to
an $\Omega(\phi)$ lower bound on maximum s-aware pi-blocking \cite{B:11,B:14b}, where
$\phi$ corresponds to the ratio of the longest to the shortest period
of the task set (and which in general cannot be bounded in terms of
$m$ or $n$). The generalized \fmlpp \cite{B:14b}, which we discuss in \secref{saw:clust} below, thus
requires more sophisticated machinery to achieve its $O(n)$
bound under clustered scheduling.

\subsubsection{Global Scheduling} \label{sec:saw:global}

Almost all semaphore protocols designed specifically for global
scheduling rely on priority inheritance.
As already mentioned in the discussion of the
s-oblivious case (\secref{sob:global}), \citeauthor{BLBA:07}'s global FMLP for long resources \cite{BLBA:07},
based on FIFO wait queues, was the first protocol in this category, and
even though the initial analysis was s-oblivious \cite{BLBA:07}
(no s-aware analysis for global scheduling was known yet at the time of publication), the
protocol itself works well under s-aware analysis, too, and an effective
s-aware analysis has been presented by \citeauthor{YWB:15}
\cite{YWB:15} for G-FP scheduling.

The classic \emph{Priority Inheritance Protocol} (PIP) \cite{SRL:90}, which
combines priority inheritance (the progress mechanism, \secref{prog:pi}) with
priority-ordered wait queues, was initially analyzed under G-FP
scheduling by \citeauthor{EA:09} \cite{EA:09}. In more recent work, \citeauthor{EA:09}'s original
analysis has been subsumed by \citeauthor{YWB:15}'s more accurate analysis of the PIP
(and several other protocols)~\cite{YWB:15}, a  state-of-the-art blocking analysis
approach based on linear programming \cite{B:13a} applied to G-FP scheduling.

\citeauthor{NN:11} \cite{NN:11} transferred \citeauthor{EA:09}'s original
analysis of the PIP \cite{EA:09} to a variant of
the protocol that they called \emph{Immediate PIP} (I-PIP) \cite{NN:11}, which
retains the use of priority-ordered wait queues, but replaces priority
inheritance with priority boosting. They further derived bounds on the
\emph{maximum resource-hold times} under both the original PIP and the
I-PIP, as well as heuristics for reducing resource-hold times without
violating schedulability \cite{NN:11}.

Motivated by their analysis of the PIP, \citeauthor{EA:09}  also proposed a new
semaphore protocol~\cite{EA:09,EA:09a} that they called the \emph{Parallel Priority-Ceiling
Protocol} (P-PCP) \cite{EA:09}, which is also based on priority inheritance and
 priority-ordered wait queues. Additionally, inspired by the
classic uniprocessor PCP \cite{SRL:90}, the P-PCP introduces rules that prevent
jobs from acquiring \emph{available} resources to limit, at each
priority level, the maximum number of lower-priority jobs that may
simultaneously hold locks. Intuitively, such a rule can help to limit
the amount of pi-blocking caused by the progress mechanism, but it
introduces considerable complexity and has to be carefully balanced with
the extra delay introduced by withholding available resources. \citeauthor{EA:09} \cite{EA:09}
did not provide an empirical comparison of the PIP and the P-PCP; a 
later evaluation by \citeauthor{YWB:15} \cite{YWB:15} based on their more accurate re-analysis of both
protocols  found that the P-PCP offers no substantial benefits over
the (much simpler) PIP and FMLP protocols.

\citeauthor{YWB:15} \cite{YWB:15} also compared the long FMLP and the PIP, and found the two
protocols to be incomparable: fundamentally, some real-time workloads
require the non-starvation guarantees of the FMLP's FIFO queues, whereas
other workloads require that urgent jobs are prioritized over less-urgent
jobs. In practice, it is thus preferable for a system to offer
\emph{both} FIFO and priority-ordered wait queues, and it would not be
difficult to combine \citeauthor{YWB:15}'s analyses of the FMLP and the PIP to
analyze such a hybrid protocol.

\subsubsection{Partitioned Scheduling} \label{sec:saw:part}
Across all combinations of multiprocessor scheduling and synchronization approaches,
the category considered next---multiprocessor real-time semaphore protocols for partitioned scheduling with in-place
execution of critical sections---has received the most attention in prior work.
The classic, prototypical protocol in
this domain is \citeauthor{R:90}'s \emph{Multiprocessor Priority Ceiling Protocol}
(MPCP) for P-FP scheduling~\cite{R:90,R:91}.
The MPCP is a natural extension of uniprocessor synchronization
principles, is appealingly simple, and has served
as a template for many subsequent protocols. 
To ensure lock-holder
progress, the MPCP relies on priority boosting (\secref{prog:boost}). Specifically,
for each resource, the protocol determines a \emph{ceiling priority}
that exceeds the priority of any regular task, and the \emph{effective priority} of a resource-holding
task is unconditionally boosted to the ceiling priority of the resource that it holds.

To resolve contention, each shared resource is associated with a
priority-ordered wait queue, in which blocked tasks wait in
order of their regular scheduling priority. From a blocking optimality
point of view, this choice prevents asymptotic optimality \cite{BA:10,B:11}. However,
empirically, the MPCP is known to perform well for many (but not all)
workload types \cite{B:13a}.

Priority-boosted tasks remain
preemptable under the MPCP. Resource-holding jobs can thus be preempted by other
resource-holding jobs. The choice of ceiling priorities therefore has a
significant impact on blocking bounds. \citeauthor{R:90}'s original proposal \cite{R:90,R:91}
did not provide a specific rule for determining ceiling priorities;
rather, it specified certain conditions for acceptable ceiling
priorities, which left some degree of choice to the implementor. Later
works \cite{LNR:09,MDPL:14} have simply assumed that the priority ceiling of a
global resource is the maximum priority of any task accessing the
resource, offset by a system-wide constant to ensure priority-boosting
semantics (as discussed in \secref{prog:boost}).

Since the MPCP is the original shared-memory multiprocessor real-time
locking protocol, it unsurprisingly has received considerable attention
in subsequent works.
\citeauthor{LS:95} \cite{LS:95} studied the choice of queue order in the MPCP and observed that
assigning tasks explicit \emph{synchronization priorities} (unrelated to their
scheduling priorities) that reflect each task's \emph{blocking
tolerance} can significantly improve schedulability \cite{LS:95}. Furthermore, \citeauthor{LS:95}
observed that simply using FIFO  queues instead of priority-ordered wait queues
can yield substantial schedulability improvements \cite{LS:95}, which is consistent with
observations made later in the context of the FIFO-based FMLP and \fmlpp
\cite{B:13a,BA:08,B:11,B:14b}.
A variant of the MPCP with FIFO queues was later also studied by
\citeauthor{CO:12a}  \cite{CO:12a}, as well as a variant in which priority boosting is replaced with
non-preemptive execution of critical sections~\cite{CO:12a}. 
\citeauthor{YLLR:14}~\cite{YLLR:14} proposed to exploit knowledge 
of each task's \textit{best-case execution time}~(BCET) to refine the blocking
bounds for the MPCP under P-FP scheduling by obtaining a more realistic bound on the worst-possible
contention possible at runtime.

In work primarily aimed at the task mapping problem (discussed in \secref{placement}),
\citeauthor{LNR:09}~\cite{LNR:09} also proposed a variant of the MPCP based on \emph{virtual
spinning}, in which blocked jobs do not actually spin, but which can be
analyzed using a WCET-inflation approach as commonly used in the
analysis of spin locks (recall \secref{spin:part}). 
While the term was not yet in widespread use at
the time, \citeauthor{LNR:09}'s ``virtual spinning'' approach is in fact an
s-oblivious analysis of the MPCP, together with a protocol tweak to
simplify said analysis. Specifically,  at most one job per core may issue a request
for a global resource at any time (as is the case with non-preemptive
spin locks) \cite{LNR:09}. Contention for global resources is thus first
resolved locally on each core, similar to the token mechanism in the partitioned
OMLP \cite{BA:10} already discussed in \secref{sob:part}, which limits global contention. 
However, in both \citeauthor{LNR:09}'s own evaluation \cite{LNR:09} as
well as in later comparisons \cite{B:11,BA:11,BA:13}, it was observed that the ``virtual
spinning'' variant of the MPCP performs poorly compared to both the
regular MPCP (under s-aware analysis) and the partitioned and clustered OMLP variants,
which are optimized for s-oblivious analysis \cite{B:11,BA:13,BA:11}.

A number of blocking analyses and integrated schedulability analyses of the MPCP 
have been proposed over the years \cite{LNR:09,R:90,R:91,B:13a}, including
analyses for arbitrary activation
models based on arrival curves~\cite{NSE:09,SNE:09}. It should also be noted that over the years a number of
misconceptions had to be
corrected  related to the critical instant \cite{YCH:17},
use of the \emph{period enforcer} technique~\cite{R:91a} to
shape locking-induced self-suspensions~\cite{CB:17}, and the proper
accounting of self-suspensions in response-time analyses \cite{CNHY:19}. These corrections should be
consulted before reusing or extending existing analyses.

The most accurate and most extensible blocking analysis of
the MPCP available today \cite{B:13a} models the blocking analysis problem as a
linear program, which allows for the systematic avoidance of structural pessimism
such as the repeated over-counting of long critical sections in
blocking bounds \cite{B:13a} (as already discussed in \secref{spin:part}). In particular, the
LP-based approach allows for a much more accurate analysis of
critical sections that contain self-suspensions (\eg, due to accesses
to devices such as GPUs) by cleanly separating the time that a job holds
a resource from the time that a job executes while being priority
boosted \cite{B:13a}. Recently, \citeauthor{PBKR:18}~\cite{PBKR:18} expressed
very similar ideas using more conventional notation,
but unfortunately did not compare their proposal with the earlier LP-based analysis
of the MPCP~\cite{B:13a}.

A substantially different, early proposal for a predictable
multiprocessor real-time locking protocol---almost as old as the MPCP,
but known not nearly as well---is due to \citeauthor{Z:92} \cite{Z:92}, who 
developed a real-time threading package \cite{SZ:92,Z:92} on top of the
\emph{Mach} microkernel's thread interface. In \citeauthor{Z:92}'s protocol, critical sections
are executed non-preemptively and blocked threads wait in FIFO order. A unique aspect is that \citeauthor{Z:92} introduced the notion of \emph{dynamic
blocking analysis}, where threads specify a \emph{maximum acceptable
wait time} when requesting a resource that is checked by the real-time resource
management subsystem by means of an \emph{online blocking analysis} that takes
current contention conditions into account. This allows the system to
dynamically determine whether the specified maximum acceptable wait time can be
exceeded, and to reject the request \emph{before} any delay is actually incurred (in contrast to a regular \emph{timeout}, which kicks in only \emph{after} the maximal wait time has been exceeded).
Such a mechanism of course comes with non-negligible
runtime overheads and has some impact on system complexity. It has been
absent from later proposals.

In work targeting P-EDF, \citeauthor{CTB:94} developed the \emph{Multiprocessor
Dynamic Priority Ceiling Protocol} (MDPCP) \cite{CTB:94}, which despite its
name is quite different from the earlier MPCP. The MDPCP ensures
progress by letting jobs holding global resources execute
non-preemptively, and orders jobs in per-resource wait queues by
decreasing priority (\ie, increasing deadlines). Additionally, the
MDPCP defines for each resource a \emph{current} priority ceiling, which is defined
``as the maximum priority of all jobs that are currently locking or
will lock'' the semaphore \cite{CTB:94}. As a result, the MDPCP is fundamentally
tied to the \emph{periodic} task model, where the arrival times of future jobs are known \apriori.
In contrast, in a sporadic setting under P-EDF scheduling,
it is impossible to precisely determine the set of jobs and their
priorities that will lock a resource in the future.

The MDPCP includes a non-work-conserving rule akin to the uniprocessor PCP \cite{SRL:90} that prevents
jobs with insufficiently high priorities (\ie, insufficiently urgent deadlines) from acquiring resources
that might still be needed by higher-priority jobs (\ie, jobs with earlier deadlines). 
More precisely, a job on processor $P$ is allowed to lock a global resource only if its
own priority exceeds (\ie, its deadline is earlier than) the
maximum priority ceiling of any resource currently in use on any of the
processors that $P$ might conflict with, where two processors ``might conflict with'' each other 
if there exists some resource that is accessed by
tasks on both processors \cite{CTB:94}. 
This rule is required to
avoid deadlock in the presence of nested requests, as will be discussed
in \secref{nested}.  In an
accompanying tech report \cite{CT:94}, \citeauthor{CT:94} further defined a second variant
of the MDPCP based on priority boosting (rather than non-preemptive
execution); this MDPCP version is also tied to the periodic task model.

\citeauthor{BLBA:07}'s partitioned FMLP \cite{BLBA:07} (previously discussed in \secref{sob:part}) was also
applied to P-FP scheduling and analyzed in an s-aware manner
\cite{BA:08a}. Recall that the FMLP variant for long resources under partitioned
scheduling relies on per-resource FIFO queues to resolve contention and
on priority boosting to ensure lock-holder progress. While priority
boosting is conceptually simple, a key detail is how to order
simultaneously priority-boosted jobs (\ie, what to do if multiple tasks
assigned to the same processor hold a resource at the same time). The
original FMLP \cite{BLBA:07} pragmatically gives priority to whichever
job acquired its resource first (which greedily minimizes the number of preemptions).
This choice, however, turned out to
be problematic from a blocking-optimality point of view \cite{BA:10,B:11}, and a
refined version of the partitioned FMLP for long resources, called the
\emph{partitioned \fmlpp}, was introduced~\cite{B:11}.

Like its predecessor, the partitioned \fmlpp uses per-resource
FIFO queues to resolve contention and priority boosting to ensure
lock-holder progress. However, it uses a subtly different tie-breaking
rule: among priority-boosted jobs, the job with the first-\emph{issued}
(rather than the first-\emph{granted}) lock request is given priority
(\ie, priority-boosted jobs are scheduled in order of increasing
lock-request times)~\cite{B:11}. To avoid preemptions in the middle of a
critical section, the \fmlpp optionally also supports non-preemptive critical
sections \cite{B:11}.

In contrast to all prior semaphore protocols for partitioned
multiprocessor scheduling, the partitioned \fmlpp (with either preemptive
and non-preemptive critical sections) ensures asymptotically optimal maximum 
s-aware pi-blocking \cite{B:11}. Specifically, due to the use of FIFO queues and
the FIFO-based priority-boosting order, the \fmlpp 
ensures that a job is delayed by at most $n-1 = O(n)$ earlier-issued requests for \emph{any} resource
each time it executes a critical section  (assuming preemptive critical sections).
Additionally, prior to a job's arrival or while it is suspended, lower-priority jobs may issue
resource requests, which may be priority-boosted at a later time and thereby cause additional pi-blocking.
Hence, whenever a job arrives or resumes from a self-suspension, there may be up to
$n-1 = O(n)$ earlier-issued, incomplete requests that can cause additional 
pi-blocking. 
Assuming non-preemptive critical sections adds only a constant amount of additional blocking \cite{B:11}. 
The \fmlpp hence ensures maximum s-aware pi-blocking within
a factor of roughly two of the known lower bound \cite{B:11}.

In addition to the optimality result, several fine-grained (\ie,
non-asymptotic) s-aware blocking analyses of the \fmlpp have been
presented \cite{B:11,B:13a,B:14b}, with an LP-based analysis \cite{B:13a}  again yielding
the most accurate results and offering the greatest flexibility,
including support for self-suspensions within critical sections~\cite[Appendix F]{B:13a}.
Recently, \citeauthor{MKZT:16}~\cite{MKZT:16} further refined the LP-based analysis and proposed additional constraints to increase
analysis accuracy. 

Overall, the \fmlpp is simple, requires no configuration or \emph{a
priori} knowledge (such as priority ceilings), has been implemented
in \litmus \cite{B:11}, and shown to be practical \cite{B:13a}.
In an empirical comparison with the MPCP, the two protocols were shown to be incomparable: 
the \fmlpp outperforms the
MPCP for many (but not all) workloads, and vice versa \cite{B:13a}.
\citeauthor{PBKR:18}~\cite{PBKR:18} observed similar trends in their comparison of the two protocols.

As in the global case (\secref{saw:global}),
it would be desirable to develop a hybrid protocol that integrates the
advantages of the FIFO-based \fmlpp \cite{B:11} with optional prioritization as in the MPCP
\cite{R:90,R:91} for the most blocking-sensitive tasks~\cite{LS:95}. Such a protocol could be easily analyzed
by combining the existing LP-based analyses \cite{B:13a} of the MPCP and the \fmlpp.

Targeting P-FP scheduling in the context of a somewhat different system model,
\citeauthor{NBN:11a} \cite{NBN:11a} considered the consolidation of legacy uniprocessor
systems onto shared multicore platforms,
where each core is used to  host a mostly independent (uniprocessor) application
consisting of multiple tasks. Whereas intra-application synchronization
needs can be resolved with existing uniprocessor protocols (as
previously employed in the individual legacy systems), the move to a
shared multicore platform can create new inter-application
synchronization needs (\eg, due to shared platform resources in the
underlying RTOS).
To support such inter-application resource sharing,
\citeauthor{NBN:11a} \cite{NBN:11a} developed the \emph{Multiprocessors Synchronization
Protocol for Real-Time Open Systems} (MSOS) protocol \cite{NBN:11a}, with the primary
goal of ensuring that the temporal correctness of each application can
be assessed without requiring insight into any other application (\ie,
applications are assumed to be opaque and may be developed by independent teams or
organizations). To this end, the MSOS protocol uses a
two-level, multi-tailed hybrid queue for each resource. Similar to the partitioned OMLP
\cite{BA:10} (discussed in \secref{sob:part}), contention for global (\ie, inter-application)
resources is first resolved on each core, such that at most one job
per core and resource can contend for global resources. Since the
MSOS protocol resolves inter-application contention with FIFO queues,
this design allows for the derivation of blocking bounds without
knowledge of any application internals, provided the maximum
per-application resource-hold time is known (for which \citeauthor{NBN:11a}
provide a bound \cite{NBN:11a}). The intra-application queues can be either
FIFO or priority-ordered queues \cite{NBN:11a}, and lock-holder progress is
ensured via priority boosting as in the MPCP \cite{R:90}. Subsequently, \citeauthor{CNHY:19}
\cite{CNHY:19} corrected an oversight in the analysis of the MSOS protocol \cite{NBN:11a} related
to the worst-case impact of self-suspensions.

\subsubsection{Semi-Partitioned Scheduling}

\emph{Semi-partitioned multiprocessor scheduling} \cite{ABD:05} is a
hybrid variant of partitioned scheduling, where most tasks are assigned
to a single processor each (as under partitioned scheduling) and a few
\emph{migratory} tasks receive allocations on two or more processors
(\ie, their processor allocations are effectively \emph{split} across
processors). Semi-partitioned scheduling has been shown to be an
effective and highly practical technique to
circumvent bin-packing limitations without incurring the complexities and
overheads of global or clustered scheduling \cite{BBA:11,BG:16}.
From a synchronization point of view, however, semi-partitioned
scheduling has not yet received much attention. 

A notable exception is
\citeauthor{ANN:12a}'s work  \cite{ANN:12a} on semaphore protocols for semi-partitioned
fixed-priority (SP-FP) scheduling.
Since known techniques for partitioned scheduling are readily applicable to non-migratory tasks, 
 the  novel challenge that must be
addressed when targeting semi-partitioned systems is migratory tasks.
To this end, \citeauthor{ANN:12a} \cite{ANN:12a} proposed two protocol variants, both using
priority-ordered wait queues and priority boosting. 

In the first protocol variant, the \emph{Migration-based Locking
Protocol under Semi-Partitioned Scheduling} (MLPS) \cite{ANN:12a}, each task is
assigned a \emph{marked} processor on which it must execute all its
critical sections. This approach simplifies the problem, as it ensures
that all of a task's critical sections are
executed on a statically known processor, which reduces the analysis
problem to the partitioned case. However, it also introduces additional task migrations,
as a migratory task that currently
resides on the ``wrong'' (\ie, non-marked) processor must first migrate
to its marked processor before it can enter a critical section, and
then  back again to its non-marked processor when it
releases the lock.

As an alternative, \citeauthor{ANN:12a}'s \emph{Non-Migration-Based Locking
Protocol under Semi-Partitioned Scheduling} (NMLPS) \cite{ANN:12a} lets migratory tasks
execute their critical sections on their current processor (\ie, on whichever processor they
happen to be executing at the time of lock acquisition). This avoids any
superfluous migrations, but causes greater analysis uncertainty as it is
now unclear on which processor a critical section will be executed.

Additional complications arise when a resource-holding migratory task
should, according to the semi-partitioning policy, be migrated in the
middle of a critical section. At this point, there are two
choices: either let the task finish its critical section before enacting
the migration, which may cause it to overrun its local budget, or
instead preempt the execution of the critical section, which causes
extra delays and makes the analysis considerably more pessimistic.
\citeauthor{ANN:12a} \cite{ANN:12a} chose the former approach in the NMLPS, which means that
budget overruns up to the length of one critical section must be
accounted for in all but the last segments of migratory tasks.

\subsubsection{Clustered Scheduling} \label{sec:saw:clust}

Like the semi-partitioned case, the topic of s-aware semaphore protocols
for clustered scheduling has not received much attention to date. The
primary works are an extension of \citeauthor{NBN:11a}'s MSOS
protocol~\cite{NBN:11a}, called the \emph{clustered MSOS} (C-MSOS) protocol~\cite{NN:13}, and the
\emph{generalized \fmlpp}~\cite{B:14b}, which establishes asymptotic tightness
of the known lower bound on s-aware pi-blocking (recall \secref{saw:opt}).

Under the C-MSOS, legacy applications are allowed to span multiple cores (\ie, there is 
one application per cluster). Local (\ie, intra-application) resources
are managed using the PIP \cite{SRL:90,EA:09} (as discussed in \secref{saw:global}). Global
(\ie, inter-application) resources are managed using a two-stage queue
as in the MSOS protocol \cite{NBN:11a}. However, in the C-MSOS protocol,
each resource's global queue can be either a FIFO queue (as in the MSOS
protocol \cite{NBN:11a}) or a round-robin queue. As before, the per-application
queues can be either FIFO or priority queues, and lock-holder progress is
ensured via priority boosting (which prevents asymptotic optimality \wrt maximum s-aware pi-blocking
under global and clustered scheduling  \cite{B:11,B:14b}). 

The generalized \fmlpp \cite{B:14b} was designed specifically to close the
``s-aware optimality gap'' \cite{B:11}, \ie, to provide a matching upper bound
of $O(n)$ s-aware pi-blocking under clustered (and hence also global)
scheduling, thereby establishing the known $\Omega(n)$ lower
bound \cite{BA:10a,B:11} to be asymptotically tight~\cite{B:14b}. The name derives from the
fact that the generalized \fmlpp produces the same schedule as the
partitioned \fmlpp when applied to partitioned scheduling \cite{B:14b}. However,
despite this lineage, in terms of protocol rules, the generalized
\fmlpp \cite{B:14b} differs substantially from the (much simpler) partitioned
\fmlpp \cite{B:11}.

The generalized \fmlpp \cite{B:14b}
resolves contention with simple per-resource FIFO queues, 
as in the prior FMLP \cite{BLBA:07} and the partitioned \fmlpp \cite{B:11}. 
The key challenge is to ensure lock-holder
progress, since neither priority inheritance nor (unrestricted) priority
boosting can yield asymptotically optimal s-aware pi-blocking bounds
under global and clustered scheduling \cite{B:11,B:14b}. Intuitively, the main
problem is that raising the priority of a lock holder (via either
inheritance or boosting) can cause other, unrelated higher-priority jobs
to be preempted. Furthermore, in pathological cases, it can cause the
\emph{same} job to be \emph{repeatedly} preempted, which gives rise to
asymptotically non-optimal s-aware pi-blocking \cite{B:11,B:14b}. The
generalized \fmlpp overcomes this effect by employing a progress
mechanism tailored to the problem, called \emph{restricted segment boosting} (RSB) \cite{B:14b}.
Under the RSB rules, in each cluster, only the (single) job with the earliest-issued request
benefits from priority boosting (with any ties in request-issue time broken arbitrarily).
In addition to this single boosted lock
holder, certain \emph{non-lock-holding} jobs are
\emph{co-boosted}, specifically to prevent repeated preemptions in the pathological scenarios
that cause the non-optimality of priority inheritance and priority boosting \cite{B:11,B:14b}. 

Based on RSB, the generalized \fmlpp ensures asymptotically optimal
s-aware pi-blocking under clustered scheduling \cite{B:14b}, and hence also
under global scheduling, which closes the s-aware optimality gap \cite{B:11}.
However, in an empirical comparison under
global scheduling \cite{YWB:15}, the generalized \fmlpp performed generally
worse than protocols specifically designed for global scheduling, which
indicates that the \emph{generalized} \fmlpp \cite{B:14b} is primarily of
interest from a blocking optimality point of view. In contrast, the
simpler \emph{partitioned} \fmlpp \cite{B:11}, which is designed specifically
for partitioned scheduling and hence avoids the complexities resulting from
clustered and global scheduling, is known to perform empirically very well and to be
practical \cite{B:13a}.

\section{Centralized Execution of Critical Sections}\label{sec:rpc}

In the preceding two sections, we have considered protocols for in-place
execution of critical sections, where jobs \emph{directly} access shared
resources. Under in-place protocols, the critical sections pertaining to
each (global) resource are spread across multiple processors (\ie,
wherever the tasks that share the resource happen to be executing). For
spin locks (\secref{spin}), this is the natural choice. 
In the case of semaphores, however, this is not
the only possibility, nor is it necessarily the best.
Instead, as discussed in \secref{prob:loc}, it is
also possible to \emph{centralize} the execution of all critical
sections onto a designated processor, the
\emph{synchronization processor} of the resource.
In fact, the very
first multiprocessor real-time semaphore protocol, namely the DPCP
\cite{RSL:88}, followed exactly this approach.

Protocols that call for the centralized execution of critical sections
are also called \emph{distributed} multiprocessor real-time locking
protocols, because the centralized approach does not necessarily require
shared memory (as opposed to the in-place execution of critical sections,
which typically relies on cache-consistent shared memory).
From a systems point of view, there are three ways to interpret such
protocols. In the following discussion, let a job's
\emph{application processor} be the processor on
which it carries out its regular execution (\ie, where it executes its non-critical
sections).

In the first interpretation, which is consistent with a distributed-systems
perspective, each critical section of a task is seen as a
synchronous \emph{remote procedure call} (RPC) to a \emph{resource
server} executing on the synchronization processor. The resource
server, which may be multi-threaded, is in charge of serializing
concurrent RPC calls. A job that issues an RPC to the synchronization
processor self-suspends after sending the RPC request and resumes again when the
resource server's response is received by the application processor. The
job's self-suspension duration thus includes both the time required to
service its own request plus any delays due to contention for the
resource (\ie, blocking due to earlier-serviced requests).
Additionally, in a real system, any communication overheads contribute
to a job's self-suspension time (\eg, transmission delays if the RPC request is communicated over a shared
interconnect, argument marshalling and unmarshalling costs, \etc).

In the second interpretation, which is typically adopted in a
shared-memory context, jobs are considered to \emph{migrate} from the
application processor to the synchronization processor when they attempt
to lock a shared resource, and to migrate back to their application
processor when unlocking the resource. All resource contention is
hence reduced to a uniprocessor locking problem. However, from a
schedulability analysis point of view, the time that the job resides on
the synchronization processor still constitutes a self-suspension \wrt to 
the analysis of the application processor.

Finally, the third interpretation, which is appropriate for both
shared-memory and distributed systems, is to see each job as a sequence
(\ie, as a linear DAG) of subjobs with precedence constraints and an
end-to-end deadline \cite{SBL:94}, where different subjobs are spread out across
multiple processors. In this view, the synchronization problem is again
reduced to a uniprocessor problem. The end-to-end analysis, however,
must deal with the fact that each DAG visits the application processor
multiple times, which can give rise to pessimism in the analysis.

From an analytical point of view---\ie, for the purpose of
schedulability and blocking analysis---the first two interpretations are
equivalent, that is, identical analysis problems must be solved and, ignoring overheads, 
identical bounds are obtained, regardless of how the
protocol is actually implemented. The third approach provides some
additional flexibility \cite{TL:94,SBL:94} and has recently
been exploited to enable modern analyses and heuristics~\cite{HCR:16,HYC:16,BCHY:17,DLBC:18,CLYL:19}.

\subsection{Advantages and Disadvantages}

Multiprocessor real-time locking protocols that centralize the execution
of critical sections offer a number of unique advantages. For one, they
can be easily applied to heterogenous multiprocessor platforms, where
certain critical sections may be inherently restricted to specific cores 
 (\eg, compute kernels can run only on GPUs, high-performance signal processing may need to take place on a DSP,
  low-power cores may lack floating-point support, \etc).
 Similarly, if
certain shared devices are accessible only from specific processors
(\eg, if there is a dedicated I/O processor), then those processors
naturally become synchronization processors for critical sections
pertaining to such devices. Furthermore, centralizing all 
critical sections is also attractive in non-cache-coherent systems,
since it avoids the need to keep a shared resource's state consistent
across multiple local memories. In fact, even in cache-coherent shared-memory
systems it can be beneficial to centralize the execution of critical
sections to avoid cache-line bouncing \cite{LDTL:12}. And last but not least,
from a real-time perspective, the centralized approach allows for the reuse
of well-established uniprocessor protocols, which for some
workloads can translate into significant schedulability improvements
over in-place approaches \cite{B:13a}.

Centralized protocols, however, also come with a major downside. Whereas
schedulability and blocking analysis is typically concerned with
worst-case scenarios, many systems also require excellent \emph{average-case}
performance, and this is where in-place execution of critical sections
has a major advantage. In well-designed
systems, resource contention is usually rare, which means that uncontested lock acquisitions are
the \emph{common case} that determines average-case performance. In a
semaphore protocol based on in-place execution, uncontested lock
acquisitions do not cause self-suspensions and can be optimized
to incur very low acquisition and release overheads (see ``futexes,'' discussed in 
\secref{futexes}). In contrast, in protocols based on the centralized approach, every
remote critical section necessarily involves a self-suspension regardless of whether the
shared resource is actually under contention,
which is likely to have a
significant negative impact on average-case performance.

\subsection{Centralized Protocols}

The original protocol for the centralized execution of critical
sections, and in fact the first multiprocessor real-time locking
protocol altogether, is the \emph{Distributed Priority Ceiling
Protocol} (DPCP) \cite{RSL:88,R:91}. Unfortunately, there is some confusion
regarding the proper name of the DPCP. The protocol was originally
introduced as the ``Multiprocessor Priority Ceiling Protocol'' and
abbreviated as ``MPCP''~\cite{RSL:88}, but then renamed to ``Distributed
Priority Ceiling Protocol,'' properly abbreviated as ``DPCP,'' shortly
thereafter~\cite{R:91}. To make matters worse, the shared-memory protocol \emph{now} known as
the MPCP (discussed in \secref{saw:part}) was introduced in the meantime \cite{R:90}.
However, for some time, the authors of several subsequent works remained unaware of the
name change, and hence a number of later publications,
including a popular textbook on real-time systems \cite{L:00}, refer to the
DPCP~\cite{RSL:88} by the name ``MPCP.'' We follow the modern terminology~\cite{R:91} and
denote by ``DPCP'' the original protocol~\cite{RSL:88}, which is based on the
centralized execution of critical sections, and reserve the abbreviation
``MPCP'' to refer to the later shared-memory protocol~\cite{R:90}, which is
based on the in-place execution of critical sections (as discussed in \secref{saw:part}).

The DPCP has been designed for P-FP scheduling. As the name suggests,
the DPCP relies on the classic PCP~\cite{SRL:90} to arbitrate conflicting
requests on each synchronization processor. To ensure resource-holder
progress, that is, to prevent lock-holding jobs from being preempted by
non-lock-holding jobs if the sets of synchronization and application
processors are not disjoint, the DPCP relies on priority boosting. As a
result of reusing the uniprocessor PCP, the DPCP effectively uses a
priority-ordered wait queue (\ie, conflicting requests from two remote
jobs are served in order of their regular scheduling priorities). This
simple design has proven to be highly effective and practical even in modern systems \cite{B:13a}.

A number of s-aware blocking analyses of the DPCP have been presented in
the literature \cite{RSL:88,R:91,B:13a,HYC:16}, with an LP-based approach \cite{B:13a}
yielding the most accurate bounds. Recent works \cite{YCH:17,CB:17} documented
some misconceptions in the original analyses \cite{RSL:88,R:91}.

\citeauthor{RC:08} \cite{{RC:08}} investigated a variant of the DPCP that uses the SRP \cite{B:91}
instead of the PCP~\cite{SRL:90} on each core. The resulting protocol, which
they called the \emph{Distributed Stack Resource Policy} (DSRP)~\cite{RC:08},
has the advantage of integrating better with P-EDF scheduling (since the
underlying SRP is well-suited for EDF).

Just as the FIFO-ordered \fmlpp complements the priority-ordered
MPCP in case of in-place critical sections, the
\emph{Distributed FIFO Locking Protocol} (DFLP) \cite{B:13a,B:14a} is a
FIFO-ordered protocol for centralized critical-section execution that
complements the priority-ordered DPCP. The DFLP works in large parts just like the DPCP,
with the exception that it does not use the PCP to manage access to
global resources. Instead, it adopts the design first introduced with
the partitioned \fmlpp \cite{B:11} (discussed in \secref{saw:part}): 
conflicting lock requests are served in FIFO order,
lock-holding jobs are (unconditionally) priority-boosted, and
jobs that are priority-boosted simultaneously on the same synchronization processor
are scheduled in order of
their lock requests (\ie, the tie-breaking rule favors earlier-issued
requests).

Blocking under the DFLP has been analyzed using an s-aware, LP-based
approach \cite{B:13a}. In an empirical comparison under P-FP scheduling based
on an implementation in \litmus \cite{B:13a}, the DFLP and DPCP were
observed to be incomparable: the DFLP performs better than the DPCP for
many (but not all) workloads, and vice versa. Similarly, both protocols
were also observed to each be incomparable with their in-place counterparts
(\ie, the partitioned \fmlpp \cite{B:11} and the MPCP \cite{R:90},
respectively).

In contrast to the DPCP, which is defined only for P-FP scheduling, the
DFLP can also be combined with P-EDF or clustered scheduling \cite{B:14a}.

\subsection{Blocking Optimality}

Centralized locking protocols have also been studied from the point of
view of blocking optimality~\cite{BA:10,B:11}, and asymptotically tight bounds on
maximum pi-blocking have been obtained for both the s-aware and s-oblivious cases~\cite{B:14a}.
Interestingly, the way in which resources and tasks are assigned to
synchronization and application processors, respectively, plays a major
role. If some tasks and resources are \emph{co-hosted}, that is, if the
sets of synchronization and application processors are not disjoint, then
maximum pi-blocking is asymptotically worse than in the shared-memory case:
a
lower bound of $\Omega(\Phi \times n)$ maximum pi-blocking  has been
established \cite{B:14a}, where $\Phi$ denotes the ratio of the maximum
response time and the minimum period of any task. Notably, this bound
holds under both s-aware and s-oblivious analysis due to the existence
of pathological corner cases in which jobs are repeatedly preempted \cite{B:14a}. Both the
DPCP \cite{RSL:88,R:91} and the DFLP \cite{B:13a,B:14a} ensure $O(\Phi \times n)$
maximum s-aware pi-blocking, and are hence asymptotically optimal in the
case with co-hosted tasks and resources \cite{B:14a}.

In contrast, if the sets of synchronization and application processors are \emph{disjoint}
(\ie, if no processor serves both regular tasks and critical
sections), then the same lower bounds as in the case of in-place
critical sections apply \cite{B:14a}: $\Omega(n)$ under s-aware analysis,
and $\Omega(m)$ under s-oblivious analysis.

The DFLP \cite{B:13a,B:14a} ensures $O(n)$ maximum s-aware pi-blocking in the
disjoint case under clustered (and hence also partitioned) scheduling,
and is thus asymptotically optimal under s-aware analysis \cite{B:14a}.
Asymptotic tightness of the $\Omega(m)$ bound on maximum s-oblivious 
pi-blocking was established with the \emph{Distributed OMLP} (D-OMLP)
\cite{B:14a}, which was obtained by transfering techniques introduced with the OMLP family for
in-place critical sections to the centralized setting.
\section{Independence Preservation: Avoiding the Blocking of Higher-Priority Tasks}\label{sec:ip}

All locking protocols discussed so far use either priority inheritance,
non-preemptive sections, unconditional priority boosting, or a restricted variant of the latter
(such as priority donation or RSB). Of these, priority
inheritance has a unique and particularly attractive property: independent
higher-priority jobs are not affected by the synchronization behavior of
lower-priority jobs.

For example, consider three tasks $\tau_1$, $\tau_2$, and $\tau_3$
under uniprocessor fixed-priority scheduling, and suppose that
$\tau_2$ and $\tau_3$ share a resource $\res_1$ that the
higher-priority task $\tau_1$ does not require. If the tasks follow a
protocol based on priority inheritance, then the response time of
$\tau_1$ is completely independent of the lengths of the critical sections  of
$\tau_2$ and $\tau_3$, which is obviously desirable. In contrast, if
$\tau_2$ and $\tau_3$ synchronize by means of non-preemptive
critical sections, then $\tau_1$'s response time, and ultimately
its temporal correctness, depends on the critical section lengths of
lower-priority tasks. In other words, the use of non-preemptive sections
induces a \emph{temporal dependency} among logically independent tasks.

Unfortunately, priority boosting, priority donation, and RSB similarly
induce temporal dependencies when they force the execution of
lock-holding lower-priority jobs. Since on multiprocessors priority inheritance is
effective only under global scheduling (recall \secref{prog:pi}), this poses a
significant problem for multiprocessor real-time systems that do not use
global scheduling (of which there are many in practice). 
In response, a
number of multiprocessor real-time locking protocols have been proposed
that avoid introducing temporal dependencies in logically
unrelated jobs. 
We use the term
\emph{independence preservation}~\cite{B:12,B:13,B:14} to generally refer to the desired
isolation property and this class of protocols.

\subsection{Use Cases}

Independence preservation is an important property in practice, but to date it has received relatively 
little attention compared to the classic spin and semaphore protocols discussed in \secrefr{spin}{rpc}.
To highlight the concept's significance,
we briefly sketch four contexts in which independence preservation is essential.

First, consider multi-rate systems with a wide range of
activation frequencies~\cite{B:13}. For instance, in automotive systems, it is not
uncommon to find tasks periods ranging from as low as 1\ms to as high as
1,000\ms or more~\cite{KZH:15}. 
Now, if a 1,000\ms task has a utilization of only 10\%, and if
each job spends only 1\% of its execution time in a critical section, then a
single such critical section is already long enough (1\ms) to render
 any 1\,ms task on the same core infeasible. This shows the importance of
independence preservation in the face of highly heterogeneous timing
requirements.

As a second example, consider an infrequently triggered sporadic event
handler that must react within, say, 100\mus (\eg, a critical
interrupt handler with a tight latency constraint). Now assume the system is deployed on an eight-core
platform and consider a shared-memory object accessed by all cores
(\eg, an OS data structure) that is protected with a non-preemptive
FIFO spin lock (\eg, as used in the MSRP~\cite{GLD:01}, recall \secref{spin:part}). 
Even if each
critical section is  only at most  20\mus long, when accounting for the transitive impact of
spin delay, the worst-case latency on every core is at least
160\mus, which renders the \emph{latency-sensitive} interrupt handler
infeasible. Generally speaking, if job release latency is a concern,
then non-independence-preserving synchronization methods must be
avoided.
 Case in point: the PREEMPT\_RT
real-time patch for the Linux
kernel converts most non-preemptive spin locks in the kernel to
suspension-based mutexes for precisely this reason. In other words, \emph{none} of the protocols discussed in \secrefr{spin}{rpc}
based on priority boosting or non-preemptive execution is appropriate for general use in the Linux kernel.
\citeauthor{TS:97} highlighted the negative impact of delays due to real-time synchronization
on interrupt latency as a major problem in multiprocessor RTOS kernels already more than 20 years ago~\cite{TS:94,TS:95,TS:96,TS:97,T:96}.

Third, consider \emph{open systems},
where at design time it is not (fully) known which applications
will be deployed and composed at runtime. Non-preemptive sections and priority boosting
are inappropriate for such systems, because the pi-blocking that
they induce is a global property, in the sense that it affects all applications, and
because the maximum critical section length in newly added applications
is not always known. Independence preservation ensures \emph{temporal isolation} among independent applications,
which greatly simplifies the  online admission and composition problem. 
\citeauthor{FLC:10} argue this point in detail \cite{FLC:10,FLC:12}. 

As a fourth and final example, independence preservation is also
important in the context of \emph{mixed-criticality systems} \cite{BD:18}, where it
is highly desirable to avoid any dependencies from critical on
non-critical components. Specifically, if the temporal correctness of a
highly critical task depends on the maximum critical section length in a
lower-criticality task, then there exists an implicit
\emph{trust} relationship: the temporal correctness of higher-criticality tasks
is guaranteed only as long as the lower-criticality task does not violate the
worst-case  timing behavior \emph{assumed}  in the analysis of higher-criticality tasks.
Independence preservation can help to avoid such
dependencies,  which violate
the freedom-from-interference principle at the heart of mixed-criticality systems~\cite{BD:18}. A detailed argument along these lines has been presented
in prior work \cite{B:14}.

\subsection{Fully Preemptive Locking Protocols for Partitioned and Clustered Scheduling}

Since priority inheritance ensures independence preservation under
global scheduling, we focus on partitioned and clustered
scheduling (and in-place critical sections).

Recall from \secref{progress} that the fundamental challenge under partitioned
scheduling can be described as follows: a lock-holding task $\tau_l$
on processor $P_1$ is preempted by a higher-priority task $\tau_h$
while $\tau_l$ blocks a remote task $\tau_b$ located on processor $P_2$.
There are fundamentally only three choices:
\begin{enumerate}
	\item priority-boost $\tau_l$ to expedite the completion of its
critical section, in which case $\tau_h$ is delayed; 
	\item do
nothing and accept that $\tau_b$'s blocking bound depends on
$\tau_h$'s execution cost; or
	\item use processor time originally
allocated to $\tau_b$ on processor $P_2$ to finish $\tau_l$'s critical section: allocation inheritance, as discussed in \secref{prog:ai}.
\end{enumerate}
Option (1) violates independence preservation,
option (2) results in potentially ``unbounded'' pi-blocking,
and hence all protocols considered in this section rely on option (3).

Allocation inheritance can be combined with both spin- and
suspension-based waiting. In both cases, the key property is that
lock-holding tasks remain fully preemptable at all times, and that they
continue to execute with their regular (\ie, non-boosted priorities),
which ensures the desired independence-preservation property.

\citeauthor{HP:01} \cite{HP:01} were the first to describe an independence-preserving 
multiprocessor real-time synchronization protocol, which they realized in the Fiasco L4
microkernel under the name \emph{local helping}. Given that microkernels in the L4 family rely
exclusively on IPC, the shared resource under contention is in fact a
single-threaded \emph{resource server} that synchronously responds to
invocations from client tasks, thereby implicitly
sequencing concurrent requests (\ie, the execution of the server's
response handler forms the ``critical section'').

\citeauthor{HP:01}'s solution \cite{HP:01} is based on earlier work by
\citeauthor{HH:01} \cite{HH:01}, who described an elegant way to
realize temporally predictable resource servers on uniprocessors  that is analytically
equivalent to the better-known (uniprocessor) \emph{Bandwidth Inheritance} (BWI) protocol
\cite{LLA:01} (which was independently proposed in the same year). 
Specifically, \citeauthor{HH:01} proposed a mechanism that they termed \emph{helping}:whenever a
blocked client (\ie, a client thread that seeks to rendezvous with a
server thread that is not waiting to accept a synchronous IPC
message) is selected by the scheduler, the server process is dispatched
instead \cite{HH:01} (see also \emph{time-slice donation} \cite{SBK:10}).\footnote{
\citeauthor{HH:01}'s helping mechanism~\cite{HH:01} is named in analogy to the ``helping''
employed in wait-free algorithms~\cite{H:91}, but fundamentally a different mechanism. 
``Helping'' in wait-free algorithms does not rely on any support by the OS~\cite{H:91}; rather, it
is realized exclusively with a processor's atomic operations (such as an atomic compare-and-swap instruction).
}

\citeauthor{HP:01} extended \citeauthor{HH:01}'s helping approach to
 multiprocessors (under P-FP scheduling) and systematically considered key
design choices and challenges. Specifically, with local
helping,\footnote{\citeauthor{HP:01} also describe a variant called
  \emph{remote helping} where a blocked client migrates to the core
  assigned to the server \cite{HP:01}, which however is not an attractive
  solution from an analytical point of view and thus not further
  considered here.} a preempted resource server is migrated (\ie, \emph{pulled})
  to the core of the blocked client, at which point
the uniprocessor helping mechanism \cite{HH:01} can be applied---an instance of the
allocation inheritance principle (\secref{prog:ai}). 
However, two interesting challenges arise:

\begin{enumerate}
\item
  What should blocked clients do when the resource server is already
  executing on a remote core?
\item
  How does a blocked client learn that the resource server was preempted
  on a remote core?
\end{enumerate}

\citeauthor{HP:01} considered two fundamental approaches. In the first approach,
which they termed \emph{polling} \cite{HH:01}, the blocked client simply
executes a loop checking whether the resource server has become
available for dispatching (\ie, whether it has been preempted), which
addresses both questions. This polling approach is equivalent to
preemptable spinning (\ie, it is conceptually a busy-wait loop that
happens to spin on the process state of the server process), with the
typical advantage of avoiding self-suspensions and the typical
disadvantage of potentially wasting many processor cycles.

As an alternative, \citeauthor{HP:01} considered a \emph{sleep and callback} \cite{HP:01}
approach, where the blocked client registers its willingness to help in
a data structure and then self-suspends. When the server process is
preempted, the register of potential helpers is consulted and one or
more blocked clients are woken up by triggering their callback
functions, which requires sending \emph{inter-processor interrupts}
(IPIs) to the cores on which they are hosted. The sleep and callback
approach is equivalent to self-suspending clients, which comes with the typical
advantage that blocked clients yield the processor to lower-priority
tasks,  but which also introduces the typical analytical challenges and overhead issues. Since
\citeauthor{HP:01} expected critical sections (\ie, server request handlers)
in their system to be quite short, and due to implementation
challenges associated with the sleep and callback approach, \citeauthor{HP:01}
chose the polling approach in their implementation \cite{HP:01}.

Given that synchronous IPC (with single-threaded processes) and mutual
exclusion are duals of each other, \citeauthor{HP:01}'s work \cite{HP:01} directly
applies to the multiprocessor real-time locking problem, and in fact
their combination of local helping and synchronous IPC can be considered
a multiprocessor real-time locking protocol that combines
priority-ordered wait queues with allocation inheritance under P-FP
scheduling.

Not long after \citeauthor{HP:01} \cite{HH:01},
\citeauthor{HA:02a} \cite{HA:02a,H:04,HA:06} proposed the use
of allocation inheritance to realize a predictable suspension-based
locking protocol for Pfair scheduling~\cite{BCPV:96,SA:06}. While Pfair is a global scheduler, 
it is not compatible with priority inheritance due to its sophisticated scheduling rules and more nuanced notion of ``priority.'' 
\citeauthor{HA:02a} hence proposed  allocation inheritance as a generalization of the priority-inheritance principle 
that neither assumes priority-driven scheduling nor requires a priority concept. As already mentioned in \secref{prog:ai},
\citeauthor{HA:02a} coined the term ``allocation inheritance,'' which we
have adopted in this survey to refer to the general idea of dynamically re-purposing processor-time allocations
to ensure lock-holder progress.
\citeauthor{HA:02a} further
considered two alternatives to allocation inheritance named \emph{rate
inheritance} and \emph{weight inheritance} \cite{HA:02a,HA:06,H:04}, which are
both specific to Pfair scheduling and not further considered herein.

Much later, \citeauthor{FLC:10} \cite{FLC:10,FLC:12} extended the uniprocessor
bandwidth inheritance protocol \cite{LLA:01} to multiprocessors, targeting in
particular multiprocessors under reservation-based scheduling. The
resulting protocol, the \emph{Multiprocessor Bandwidth Inheritance}
(MBWI) protocol \cite{FLC:10,FLC:12}, combines allocation inheritance with FIFO-ordered wait
queues and busy-waiting. 

\citeauthor{FLC:10} observed that,
since the allocation inheritance principle is not specific to any
particular scheduling algorithm, the MBWI protocol can be employed
\emph{without any modifications or adaptations} under partitioned,
global, or clustered scheduling \cite{FLC:10,FLC:12}. In fact, it can even be
used in unchanged form under semi-partitioned scheduling or in the
presence of tasks with \emph{arbitrary processor affinities}~(APAs)~\cite{GCB:13,GCB:15}.

Like \citeauthor{HP:01} \cite{HP:01}, \citeauthor{FLC:12} chose to follow the polling
approach in their implementation of the MBWI protocol in \litmus
\cite{FLC:12}. As an interesting practical tweak, in \citeauthor{FLC:12}'s implementation,  polling jobs detect when the
lock-holding job self-suspends (\eg, due to I/O) and then self-suspend
as well, to prevent wasting large amounts of processor time when
synchronizing access to resources that induce self-suspensions within
critical sections (\eg, such as GPUs) \cite{FLC:10,FLC:12}. Nonetheless, the
MBWI protocol is fundamentally a spin-based protocol \cite{FLC:10,FLC:12}.

In work targeting Linux  with the
PREEMPT\_RT patch, \citeauthor{BB:12} \cite{BB:12} proposed to replace Linux's
implementation of priority inheritance with allocation
inheritance (which they referred to as \emph{migratory priority
inheritance} \cite{BB:12}) because priority inheritance is
ineffective in the presence of tasks with disjoint APAs, which Linux supports~\cite{GCB:13,GCB:15}. 

In
contrast to \citeauthor{FLC:10}'s MBWI protocol \cite{FLC:10,FLC:12} and
\citeauthor{HP:01}'s local helping implementation in Fiasco \cite{HP:01}, \citeauthor{BB:12} \cite{BB:12}
proposed to retain Linux's usual semaphore semantics, wherein blocked
tasks self-suspend. Similarly to \citeauthor{HP:01}'s work \cite{HP:01}, and unlike
the MBWI protocol \cite{FLC:10,FLC:12}, \citeauthor{BB:12}'s proposal \cite{BB:12} uses priority-ordered wait queues.

The \emph{$O(m)$ Independence-preserving Protocol} (OMIP) \cite{B:13,B:12}
for clustered scheduling is the only protocol based on allocation
inheritance that achieves asymptotic blocking optimality under
s-oblivious analysis. Recall that the only other protocol for clustered
scheduling that is asymptotically optimal \wrt maximum s-oblivious pi-blocking
is the clustered OMLP \cite{BA:11,B:11,BA:13}, which relies on
priority donation, a restricted variant of priority boosting, and which
hence is not independence preserving. 
The OMIP improves upon the clustered OMLP by
replacing priority donation with allocation inheritance, which ensures
that lock-holding tasks remain preemptable at all times. As a result of
this change in progress mechanism, the OMIP requires a multi-stage
hybrid queue \cite{B:13,B:12} similar to the one used in the global OMLP \cite{BA:10},
in contrast to the simple FIFO queues used in the clustered OMLP
\cite{BA:11}. In fact, in the special case of global scheduling, the OMIP
reduces to the global OMLP, and hence can be understood as a
generalization of the global OMLP \cite{B:13,B:12}. This also underscores
that allocation inheritance is a generalization of priority inheritance (\secref{prog:ai}).

The OMIP, which has been prototyped in \litmus \cite{B:13}, is
suspension-based and hence requires the implementation to follow a sleep
and callback approach \cite{HP:01}. However, because the available blocking
analysis is s-oblivious \cite{B:13}, which already accounts for suspension times as
processor demand (\secref{sob}), it can be trivially (\ie, without any changes to the
analysis) changed into a spin-based protocol. Similarly, the
OMIP's multi-stage hybrid queue could be combined with the
MBWI protocol~\cite{FLC:10,FLC:12} to lower the MBWI protocol's bounds on
worst-case s-blocking (\ie, $O(m)$ bounds as in the OMIP rather than the MBWI
protocol's $O(n)$ bounds).

One unique feature of the OMIP worth noting is that, since the blocking bounds
are completely free of any terms depending on the number of tasks $n$,
it does not require any \emph{trust} on the maximum number of tasks
sharing a given resource. This makes it particularly interesting for
open systems and mixed-criticality systems, where the final
workload composition and resource needs are either not known or not trusted.

Exploiting
this property, as well as a close correspondence between s-oblivious analysis
and certain processor reservation techniques, the OMIP has been used to derive a locking protocol for
\emph{Virtually Exclusive Resources} (VXR) \cite{B:12} and a synchronous \emph{Mixed-Criticality IPC}
(MC-IPC) protocol \cite{B:14}. The VXR
and MC-IPC protocols exhibit three key features that aid system integration in a
mixed-criticality context:
\begin{enumerate}
	\item the number of tasks
sharing a given resource need not be known for analysis purposes and no
trust is implied,
	\item different maximum critical section
lengths may be assumed in the analysis of high- and low-criticality
tasks, and
	\item even non-real-time, best-effort background
tasks may access shared resources in a mutually exclusive way without
endangering the temporal correctness of high-criticality tasks \cite{B:14}.
\end{enumerate}

Concurrently to the OMIP \cite{B:13}, \citeauthor{BW:13a} \cite{BW:13a} proposed the
\emph{Multiprocessor Resource Sharing Protocol} (MrsP) for P-FP
scheduling. MrsP combines allocation inheritance with FIFO-ordered
spin locks and local per-processor (\ie, uniprocessor) priority
ceilings. Specifically, each global resource is protected with a
FIFO-ordered spin lock as in the MSRP \cite{GLD:01} (recall \secref{spin:part}), but jobs
remain fully preemptable while spinning or holding a resource's spin
lock, which ensures independence preservation. 

To ensure progress
locally, each resource is further managed, independently and
concurrently on each processor, with a \emph{local} priority ceiling
protocol (either the PCP \cite{SRL:90} or SRP \cite{B:91}).
From the point of view of
the local ceiling protocol, the entire request for a global resource,
including the spin-lock acquisition and any spinning, is considered to
constitute a single ``critical section,'' which is similar to the use of
contention tokens in the partitioned OMLP \cite{BA:10}. Naturally, when
determining a resource's local, per-processor priority ceiling, only
\emph{local} tasks that access the resource are considered.

MrsP employs allocation
inheritance to ensure progress across processors. Instead of
literally spinning, waiting jobs may thus be replaced transparently with the
lock holder, or otherwise contribute towards completing the operation of
the lock-holding job (as in the SPEPP protocol \cite{TS:97}), which means that
an implementation of the MrsP can follow the simpler polling approach
\cite{HP:01}.

\citeauthor{BW:13a} \cite{BW:13a} motivate the design of the MrsP with the
observation that blocking bounds for the MrsP can be stated in a way that is syntactically
virtually identical with the classic uniprocessor response-time analysis
equation for the PCP and SRP. For this reason, \citeauthor{BW:13a}
consider the MrsP to be particularly ``schedulability compatible,'' and note that the
MrsP is the first protocol to achieve this notion of compatibility.

While this is true in a narrow, syntactical sense, it should also be
noted that every other locking protocol discussed in this survey is also
``schedulability compatible'' in the sense that the maximum blocking
delay can be bounded \apriori and incorporated into a response-time analysis.
Furthermore, the ``schedulability compatible'' blocking analysis of the
MrsP presented by \citeauthor{BW:13a} \cite{BW:13a} 
is
structurally similar to \citeauthor{GLD:01}'s original analysis of the MSRP \cite{GLD:01}
and relies on execution-time
inflation (which is inherently pessimistic~\cite{WB:13a}, recall \secref{spin:part}).
More modern blocking analysis approaches avoid execution-time inflation
altogether \cite{WB:13a} and have a more detailed model of contention (\eg,
holistic blocking analyses~\cite{B:11,PBKR:18} or LP-based analyses
\cite{B:13a,WB:13a}). A less-pessimistic analysis of the MrsP using
state-of-the-art methods would similarly not resemble the classic
uniprocessor response-time equation in a one-to-one fashion;
``schedulability compatibility'' is thus less a property of the protocol, and more one of
the particular analysis (which admittedly is possible in this particular form only for the MrsP).

Recently, \citeauthor{ZGBW:17} introduced a new blocking analysis of the MrsP \cite{ZGBW:17}
that avoids execution-time inflation using an analysis setup adopted from \citeauthor{WB:13a}'s LP-based analysis framework~\cite{WB:13a}.
However, in contrast to \citeauthor{WB:13a}'s analysis, \citeauthor{ZGBW:17}'s analysis is not LP-based. 
Rather, \citeauthor{ZGBW:17} follow a notationally more conventional approach based
on the explicit enumeration of blocking critical sections, which however has been  
refined to match the accuracy of \citeauthor{WB:13a}'s LP-based analysis of the MSRP \cite{GLD:01}. 
While \citeauthor{ZGBW:17}'s new analysis~\cite{ZGBW:17} is not ``schedulability compatible'' according to \citeauthor{BW:13a}'s syntactic criterion \cite{BW:13a},
\citeauthor{ZGBW:17}'s analysis has been shown  \cite{ZGBW:17} to be much less pessimistic than \citeauthor{BW:13a}'s original inflation-based but ``schedulability compatible'' analysis \cite{BW:13a}.

\section{Protocols for Relaxed Exclusion Constraints} \label{sec:relx}

In the preceding sections, we have focused exclusively on protocols that
ensure mutual exclusion. However, while mutual exclusion is without a
doubt the most important and most widely used constraint in practice, many systems
also exhibit resource-sharing problems that call for \emph{relaxed exclusion} to allow
for some degree of concurrency in resource use. The two relaxed
exclusion constraints that have received most attention in prior work
are \emph{reader-writer} (RW) exclusion and \emph{$k$-exclusion} (KX).

RW synchronization is a classic synchronization problem \cite{CHP:71} wherein a
shared resource may be used either exclusively by a single \emph{writer}
(which may update the resource's state) or in a shared manner by any number of
\emph{readers} (that do not affect the resource's state). RW
synchronization is appropriate for shared resources that are rarely
updated and frequently queried. For instance, an in-memory data store
holding sensor values, route information, mission objectives, \etc
that is used by many subsystems and updated by few is a prime candidate for RW
synchronization. Similarly, at a lower level, the list of topic
subscribers in a publish/subscribe middleware is another example of
rarely changing, frequently queried data that must be
properly synchronized.

KX synchronization is a generalization of mutual exclusion to
\emph{replicated shared resources}, where there are multiple identical, inter-changeable 
copies (or \emph{replicas}) of a shared resource.
Replicated resources can be managed with counting
semaphores, but require special handling in multiprocessor real-time
systems to ensure analytically sound pi-blocking bounds. Examples where a need for
KX synchronization arises in real-time systems include multi-GPU systems
(where any task may use any GPU, but each GPU must be used by at most
one task at a time) \cite{EA:12}, systems with multiple DMA engines (where again
any task may program any DMA engine, but each DMA engine can carry out
only one transfer at a time), and also virtual resources such as cache
partitions \cite{WHKA:13}.

Since both RW and KX synchronization generalize mutual exclusion, any of
the locking protocols discussed in the previous sections may be used to
solve RW or KX synchronization problems. This, however, would be
needlessly inefficient. The goal of locking protocols designed
specifically for RW and KX synchronization is thus both \textbf{(i)} to
increase parallelism (\ie, avoid unnecessary blocking) and
\textbf{(ii)} to reflect this increase in parallelism as improved
\emph{worst-case} blocking bounds. Goal (ii) sets real-time RW and KX
synchronization apart from classic (non-real-time) RW and KX solutions,
since in a best-effort context it is sufficient to achieve a decrease in
blocking \emph{on average}.

We acknowledge that there is a large body of prior work on
relaxed exclusion protocols for non-real-time and uniprocessor systems,
a discussion of which is beyond the scope of this survey,
and in the following focus exclusively on work targeting multiprocessor real-time systems.

\subsection{Phase-Fair Reader-Writer Locks}\label{sec:rw}

The first multiprocessor real-time RW protocol achieving both
goals (i) and (ii) was proposed by \citeauthor{BA:09} \cite{BA:09}. Prior work on RW
synchronization for uniprocessors or general-purpose multiprocessor systems had
yielded three general classes of RW locks:

\begin{enumerate}
\item
  \emph{reader-preference locks}, where pending writers gain access to a
  shared resource only if there are no pending read requests;
\item
  conversely \emph{writer-preference locks}; and
\item
  \emph{task-fair locks} (or FIFO RW locks), where tasks gain access to
  the shared resource in strict FIFO order, but consecutive readers may
  enter their critical sections concurrently.
\end{enumerate}

From a worst-case perspective, reader-preference locks are undesirable
because reads are expected to be frequent, which gives rise to prolonged writer
starvation, which in turn manifests as extremely pessimistic
blocking bounds \cite{BA:10a}. Writer-preference locks are better suited to
real-time systems, but come with the downside that, if there are
potentially multiple concurrent writers, the worst-case blocking bound
for each reader will
pessimistically account for rare corner-case scenarios in which a reader is
blocked by multiple consecutive writers. Finally, task-fair locks degenerate to regular mutex locks in the pathological
case when readers and writers are interleaved in the queue;
consequently, task-fair locks improve average-case parallelism,
but their worst-case bounds do not reflect the desired gain in parallelism.

\citeauthor{BA:09} introduced
\emph{phase-fair locks} \cite{BA:09,BA:10a}, a new category of RW locks 
better suited to worst-case analysis. In a phase-fair lock, reader and
writer \emph{phases} alternate, where each reader phase consists of any
number of concurrent readers, and a writer phase consists of a single
writer. Writers gain access to the shared resource in FIFO order \wrt
other writers. Importantly, readers may join an ongoing reader phase
only if there is no waiting writer; otherwise newly arriving readers
must wait until the \emph{next} reader phase, which starts after the
next writer phase. 

These rules ensure that writers do not starve (as in
a writer-preference or task-fair lock), but also ensure $O(1)$
blocking for readers as any reader must await the completion of at most
one reader phase and one writer phase before gaining access to the shared resource
\cite{BA:09,BA:10a}. As a result, phase-fair locks yield much improved blocking
bounds for both readers and writers if reads are significantly more
frequent than updates \cite{BA:09,BA:10a}.

Several phase-fair RW spin-lock algorithms have been presented,
including compact (\ie, memory-friendly) spin locks \cite{BA:10a,B:11}, ticket
locks \cite{BA:09,BA:10a,B:11}, and cache-friendly scalable queue locks
\cite{BA:10a,B:11}. Concerning RW semaphores, the clustered OMLP \cite{BA:10,B:11,BA:13} based
on priority donation includes a phase-fair RW variant, which also achieves asymptotically optimal maximum
s-oblivious pi-blocking~\cite{BA:10,B:11,BA:13}.

\subsection{Multiprocessor Real-Time $k$-Exclusion Protocols}\label{sec:kex}

As already mentioned, in best-effort systems, KX synchronization can be
readily achieved with counting semaphores. Furthermore, in the case of
non-preemptive spin locks, classic ticket
locks can be trivially generalized to KX locks. We hence focus in the following on semaphore-based
protocols for multiprocessor real-time systems.

Given the strong progress guarantees offered by priority donation (discussed in \secref{sob:clust}),
it is not difficult to generalize the clustered OMLP to KX
synchronization \cite{BA:11,B:11,BA:13}, which yields a protocol that is often abbreviated
as CK-OMLP in the literature. Since it derives from the clustered OMLP,
the CK-OMLP applies to clustered scheduling, and hence also supports global
and partitioned scheduling. Furthermore, under s-oblivious analysis, it
ensures asymptotically optimal maximum pi-blocking \cite{BA:11,B:11,BA:13}. As such, it
covers a broad range of configurations. However, as it relies on
priority donation to ensure progress, it is not independence-preserving (recall \secref{ip}),
which can be a significant limitation especially when dealing with
resources such as GPUs, where critical sections are often naturally
quite long. Subsequent protocols were  specifically designed to overcome  this
limitation of the CK-OMLP.

\citeauthor{EA:11} considered globally scheduled multiprocessors and proposed
the \emph{Optimal $k$-Exclusion Global Locking Protocol} (O-KGLP)
\cite{EA:11,EA:13}. In contrast to the  CK-OMLP, their
protocol is based on priority inheritance, which is possible due to the
restriction to global scheduling, and which enables the O-KGLP to be
independence-preserving.

In the context of KX synchronization, applying priority inheritance is
not as straightforward as in the mutual exclusion case  because priorities must not be ``duplicated''. That
is, while there may be multiple resource holders (if $k > 1$), only at
most one of them may inherit a blocked job's priority at any time, as
otherwise analytical complications similar to those caused by priority
boosting arise (including the loss of independence preservation). The
challenge is thus to determine, dynamically at runtime and with low
overheads, which resource-holding job should inherit which blocked job's
priority.

To this end, \citeauthor{EA:11} \cite{EA:11,EA:13} proposed a multi-ended hybrid queue
consisting of a shared priority queue that forms the tail (as in the
global OMLP \cite{BA:10}) and a set of per-replica FIFO queues (each of length
$\frac{m}{k}$) that serve to serialize access to specific replicas. A
job $J_i$ holding a replica $\res_q^x$ inherits the priorities of
the jobs in the FIFO queue corresponding to the $x$\xth replica of
resource $\res_q$, and additionally the priority of \emph{one} of the $k$
highest-priority jobs in the priority tail queue. Importantly, if
$J_i$ inherits the priority of a job $J_h$ in the priority tail
queue, then $J_h$ is called the \emph{claimed job} of $\res_q^x$ and
moved to the FIFO queue leading to $\res_q^x$ when $J_i$ releases
$\res_q^x$.
This mechanism ensures that priorities are not ``duplicated'' while also
ensuring progress. In fact, \citeauthor{EA:11} established that the O-KGLP is
asymptotically optimal \wrt s-oblivious maximum pi-blocking
\cite{EA:11,EA:13}.

In work on predictable interrupt management in multi-GPU systems \cite{EA:12a},
\citeauthor{EA:12a} further proposed a KX variant of the FMLP (for long
resources) \cite{BLBA:07}. This variant, called $k$-FMLP, simply consists of
one instantiation of the FMLP for each resource replica (\ie, each
resource replica is associated with a replica-private FIFO queue that does not interact with other queues). When
jobs request access to a replica of a $k$-replicated resource, they
simply enqueue in the FIFO queue of the replica that ensures the minimal
worst-case wait time (based on the currently enqueued requests). While the
$k$-FMLP is not asymptotically optimal under s-oblivious analysis
(unlike the O-KGLP and the CK-OMLP), it offers the advantage of being
relatively simple to realize \cite{EA:12a} while also ensuring independence
preservation under global scheduling (unlike the CK-OMLP).

\citeauthor{WEA:12} \cite{WEA:12} realized that blocking under the O-KGLP \cite{EA:11,EA:13} could
be further improved with a more nuanced progress mechanism, which they
called \emph{Replica-Request Priority Donation} (RRPD) \cite{WEA:12}, and
proposed the \emph{Replica-Request Donation Global Locking Protocol}
(\rrdglp) based on RRPD \cite{WEA:12}. As the name suggests, RRPD transfers the
ideas underlying priority donation \cite{BA:11,B:11,BA:13} to the case
of priority inheritance under global scheduling. Importantly, whereas priority
donation applies to \emph{all} jobs (regardless of whether they
request any shared resource), RRPD applies only to jobs that synchronize (\ie, that actually
request resource replicas). This ensures that RRPD is
independence-preserving (in contrast to priority donation); however,
because RRPD incorporates priority inheritance, it is effective only under
global scheduling.

Like the O-KGLP, the \rrdglp is asymptotically optimal \wrt to maximum
s-oblivious pi-blocking. Furthermore, when fine-grained (\ie,
non-asymptotic) pi-blocking bounds are considered, the \rrdglp ensures
higher schedulability due to lower s-oblivious pi-blocking bounds (\ie, the \rrdglp
achieves better constant factors than the O-KGLP) \cite{WEA:12}.

Another CK-OMLP variant is the PK-OMLP due to \citeauthor{YLLR:13} \cite{YLLR:13}.
Priority donation as used by the CK-OMLP ensures that there is at most
one resource-holding job per processor at any time. For resources such
as GPUs, where each critical section is likely to include significant
self-suspension times, this is overly restrictive. The PK-OMLP, which is
intended for partitioned scheduling, hence improves upon the CK-OMLP by
allowing multiple jobs on the same processor to hold replicas at the
same time \cite{YLLR:13}. Furthermore, \citeauthor{YLLR:13} presented an s-aware
blocking analysis of the PK-OMLP under P-FP scheduling, which enables a more accurate
treatment of self-suspensions within critical sections (this analysis was later amended	 by
\citeauthor{YCH:17}~\cite{YCH:17}). As a result,
the PK-OMLP usually outperforms the
CK-OMLP when applied in the context of multi-GPU systems \cite{YLLR:13}.
More recently, \citeauthor{YLLC:16} presented another KX locking protocol
specifically for P-FP scheduling and s-aware analysis that forgoes
asymptotic optimality in favor of priority-ordered wait queues and
non-preemptive critical sections \cite{YLLC:16}.

Finally, all discussed KX protocols only ensure that no more than $k$
tasks enter critical sections (pertaining to a given resource) at the
same time. This, however, is often not enough: to be practical, a KX
protocol must also be paired with a \emph{replica assignment
protocol} to match lock holders to replicas. That is, strictly speaking
a KX algorithm blocks a task until it may use \emph{some} replica, but it usually
is also necessary to quickly resolve exactly \emph{which} replica
a task is supposed to use. To this end, \citeauthor{NYYE:16}~\cite{NYYE:16}
introduced several
algorithms for the \emph{$k$-exclusion replica assignment problem},
with the proposed algorithms representing different tradeoffs
\wrt optimality considerations and overheads in practice~\cite{NYYE:16}.
\section{Nested Critical Sections}\label{sec:nested}

Allowing \emph{fined-grained}, \emph{incremental nesting} of critical sections---that
is, allowing tasks already holding one or more locks to issue further lock
requests---adds
another dimension of  difficulty to the multiprocessor real-time
locking problem. 

First of all, if tasks may request locks in any order, then allowing tasks to
nest critical sections can easily result in deadlock.
However, even if programmers
take care to manually avoid deadlocks by carefully ordering all requests, the blocking analysis problem
becomes \emph{much} more challenging. In fact, in the presence of nested
critical sections, the blocking analysis problem is NP-hard even in
extremely simplified settings \cite{WB:14}, while it can be solved in
polynomial time on both uniprocessors (even in the presence of nesting) and multiprocessors
in the absence of nesting (at least in simplified settings)~\cite{WB:14,W:18}.

As a result, today's nesting-aware blocking analyses 
either are computationally highly expensive or yield
only coarse, structurally pessimistic bounds. Furthermore, authors frequently
exclude nested critical sections from consideration altogether
(or allow only coarse-grained nesting via group locks, see \secref{nested:groups} below).
In the words of \citeauthor{R:90} in
his original analysis of the MPCP \cite{R:90}: ``{[}s{]}ince nested global
critical sections can potentially lead to large increases in blocking
durations, {[}\ldots{}{]} global critical sections cannot nest other
critical sections or be nested inside other critical sections.'' In the
subsequent decades, many authors have adopted this expedient assumption.

The aspect unique to nesting that makes it so difficult to derive accurate blocking bounds 
is \emph{transitive blocking}, where
jobs are indirectly delayed due to contention for resources that they
(superficially) do not even depend on. For example, if a job $J_1$
requires only resource $\res_1$, but another job $J_2$ holds $\res_1$
while trying to acquire a second resource $\res_2$ in a nested fashion,
then $J_1$ is exposed to delays due to contention for $\res_2$
even though it does not require $\res_2$ itself. 

While this is a trivial
example, such transitive blocking can arise via long \emph{transitive blocking
chains} involving arbitrarily many resources and jobs   on potentially all
processors. As a result, characterizing the effects of such chains in a safe way and
without accruing excessive pessimism is a very challenging analysis
problem. \citeauthor{BBW:16} \cite{BBW:16} provide more detailed examples of
some of the involved challenges.

Nonetheless, despite all difficulties,  fine-grained lock nesting arises naturally in many systems
\cite{BBW:16} and is usually desirable (or even unavoidable) from an
average-case perspective. That is, even though fine-grained locking
may not be advantageous from a worst-case blocking perspective, the
alternative---\emph{coarse-grained locking}, where lock scopes are
chosen to protect multiple resources such that tasks must never  acquire
more than one lock at any time---is usually much worse in terms of average-case
contention, attainable parallelism, scalability, and ultimately
throughput.
Robust and flexible support for fine-grained nesting is thus
indispensable. While the current state of the art, as discussed in
the following, may not yet fully meet all requirements in practice,
 support for nesting in analytically sound multiprocessor real-time
 locking protocols is an active area of research and we expect
capabilities to continue to improve in the coming years.

\subsection{Coarse-Grained Nesting with Group Locks}\label{sec:nested:groups}

One easy way of allowing at least some degree of ``nested'' resource
usage, without incurring the full complexity of fine-grained locking, is
to (automatically) \emph{aggregate} fine-grained resource
requests into coarser \emph{resource groups} protected by \emph{group locks}.
That is, instead of associating a lock with each resource (which is the usual
approach), the set of shared resources is partitioned into disjoint
resource groups and each such resource group is associated with a group lock.

Under this approach, prior to using a shared resource, a task must first
acquire the corresponding group
lock. Conversely, holding a resource group's lock entitles a task to use any
resource in the group. To eliminate lock nesting, resource groups are defined such that, if
any task ever requires access to two resources $\res_a$ and $\res_b$
simultaneously, then $\res_a$ and $\res_b$ are part of the same
resource group. More precisely, resource groups are defined by the transitive
closure of the ``may be held together'' relation
\cite{BLBA:07}. As a result, no task ever holds more than one group lock.

The use of group locks was first proposed by \citeauthor{R:90} \cite{R:90} in the
context of the MPCP, and re-popularized in recent years by the FMLP
\cite{BLBA:07}. Both protocols rely exclusively on group locks, in the sense
that to date no analysis with support for fine-grained nesting has been
presented for either protocol.

From an analysis point of view, group locks are extremely
convenient---the synchronization problem (at runtime) and the blocking
analysis problem (at design time) both fully reduce to the non-nested
cases. As a result, \emph{any} of the protocols and analyses surveyed in
the preceding sections can be \emph{directly} applied to the analysis of
group locks.

However, there are also obvious downsides in practice. For one,
resource groups must be explicitly determined at design time and group membership must be known
at runtime (or compiled into the system), so that tasks may acquire the appropriate group lock when
requesting a resource, which from a
software development point of view is at least inconvenient, and may actually pose
significant engineering challenges in complex systems. Furthermore, since its
very purpose is to eliminate incremental lock acquisitions, group
locking comes with all the scalability and performance problems
associated with coarse-grained synchronization.

Last but not least, for certain resources and systems, it may not be
possible to define appropriate resource groups. As a pathological
example, assume a UNIX-like kernel and consider the file system's
\emph{inode} objects, which are typically arranged in a tree that
reflects the file system's hierarchy. Importantly, certain file system procedures
operate on multiple inodes at once (\eg, the inodes for a file and its
parent directory), and since files may be moved dynamically at runtime
(\ie, inodes may show up at any point in the tree), virtually any two
inodes could theoretically be held simultaneously at some point.
As a result, the set of all
inodes collapses into a single resource group, with obvious performance
implications.

Thus, while group locks can help to let programmers
express resource use in a fine-grained manner, clearly more flexible
solutions are needed. Specifically, for performance and scalability reasons, non-conflicting requests for
different resources should generally be allowed to proceed in parallel, even if some task may
simultaneously hold both resources at some other time. We next discuss multiprocessor real-time locking 
protocols that realize this to varying degrees. 

\subsection{Early Protocol Support for Nested Critical Sections} \label{sec:spin:early}

Research on support for fine-grained nesting in real-time multiprocessor
locking protocols can be grouped into roughly two eras: a period of initial
results that lasted from the late 1980s until the mid 1990s, and a recently
renewed focus on the topic, which started to emerge in 2010. We discuss protocols from the
initial period next and then discuss the more recent developments in
\secref{nested:recent} and asymptotically optimal fine-grained nesting in \secref{nested:rnlp} below.

The first multiprocessor real-time locking protocol, the DPCP \cite{RSL:88},
was in fact also the first protocol to include support for fine-grained
nesting, albeit with a significant restriction. Recall from \secref{rpc} that
the DPCP executes critical sections centrally on designated
synchronization processors. Because the DPCP relies on the uniprocessor
PCP on each synchronization processor,
and since the PCP supports nested critical sections (and prevents
deadlock) \cite{SRL:90}, it is in fact trivial for the DPCP to support nested
critical sections \emph{as long as nesting occurs only among resources
assigned to the same processor}. Resources assigned to different
synchronization processors, however, are not allowed to be nested under
the DPCP~\cite{RSL:88,R:91}.

Consequently, the DPCP's support for fine-grained nesting is actually
not so different from group locks---all nesting must be taken into
account up front, and resources assigned to the same synchronization
processor form essentially a resource group. In fact, just as there is
no parallelism among non-conflicting requests for resources protected by
the same group lock, under the DPCP, there is no parallelism in case of
(otherwise) non-conflicting requests for resources assigned to the same
synchronization processor. (Conversely, group locks can also be
thought of as a kind of ``virtual synchronization processors.'') The approach
followed by the DPCP thus is attractively simple, but not substantially more
flexible than group locks. 

The DPCP's same-processor restriction was later removed by \citeauthor{RM:95}
\cite{RM:95}, who in 1995 proposed a protocol that fully generalizes the DPCP
and supports fine-grained nesting for all resources. As with the DPCP,
for each resource, there is a dedicated synchronization processor
responsible for sequencing conflicting requests.
However, unlike the DPCP, \citeauthor{RM:95}'s protocol
\cite{RM:95} does not require all critical sections pertaining a
resource to execute on the synchronization processor; rather, the protocol allows
for full flexibility:  any critical section may reside on any processor.
As a result, nested sections that access multiple resources managed by different
synchronization processors become possible.

\citeauthor{RM:95}'s protocol \cite{RM:95} works as follows. To ensure mutual
exclusion among distributed critical sections and to prevent deadlock,
\citeauthor{RM:95} introduced a \emph{pre-claiming} mechanism that realizes
\emph{conservative two-phase locking}: when a task seeks to enter a critical
section, it first identifies the set of all resources that it
\emph{might} require while executing the critical section,
and then for each such resource sends a
\textsc{request} message to the  corresponding synchronization processor.
Each synchronization processor replies with a \textsc{grant}
message when the resource is available, and once \textsc{grant} messages
have been received for all requested resources, the task enters
its critical section.
As resources are no longer required, the task sends
\textsc{release} messages to the respective synchronization processors; the critical
section ends when all resources have been released.

To avoid deadlock, synchronization processors further send
\textsc{preempt} messages if a \textsc{request} message from a higher-priority task is
received after the resource has been already granted to a lower-priority
task (and no matching \textsc{release} message has been received yet). There are
two possibilities: either the lower-priority task has already commenced
execution of its critical section, in which case the \textsc{preempt} message is safely ignored
as it will soon release the resource anyway,
or it has not yet commenced execution, in which case it
releases the resource immediately and awaits another \textsc{grant}
message for the just-released resource.
Deadlock is impossible because of the protocol's all-or-nothing semantics:
tasks request all resources up front, commence execution only when they have acquired all
resources, and while executing a critical section only release
resources (\ie, conservative two-phase locking).

Compared to the DPCP \cite{RSL:88}, \citeauthor{RM:95}'s protocol \cite{RM:95} is a
significant improvement in terms of flexibility and versatility.
However, while their protocol allows tasks to use multiple
shared resources at once, it does not allow tasks to lock multiple
resources \emph{incrementally}. From a programmer's point of
view, it can be cumbersome (or even impossible) to determine all
resources that will be required prior to commencing a critical section.
Specifically, if the current state of one of the requested
resources impacts which other resources are also needed (\eg, if a shared
object contains a pointer to another, \apriori unknown resource), then a task
must initially lock the superset of all resources that it
\emph{might} need, only to then immediately release whichever
resources are not actually needed. Such conservative two-phase locking
leads to a lot of unnecessary blocking in the worst case, and is
known to suffer from performance penalties even in the average case.

The first \emph{work-conserving} protocol---in the sense that it
\emph{always} allows non-conflicting requests
to proceed in parallel---was developed already in
1992 and is due to \citeauthor{SZ:92} \cite{Z:92,SZ:92}. As previously discussed in
\secref{saw:part}, \citeauthor{SZ:92}'s protocol includes an online admission
test, which rejects lock requests that cannot be shown (at
runtime, based on current
contention conditions) to be satisfied within a specified waiting-time bound \cite{Z:92}.
As a result, \citeauthor{SZ:92}'s protocol
prevents deadlock---even if tasks request resources incrementally and in arbitrary order---as
any request that would cause deadlock will certainly be denied by the
admission test.

It should be noted that, from the programmer's point of view, this
notion of deadlock avoidance is significantly different from deadlock
avoidance as realized by the classic PCP \cite{SRL:90} and SRP \cite{B:91} uniprocessor
protocols: whereas the PCP and SRP \emph{defer} potentially
deadlock-causing resource acquisitions, which is logically \emph{transparent} to
the task, \citeauthor{SZ:92}'s protocol outright \emph{rejects} such lock requests \cite{Z:92}, so
that lock-acquisition failures must be handled in the task's logic.

Concerning blocking bounds, \citeauthor{Z:92}'s analysis \cite{Z:92} requires that
the bound on the maximum length of \emph{outer} critical sections must
include all blocking incurred due to \emph{inner} (\ie, nested)
critical sections. This assumption, common also in later analyses \cite{BW:13a,GZBW:17},
leads unfortunately to substantial structural pessimism.

For example, consider a scenario in which a job $J_1$ repeatedly
accesses two resources $\res_1$ and $\res_2$ in a nested fashion
(\ie, $J_1$ locks $\res_1$ first and then locks $\res_2$, and
does so multiple times across its execution). Now suppose there's
another job $J_2$ on a remote core that accesses $\res_2$ just once.
Since $J_1$ can incur blocking due to $J_2$'s infrequent critical
section when it tries to acquire $\res_2$ while already holding
$\res_1$, it follows that $J_1$'s maximum critical section length
\wrt $\res_1$ must include $J_2$'s maximum critical section length
\wrt $\res_2$. Thus, if $J_1$ accesses $\res_1$ and $\res_2$
in a nested fashion $r$ times, then $J_2$'s critical section will be
over-represented in $J_1$'s response-time bound by a factor of
$r$---a safe, but pessimistic bound that grossly overstates the
actual blocking penalty due to transitive blocking.

Another early protocol that supports fine-grained nesting is
\citeauthor{CTB:94}'s MDPCP \cite{CTB:94} for periodic tasks. Due to the underlying
careful definition of inter-processor priority ceilings
(which, as discussed in \secref{saw:part}, rests on the restriction to periodic tasks), the MDPCP is able to
prevent transitive blocking and deadlocks \cite{CT:94} analogously to the PCP
\cite{SRL:90}. Furthermore, for the priority ceilings to work correctly, the MDPCP requires
that any two resources that might be held together must be shared by
exactly the same sets of processors \cite{CTB:94}.

Finally, in 1995, \citeauthor{TS:95} \cite{TS:95} made an important
observation concerning the worst-case s-blocking in the presence of nested
non-preemptive FIFO spin locks. Let $d$ denote the \emph{maximum
nesting depth} (\ie, the maximum number of FIFO spin locks that any
task holds at a time), where $2 \leq d$. \citeauthor{TS:95} \cite{TS:95} showed that, under maximum
contention, tasks may incur
s-blocking for the combined duration of $\Omega(m^d)$ critical
sections \cite{TS:95}. That is, 
the potential
for \emph{accumulated transitive blocking} makes the bound on worst-case s-blocking
in the presence of nesting \emph{exponentially worse} (\wrt the
maximum nesting depth $d$) relative to the bound on maximum s-blocking in the
non-nested case, which is simply $O(m)$ (recall \secref{spin:part}).

Intuitively, this effect arises as follows. Let us say that an
\emph{outermost} (\ie, non-nested) critical section is of level one, and
that any critical section \emph{immediately} nested in a level-one
critical section is of level two, and so on. Consider a tower of critical
sections, that is, a level-1 critical section, containing exactly one
level-2 critical section, containing exactly one level-3 critical
section, and so on up to level $d$. Observe that:

\begin{enumerate}
\item
  a job $J$ blocked on the level-1 lock can be directly delayed by
  $m-1$ earlier-enqueued jobs;
\item
  \emph{each} of which can be directly delayed by $m-2$
  earlier-enqueued jobs when trying to enter the level-2 nested critical
  section;
\item
  \emph{each} of which can be directly delayed by $m-3$
  earlier-enqueued jobs when trying to enter the level-3 nested critical
  section;
\item
  and so on up to level $d$.
\end{enumerate}

Crucially, all the direct s-blocking incurred by various jobs in steps (2)--(4) also
\emph{transitively} delays $J$, which thus accumulates all delays and
therefore incurs s-blocking  exponential in $d$. The actual
construction that \citeauthor{TS:95} \cite{TS:95} used to establish the $\Omega(m^d)$
lower bound is more nuanced than what is sketched here because non-preemptive
execution can actually reduce blocking (a job that occupies a processor while
spinning on a level-$x$ lock prevents the processor from serving jobs that
generate contention for level-$y$ locks, where $y > x$).

Even in the most basic case (\ie, simple immediate nesting), where no task
ever holds more than two locks at the same time ($d=2$), nested FIFO spin locks result in an s-blocking
bound that is quadratic in the number of cores, which is 
clearly undesirable from a scalability point of view.
To address this bottleneck, \citeauthor{TS:95} \cite{TS:95} proposed an ingenious solution that
brings worst-case s-blocking back under control (especially in the special case of $d=2$) by using
priority-ordered spin locks, but with a twist. Instead of using scheduling priorities,
\citeauthor{TS:95}  let jobs use \emph{timestamps} as priorities---more precisely, a job's
locking priority is given by the time at which it issued its current
\emph{outermost} lock request (\ie, nested requests do not affect a
job's current locking priority), with the interpretation that an earlier
timestamp implies higher priority (\ie, FIFO \wrt outermost request timestamps).
The necessary timestamps need not
actually reflect ``time'' and can be easily obtained from an atomic
counter, such as those used in ticket locks.

For the outermost lock (\ie, the level-one lock), the timestamp order is actually equivalent to FIFO.
The key difference is the effect on the queue order in  nested locks: when using FIFO-ordered spin locks, a job's ``locking priority''
is effectively given by the time of its \emph{nested} lock request; with \citeauthor{TS:95}'s scheme \cite{TS:95},
the job's locking priority is instead given by the time of its outermost,  \emph{non-nested} lock request and remains invariant
throughout all nested critical sections (until the outermost lock is released).

Because at most $m$ jobs are running at any time, \citeauthor{TS:95}'s definition of
locking priorities ensures that there are never more than $m-1$ jobs
with higher locking priorities (\ie, earlier timestamps). In the
special case of $d=2$, this highly practical technique suffices to restore an $O(m)$ upper
bound on maximum s-blocking.
However, in the general case ($d > 2$), additional
heavy-weight techniques are required to ensure progress~\cite{TS:95}, and even then \citeauthor{TS:95}'s
method \cite{TS:95} unfortunately does not achieve an $O(m)$ bound in the general case. 
Interestingly, \citeauthor{TS:95} report that they were able to realize a
multiprocessor RTOS kernel with a maximum nesting depth of $d=2$~\cite{TS:95}.

\subsection{Recent Advances in Multiprocessor Real-Time Locking with Unrestricted Nesting}\label{sec:nested:recent}

Renewed interest in fine-grained locking emerged again in 2010 with
\citeauthor{FLC:10}'s MBWI protocol~\cite{FLC:10,FLC:12}, which explicitly supports
fine-grained locking and nested critical sections, albeit without any
particular rules to \emph{aid} or \emph{restrict} nesting. In particular, the MBWI \eone~does not
prevent deadlock and \etwo~uses FIFO queues even for nested requests. 

As a result of (ii), %
\citeauthor{TS:95}'s observation regarding the exponential growth of maximum
blocking bounds \cite{TS:95} also transfers to nesting under the MBWI protocol. However, the impact
of \citeauthor{TS:95}'s observation \cite{TS:95} is lessened somewhat in practice once
fine-grained blocking analyses (rather than coarse, asymptotic bounds) are
applied to specific workloads, since \citeauthor{TS:95}'s lower bound is based
on the assumption of extreme, maximal contention, whereas lock
contention in real workloads (and hence worst-case blocking) is constrained
by task periods and the number of critical sections per job.

Concerning (i), programmers are expected to arrange all critical sections such
that the ``nested in'' relation among locks forms a \emph{partial
order}---which prevents cycles in the \emph{wait-for graph} and thus
prevents deadlock. This is a common approach and widely used in
practice. For instance, the Linux kernel relies on this
\emph{well-ordered nesting principle} to prevent deadlock, and (as a
debugging option) employs a locking discipline checker called
\texttt{lockdep} to validate at runtime that all observed lock acquisitions are
compliant with \emph{some} partial order.

Notably, to prevent deadlock, it is sufficient for such a partial order
to exist; it need not be known (and for complex systems such as Linux it
generally is not, at least not in its entirety). For worst-case blocking analysis
purposes, however, all critical sections and all nesting relationships
must of course be fully known, and based on this information it is trivial to infer the
nesting partial order. For simplicity, we assume that resources are indexed
in accordance with the partial order (\ie, a job may lock $\res_x$ while already 
holding $\res_q$ only if $q < x$).

Assuming that all nesting is well-ordered,
\citeauthor{FLC:12} \cite{FLC:12}  presented a novel blocking-analysis
algorithm for nested critical sections that characterizes the
effects of transitive blocking much more accurately than the crude,
inflation-based bounds used previously (\eg, recall the discussion of
\citeauthor{SZ:92}'s protocol \cite{Z:92,SZ:92} and \citeauthor{Z:92}'s analysis \cite{Z:92} in \secref{spin:early} above). Aiming for this level
of accuracy was a major step forward, but unfortunately \citeauthor{FLC:12}'s
algorithm exhibits super-exponential runtime complexity \cite{FLC:12}. As
already mentioned, \emph{unrestricted} nesting is inherently difficult
to analyze accurately \cite{WB:14}.

\citeauthor{BW:13a}'s MrsP \cite{BW:13a} is another recent spin-based protocol that supports fine-grained, well-ordered
nesting in an otherwise unrestricted manner. While the initial version of the protocol \cite{BW:13a}
already offered basic support for nesting, the original analysis (which
heavily relies on inflation, similar to \citeauthor{Z:92}'s approach \cite{Z:92})
left some questions pertaining to the correct accounting of
transitive blocking unanswered \cite{BBW:16}. A revised and clarified version of the MrsP with
better support for nested critical sections was recently presented by
\citeauthor{GZBW:17} \cite{GZBW:17}, including a corrected analysis of worst-case
blocking in the presence of nested critical sections \cite{GZBW:17}. \citeauthor{GZBW:17}'s
revised analysis  is still based on execution-time inflation and
thus subject to the same structural pessimism as \citeauthor{Z:92}'s approach
\cite{Z:92} (as discussed in \secref{spin:early}). In particular, \citeauthor{GZBW:17}'s
revised analysis \cite{GZBW:17} does not yet incorporate \citeauthor{ZGBW:17}'s recently introduced,
less pessimistic, inflation-free analysis setup \cite{ZGBW:17}. Conversely,
\citeauthor{ZGBW:17}'s improved analysis \cite{ZGBW:17} does not yet support fine-grained
nesting.

In recent work following an alternative approach, \citeauthor{BBW:16} \cite{BBW:16} developed a MILP-based blocking analysis
of the classic MSRP \cite{GLD:01} with fine-grained, well-ordered nesting.
In \citeauthor{GLD:01}'s original definition of the MSRP \cite{GLD:01},
nesting of global resources is explicitly disallowed. However, as long
as all nesting is well-ordered, the protocol is capable of supporting
fine-grained nesting---the lack of nesting support in the original MSRP
is simply a matter of missing analysis, not fundamental incompatibility,
and can be explained by the fact that in 2001
analysis techniques were not yet sufficiently advanced to enable a reasonably accurate blocking analysis
of nested global critical sections. Leveraging a modern
MILP-based approach inspired by earlier LP- and MILP-based analyses
of non-nested protocols \cite{B:13a,WB:13a,BB:16,YWB:15}, \citeauthor{BBW:16} \cite{BBW:16} provided  the first analysis of the MRSP with support for fine-grained nesting. As a result, the MSRP may now be employed under
P-FP scheduling without any
nesting restrictions (other than the well-ordered nesting principle, which is required to prevent deadlock).

Interestingly, while the MSRP uses non-preemptive FIFO spin locks, which
is precisely the type of lock that \citeauthor{TS:95} \cite{TS:95} showed to be
vulnerable to exponential transitive blocking, \citeauthor{BBW:16}'s MILP-based
analysis \cite{BBW:16} is effective in analyzing transitive blocking
because the MILP-based approach inherently avoids
accounting for any critical section more than once \cite{BBW:16}. Thus, while
in theory FIFO-ordered spin locks cannot prevent exponential transitive
blocking in pathological corner cases with extreme levels of contention,
this is less of a concern in practice given a sufficiently accurate
analysis (\ie, if the analysis does not over-estimate contention) since
well-engineered systems are usually designed to minimize resource conflicts.

While MILP solving is computationally quite demanding,
\citeauthor{BBW:16}'s analysis \cite{BBW:16} comes with the advantage of resting on a solid
formal foundation that offers a precise, graph-based abstraction for
reasoning about possible blocking delays, and which ultimately enables
rigorous individual proofs of all MILP constraints. Given the
challenges inherent in the analysis of transitive blocking,
\citeauthor{BBW:16}'s formal foundation and MILP-based analysis
approach \cite{BBW:16}  provide a good starting
point for future analyses of fine-grained nesting in
multiprocessor real-time locking protocols.

\subsection{Asymptotically Optimal Multiprocessor Real-Time Locking}\label{sec:nested:rnlp}

The most significant recent result in  real-time locking is due to 
 \citeauthor{WA:12} \cite{WA:12}, who in 2012  presented  a surprising
breakthrough by showing that, with a few careful
restrictions, it is possible to control the occurrence of transitive blocking
and thereby ensure favorable---in fact, asymptotically
optimal---worst-case blocking bounds. Specifically, \citeauthor{WA:12}
introduced the \emph{Real-time Nested Locking Protocol} (RNLP) \cite{WA:12},
which is actually a meta-protocol that can be configured with several progress
mechanisms and lock-acquisition rules to yield either
 spin- or suspension-based protocols
that support fine-grained, incremental, and yet highly
\emph{predictable} nested locking.

Depending on the specific configuration, the 
RNLP ensures either  s-oblivious or s-aware asymptotically optimal maximum pi-blocking,
in the presence of well-ordered nested
critical sections and for \emph{any} nesting depth $d$.
 Furthermore, the RNLP is widely applicable: 
it supports clustered JLFP scheduling,
and hence also covers the
important special cases of G-EDF, G-FP, P-EDF, and P-FP scheduling.
Specifically, if applied on top of priority donation \cite{BA:11} (\resp,
RSB \cite{B:14b}), the RNLP yields an $O(m)$ (\resp,
$O(n)$) bound on maximum s-oblivious (\resp, s-aware) pi-blocking
under clustered JLFP scheduling~\cite{WA:12}. The
RNLP can also be instantiated on top of priority boosting similarly to the \fmlpp \cite{B:11} under
partitioned JLFP scheduling to ensure $O(n)$ maximum s-aware
pi-blocking \cite{WA:12}, and on top of priority inheritance under global
JLFP scheduling to ensure $O(m)$ maximum s-oblivious pi-blocking
\cite{WA:12}.

Analogously to the s-oblivious case, the RNLP can also be configured to use
non-preemptive execution and spin locks
to obtain an $O(m)$ bound on maximum s-blocking (again for \emph{any}
nesting depth $d$) \cite{WA:12}. As this contrasts nicely with \citeauthor{TS:95}'s
$\Omega(m^d)$ lower bound in the case of unrestricted nesting \cite{TS:95},
and since the spin-based RNLP is slightly
easier to understand than suspension-based configurations of the RNLP optimized for either
s-oblivious or s-aware analysis, we will briefly
sketch the spin-based RNLP variant in the following.

The RNLP does not automatically prevent deadlock and requires all tasks
to issue only well-ordered nested requests (\wrt a given partial
order). The RNLP's runtime mechanism consists of two main components: a
\emph{token lock} and a \emph{request satisfaction mechanism} (RSM).
Both are global structures, that is, all requests for any resource
interact with the same token lock and RSM.

The token lock is a $k$-exclusion lock that serves two purposes:
\textbf{(i)} it limits the number of tasks that can concurrently
interact with the RSM, and \textbf{(ii)} it assigns each job a
\emph{timestamp} that indicates when the job acquired its token (similar
to the time-stamping of outermost critical sections in \citeauthor{TS:95}'s earlier
protocol \cite{TS:95} based on  priority-ordered spin locks). If the RNLP is instantiated as a spin-based protocol or
for s-oblivious analysis, then $k=m$ \cite{WA:12}. (Otherwise, in the case of
s-aware analysis, $k=n$~\cite{WA:12}.)

Before entering an outermost critical section (\ie, when not yet holding any
locks), a job must first acquire a token from the token lock. Once it
holds a token, it may interact with the RSM. In particular, it may
repeatedly request resources from the RSM in an incremental fashion,
acquiring and releasing resources as needed, as long as nested requests are
well-ordered. Once a job releases its last resource (\ie, when it
leaves its outermost critical section), it also relinquishes its token.

In the spin-based configuration of the RNLP, jobs become non-preemptable
as soon as they acquire a token, and remain non-preemptable until they
release their token. Since non-preemptive execution already ensures that
at most $m=k$ tasks can be non-preemptable at the same time, in fact no
further KX synchronization protocol is required; \citeauthor{WA:12} refer to
this solution as a \emph{trivial token lock} (TTL) \cite{WA:12}. A TTL simply
records a timestamp when a job becomes non-preemptable, at which point it
may request resources from the RSM.

The specifics of the RSM differ in minor ways based on the exact
configuration of the RNLP, but all RSM variants share the following key
characteristics of the spin-based RSM.
Within the RSM, there is a wait queue for each resource $\res_q$,
and when a job requests a resource, it enters the corresponding wait queue.
As previously seen in \citeauthor{TS:95}'s protocol \cite{TS:95}, jobs are queued in order of
increasing timestamps. In the absence of nesting, this reduces again to
FIFO queues, but when issuing nested requests, jobs may benefit from an
earlier timestamp and ``skip ahead'' of jobs that acquired their tokens
at a later time.

However, there is a crucial deviation from \citeauthor{TS:95}'s protocol \cite{TS:95}
that makes all the difference:
whereas in \citeauthor{TS:95}'s protocol a job at the head of a queue
acquires the resource as soon as possible, the RNLP's RSM may choose to
\emph{not} satisfy a request for a resource $\res_q$ even when it is
available \cite{WA:12}. That is, the RNLP is \emph{non-work-conserving} and may
elect to withhold  \emph{currently} uncontested resources in
\emph{anticipation} of a potential later request that must not be
delayed (which is not entirely unlike the use of priority ceilings in the classic PCP \cite{SRL:90}).
Specifically, a job $J_i$
at the head of a resource $\res_q$'s queue may \emph{not} acquire
$\res_q$ if there exists another token-holding job $J_h$ with an
\emph{earlier token timestamp} that \emph{might} still request
$\res_q$ \cite{WA:12}.

As a result of this non-work-conserving behavior and the use of
timestamp-ordered wait queues, the RNLP ensures that no job is ever
blocked by a request of a job with a later token timestamp, even when
issuing nested requests. This property suffices to show $O(m)$ maximum
s-blocking per outermost critical section (because there can be at most
$m-1$ jobs with earlier token timestamps). It bears repeating that the
RNLP's bound holds for \emph{any} nesting depth $d$, whereas
\citeauthor{TS:95}'s work-conserving protocol \cite{TS:95} ensures $O(m)$ maximum
s-blocking only for $d=2$, and even then \citeauthor{TS:95}'s protocol exhibits worse constant 
factors (\ie, is subject to additional s-blocking).

The RNLP's non-work-conserving RSM behavior has two major implications:
first, while the RNLP controls \emph{worst-case} blocking
\emph{optimally} (asymptotically speaking), it does so at the price of a
potential increase in \emph{average-case} blocking when jobs are denied
access to rarely nested, but frequently accessed resources.
Second, all \emph{potential} nesting must be known at runtime (\ie, the
partial nesting order must not only exist, it must also be available to
the RNLP). This is required so that the RSM can appropriately reserve
resources that may be incrementally locked at a later time (\ie, to
deny  jobs with later
token timestamps access to resources that \emph{might} still be needed by jobs with earlier token timestamps).
In practical terms, the need to explicitly determine,
store, and communicate the partial nesting order imposes some
additional software engineering effort (\eg, at system-integration
time).

Subsequently, in work aimed at making the RNLP even more versatile and efficient for
practical use, \citeauthor{WA:13} \cite{WA:13} introduced a number of extensions and refinements.
Most significantly, they introduced the notion of
\emph{dynamic group locks} (DGLs) \cite{WA:13} to the RNLP.
As the name suggests, a DGL allows tasks to lock multiple resources in
one operation with all-or-nothing semantics, similarly to a (static) group lock
(recall \secref{nested:groups}), but without the need to define groups
\apriori, and without requiring that groups be disjoint. In a sense, DGLs are
similar to the pre-claiming mechanism of \citeauthor{RM:95}
\cite{RM:95}, but there is one important difference: whereas \citeauthor{RM:95}
enforce conservative two-phase locking semantics---once a task holds
some resources, it cannot acquire any additional locks---in the RNLP, tasks
are free to issue as many DGL requests as needed in an incremental
fashion. That is, the RNLP supports truly nested, fine-grained DGLs. 
Notably,
introducing DGLs does not negatively affect the RNLP's blocking bounds,
and the original RNLP \cite{WA:12} can thus be understood as a special case of the
DGL-capable RNLP \cite{WA:13} where each 
DGL request pertains to just a single resource (\ie, a singleton ``group'' lock).

Additionally, \citeauthor{WA:13} \cite{WA:13} introduced the possibility to apply the
RNLP as a KX synchronization protocol (also with asymptotically optimal
blocking bounds). In particular, KX synchronization is possible in
conjunction with DGLs, so that tasks can request multiple replicas of
different resources as one atomic operation.

As another practical extension of the RNLP, \citeauthor{WA:13} \cite{WA:13} 
introduced the ability to combine both spin- and
suspension-based locks in a way such that requests for spin locks are
not blocked by requests for semaphores (called ``short-on-long blocking'' \cite{WA:13}), 
since critical sections pertaining to suspension-based locks are likely to be much longer
(possibly by one or more orders of magnitude) than critical sections
pertaining to spin locks.

In 2014, in a further major broadening of the RNLP's capabilities
\cite{WA:12,WA:13}, \citeauthor{WA:14} presented the \emph{Reader-Writer RNLP}
(RW-RNLP) \cite{WA:14} for nested RW synchronization.
Building on the principles of the RNLP and phase-fair locks
\cite{BA:09,BA:10a,B:11}, \citeauthor{WA:14} derived a RW protocol that achieves
asymptotically optimal maximum pi- or s-blocking (like the RNLP) and $O(1)$
per-request reader blocking (phase-fairness), while allowing for a great deal of flexibility:
tasks may arbitrarily nest read and write critical sections, upgrade read locks to
write locks, and lock resources incrementally. While a detailed discussion is
beyond the scope of this survey, we note that integrating RW semantics
into the RNLP, in particular without giving up phase-fairness, is
nontrivial and required substantial advances in techniques and analysis \cite{WA:14}.

In 2015, \citeauthor{JWA:15} \cite{JWA:15} introduced a \emph{contention-sensitive} variant of
the RNLP \cite{WA:12}, denoted C-RNLP. In contrast to the original RNLP, and
the vast majority of other protocols considered herein, the C-RNLP
exploits knowledge of maximum critical section lengths \emph{at runtime} to
react dynamically to actual contention levels. (\citeauthor{SZ:92}'s protocol \cite{Z:92,SZ:92}
also uses maximum critical section lengths at runtime.)
At a high
level, the C-RNLP dynamically overrides the RNLP's regular queue order
to lessen the blocking caused by the RNLP's non-work-conserving behavior, but only 
if it can be shown that doing so will not violate the RNLP's guaranteed
worst-case blocking bounds. Since heavy resource contention is usually
rare in practice, contention sensitivity as realized in the C-RNLP can
achieve substantially lower blocking in many systems. As a tradeoff, the
C-RNLP unsurprisingly comes with higher lock acquisition
overheads, which however can be addressed with a novel implementation approach~\cite{NAA:18} (see \secref{impl:sys}).
Furthermore, it requires accurate information on worst-case
critical section lengths to be available at runtime, which can be inconvenient
from a software engineering perspective.

Overall, the RNLP and its various extensions represent the state of the art \wrt support for fine-grained nesting
with acquisition restrictions that prevent excessive transitive blocking. Importantly,
\citeauthor{WA:12} \cite{WA:12} established with the RNLP that $O(m)$
maximum s-blocking, $O(m)$ maximum s-oblivious pi-blocking, and
$O(n)$ s-aware pi-blocking are all possible in the presence of nested
critical sections even when faced with an arbitrary nesting depth $d > 2$, which was
far from obvious at the time given  \citeauthor{TS:95}'s prior negative results \cite{TS:95}.

\section{Implementation Aspects}\label{sec:impl}

While our focus in this survey is algorithmic properties and
analytical guarantees, there also exists a rich literature
pertaining to the implementation of multiprocessor
real-time locking protocols and their integration with programming
languages. In the following, we provide a brief overview of key topics. 

\subsection{Spin-Lock Algorithms}\label{sec:impl:spin}

The spin-lock protocols discussed in \secref{spin} assume the availability of
 spin locks with certain ``real-time-friendly''
properties (\eg, FIFO-ordered or priority-ordered locks). Spin locks algorithms widely used
 in practice include \citeauthor{MS:91}'s scalable \emph{MCS queue
locks} \cite{MS:91}, simple \emph{ticket locks}~\cite{MS:91,L:74}, and basic
TAS locks, where the former two are instances of
FIFO-ordered spin locks, and the latter is an unordered lock (\ie, it is not ``real-time-friendly,'' but easy to implement and still
analyzable \cite{WB:13a}). These lock types are well-known, not specific to real-time systems, and covered by
excellent prior surveys on shared-memory synchronization \cite{R:86,AKH:03,S:13}. We focus
here on spin-lock algorithms designed specifically for use in real-time
systems.

The most prominent example in this category are priority-ordered spin
locks, which are only rarely (if ever) used in general-purpose systems.
The first such locks are due to \citeauthor{ML:91}~\cite{M:91,ML:91,M:94}, who offered a
clear specification for ``priority-ordered spin locks'' and proposed two
algorithms that extend two prior FIFO-ordered spin locks, by respectively \citeauthor{B:78}
\cite{B:78} and \citeauthor{MS:91}~\cite{MS:91},  to respect request priorities.

Several authors continued this line of research and proposed refined priority-ordered spin locks in
subsequent years. In particular, \citeauthor{C:93} \cite{C:93} proposed several  scalable FIFO-
and priority-ordered queue lock algorithms. 
\citeauthor{C:93} also presented several extensions of the basic algorithms that add support for timeouts, 
preemptable spinning, memory-efficient lock nesting (\ie, without requiring a separate
queue element for each lock) \cite{C:93}. \citeauthor{TS:94} \cite{TS:94} similarly proposed a scheme for
spinning jobs to be preempted briefly by interrupt service routines, with the goal of  
ensuring low interrupt latencies in the kernel of a multiprocessor RTOS.

\citeauthor{WTS:96} \cite{WTS:96} considered \emph{nested} priority-ordered spin locks and observed that
they can give rise to starvation
effects that ultimately lead to unbounded priority inversion.
Specifically, they identified the following scenario: when a
high-priority job $J_h$ is trying to acquire a lock $\res_q$ that is
held by a lower-priority job $J_l$, and $J_l$ is in turn trying to
acquire a (nested) lock $\res_p$ that is \emph{continuously} used by
(at least two) middle-priority jobs (in alternating fashion) located on
other processors, then $J_l$ (and implicitly $J_h$) may remain
indefinitely blocked on $\res_p$ (respectively, on $\res_q$). To
overcome this issue, \citeauthor{WTS:96} proposed two spin-lock algorithms that
incorporate priority inheritance. The first algorithm---based on \citeauthor{ML:91}'s algorithm \cite{M:91,ML:91,M:94}---is
simpler; however, it is not scalable (\ie, it is not a local-spin algorithm). \citeauthor{WTS:96}'s second
 algorithm restores the local-spin property  \cite{WTS:96}.

To improve overhead predictability, \citeauthor{JH:97} \cite{JH:97} proposed a
priority-ordered spin lock that, in contrast to earlier algorithms,
ensures that critical sections can be exited in constant time. To
this end, \citeauthor{JH:97}'s algorithm maintains a pointer to the
highest-priority pending request, which eliminates the need to search
the list of pending requests when a lock is released.

Finally, and much more recently, \citeauthor{HJ:16} \cite{HJ:16} proposed a strengthened
definition of ``priority-ordered spin locks'' that forbids races among
simultaneously issued requests of different priorities and presented an
algorithm that satisfies this stricter specification.

Concerning FIFO-ordered spin locks that support preemptable spinning, as assumed
in \secref{spin:pre}, several authors have proposed suitable algorithms 
\cite{C:93,TS:94,KWS:97,AJJ:98}.
Furthermore, in their proposal of the SPEPP approach
(which also relies on preemptable spinning, as discussed in \secref{spin:spepp}), \citeauthor{TS:97} \cite{TS:97}
provided two implementation blueprints, one based on
MCS locks \cite{MS:91} and one based on TAS locks. Notably, even the
implementation based on TAS locks ensures FIFO-ordered execution of
critical sections because all posted operations (\ie, closures) are
processed in the order in which they were enqueued (though not
necessarily by the processor that enqueued them) \cite{TS:97}.

With regard to RW locks, \citeauthor{MS:91a} provided the canonical
implementation of task-fair (\ie, FIFO) RW locks \cite{MS:91a} as an extension
of their MCS queue locks \cite{MS:91}. Several
practical phase-fair RW lock implementations were proposed and evaluated
by \citeauthor{BA:09} \cite{BA:09,BA:10a,B:11}. \citeauthor{BJ:11} subsequently proposed a
stricter specification of ``phase fairness'' and proposed a matching
lock algorithm \cite{BJ:11}.

Finally, while not aimed specifically at real-time systems, it is worth
pointing out a recent work of \citeauthor{DH:16} \cite{DH:16} in which they aim to circumvent
the lock-holder preemption problem without resorting to non-preemptive
sections or heavy-weight progress mechanisms by leveraging emerging hardware support for
\emph{transactional memory} (HTM). With a sufficiently powerful HTM implementation,
it is possible to encapsulate entire critical sections pertaining to shared
data structures (but not I/O devices) in a HTM transaction, which
allows preempted critical sections to be simply aborted and any
changes to the shared resource to be rolled backed automatically. As a result, lock holders
can be preempted without the risk of delaying remote tasks. However, HTM support is not yet widespread 
in the processor platforms typically used in real-time systems, and it still remains to be seen whether it 
will become a \emph{de facto} standard in future multicore processors for embedded systems.

In work on predictable multicore processors, \citeauthor{SS:18}~\cite{SS:18} presented a pure hardware implementation of a predictable and analysis-friendly spin lock with round-robin semantics and non-preemptable spinning. Interestingly, while a round-robin lock does not ensure FIFO ordering of requests, round-robin access is much simpler and can be realized much more efficiently in hardware, and nonetheless provides identical worst-case guarantees as a FIFO spin lock (at most $m-1$ blocking critical sections per request). In \citeauthor{SS:18}'s implementation, uncontested lock acquisitions take only two clock cycles and lock release operations proceed in a single clock cycle~\cite{SS:18}. In another recent proposal of a processor-integrated hardware synchronization facility, \citeauthor{MHKS:19}~\cite{MHKS:19} presented an on-chip scratchpad memory with support for time-predictable atomic operations, which can be used to implement spin locks in an efficient and predictable manner amenable to WCET analysis. 

\citeauthor{B:11}~\cite{B:11} discusses how to factor non-preemptive spin-lock overheads into blocking and schedulability analyses.
\citeauthor{BAGB:17}~\cite{BAGB:17} provide an overhead-aware blocking and schedulability analysis for \citeauthor{ABBN:14}'s
 FSLM~\cite{ABBN:14}. \citeauthor{BAGB:17}'s analysis~\cite{BAGB:17} is based on execution-time inflation, similar to the original analysis of the MSRP~\cite{GLD:01}, and hence is subject to structural pessimism~\cite{WB:13a}. \citeauthor{WB:13}'s LP-based analysis~\cite{WB:13a}, which is designed to avoid such structural pessimism, can be used in conjunction with the overhead-accounting techniques proposed by \citeauthor{B:11}~\cite{B:11}.

\subsection{Avoiding System Calls}\label{sec:futexes}

In an operating system with a clear kernel-mode/user-mode separation and
protection boundary, the traditional way of implementing critical
sections in user mode is to provide lock and unlock system calls.
However, system calls typically impose non-negligible overheads
(compared to regular or inlined function calls), and hence represent a
significant bottleneck.

System-call overhead poses a problem in particular for spin-lock protocols, as one of the primary
benefits of spin locks  is their 
lower overheads compared to semaphores. If each critical section requires a system call to
indicate the beginning of non-preemptive execution, and another system
call to indicate the re-enabling of preemptions, then the overhead
advantage is substantially diminished.

To avoid such overheads, \litmus  introduced a
mechanism \cite{B:11} (in version 2010.1) that allows tasks to communicate non-preemptive sections to the kernel in a way that requires a
system call only in the infrequent case of a deferred
preemption. The approach works by letting each task share a page of
memory, called the task's \emph{control page}, with the kernel, similar
to the notion of a \emph{userspace thread control block} (UTCB) found in
L4 microkernels. 
More specifically, to enter a non-preemptive section, a task simply
sets a flag in its control page, which it clears upon exiting the
non-preemptive section. To indicate a deferred preemption, the kernel
sets another flag in the control page. At the end of each non-preemptive section,
a task checks the deferred preemption flag, and if set,
triggers the scheduler (\eg, via the \texttt{sched\_yield()} system call).

To
prevent runaway tasks or attackers from bringing down the system, the
kernel can simply stop honoring a task's non-preemptive section flag if
the task fails to call \texttt{sched\_yield()} within a pre-determined
time limit \cite{B:11}, which makes the mechanism safe to use even if
user-space tasks are not trusted. The control-page mechanism thus allows spin locks to be
implemented efficiently in userspace, requiring no kernel intervention
even when inter-core lock contention occurs.

A similar problem exists with semaphores in user mode. However,
since blocking is realized by suspending in semaphores, in the worst
case (\ie, if contention is encountered), the kernel is always
involved. Nonetheless, the avoidance of system calls in user-mode semaphores 
is still an important \emph{average-case} optimization.
Specifically, since lock contention is
rare in well-designed systems, avoiding system calls in the case of
uncontested lock and release operations (\ie, in the \emph{common
case}) is key to maximizing throughput in applications with a high frequency of critical sections.

Semaphore implementations that do not
involve the kernel in the absence of contention are commonly called
\emph{futexes} (fast userspace mutexes), a name popularized by the
implementation in Linux.
From a real-time perspective, the main challenge in realizing futexes is
maintaining a protocol's predictability guarantees (\ie, to avoid
invalidating known worst-case blocking bounds). With regard to this problem,
\citeauthor{SVBD:14} \cite{SVBD:14} distinguish between \emph{reactive} and
\emph{anticipatory} progress mechanisms \cite{SVBD:14}, where the former take
effect only when contention is encountered, whereas the latter
conceptually require actions even \emph{before} a conflicting lock
request is issued. For instance, priority inheritance is a reactive
progress mechanism, whereas priority boosting is an anticipatory
progress mechanism since a job's priority is raised unconditionally
whenever it acquires a shared resource.

It is easy to combine futexes with reactive mechanisms since the kernel
is involved anyway in the case of contention (to suspend the blocking task). In contrast, anticipatory
protocols are more difficult to support since the protocol's
unconditional actions must somehow be realized without invoking the
kernel in the uncontended case. Possibly for this reason, Linux supports
priority-inheritance futexes, but currently does not offer a futex
implementation of ceiling protocols.

Despite such complications,  it is fortunately still possible to realize many anticipatory protocols 
as futexes by \emph{deferring} task state updates until the
kernel is invoked anyway for some other reason
(\eg, a preemption due to the release of a higher-priority job), as has been shown by a number
of authors~\cite{ZBK:14,Z:13,AAB:15,SVBD:14}.

In a uniprocessor context, \citeauthor{ZBK:14} \cite{Z:13,ZBK:14,ZK:18,ZK:19} considered how to
implement predictable real-time futexes in an efficient and certifiable
way in the context of a high-assurance, resource-partitioned separation kernel. Their
approach is also relevant in a multiprocessor context because it allows
for an efficient, futex-compatible implementation of priority boosting
under partitioned scheduling by means of deferred priority changes
\cite{ZBK:14}. \citeauthor{AAB:15} \cite{AAB:15} later explored similar protocols for uniprocessor FP and
EDF scheduling and verified their correctness with a model checker.

Targeting multiprocessor systems, \citeauthor{SVBD:14} \cite{SVBD:14} systematically explored the
aforementioned classes of reactive and anticipatory real-time locking protocols, and concretely proposed real-time futex
implementations of the PCP \cite{SRL:90}, the MPCP \cite{R:90}, and the partitioned \fmlpp \cite{B:11},
which were shown to be highly efficient in practice.

\subsection{Implementations of Allocation Inheritance} \label{sec:impl:ai}

Allocation inheritance is the progress mechanism that is the most
difficult to support on multiprocessors, in particular when realized as task migration,
since it implies dynamic and rapid changes in the set of processors on which a lock-holding
job is eligible to execute. While this results in nontrivial
synchronization challenges within the scheduler, allocation inheritance has been implemented
and shown to be practical in several systems.

As already mentioned in \secrefs{spin:spepp}{impl:spin}, \citeauthor{TS:97} \cite{TS:97}
provided efficient implementations of the allocation inheritance principle that avoid
task migrations by expressing critical sections as closures. However, \citeauthor{TS:97}'s algorithms
still require critical sections to be executed non-preemptively (\ie, they
are not independence-preserving).

Concerning realizations of allocation inheritance that allow tasks to remain fully preemptable at all times,
\citeauthor{SBK:10} \cite{SBK:10} describe an elegant way of implementing allocation
inheritance on uniprocessors and mention that their
implementation extends to multiprocessors (but do not provide any details). \citeauthor{HP:01}~\cite{HP:01} discuss implementation and
design choices in a multiprocessor context, but do not report on
implementation details. Both \citeauthor{SBK:10} and \citeauthor{HP:01} consider
microkernel systems, which are particularly well-suited to allocation
inheritance due to their minimalistic kernel environment and emphasis
on a clean separation of concerns.

In work on more complex monolithic kernels, \citeauthor{BB:12} \cite{BB:12} discuss a
prototype implementation in Linux. Allocation inheritance has also been
realized several times in the Linux-based \litmus: by \citeauthor{FLC:12}
when implementing the spin-based MBWI protocol \cite{FLC:12}, by \citeauthor{B:13} for the
suspension-based OMIP \cite{B:13} and MC-IPC protocols \cite{B:14}, and by
\citeauthor{CBHM:15} for the spin-based MrsP \cite{CBHM:15}.

\citeauthor{CBHM:15} also
presented an implementation of the MrsP and allocation inheritance in
RTEMS, a static real-time OS without a kernel-mode / user-mode divide targeting
embedded multiprocessor platforms, and compared and contrasted the two
implementations in \litmus and RTEMS \cite{CBHM:15}.

\subsection{RTOS and Programming Language Integration} \label{sec:impl:sys}

Over the years, a number of authors have explored the question of how to
best integrate real-time locking protocols into RTOSs
and popular programming languages, to which extent existing
theory meets the needs of real systems, and techniques for efficient
implementations of real-time locking protocols. In the following, we
provide a high-level overview of some of the considered directions and questions.

Criticism of programming language synchronization facilities from the perspective of multiprocessor real-time
predictability dates all the way back to 1981, when
\citeauthor{REMC:81} \cite{REMC:81} reviewed the then-nascent ADA standard.
Interestingly, \citeauthor{REMC:81} argued already then in favor of introducing
spin-based synchronization  (rather than exclusively
relying on suspension-based methods) to avoid 
scheduler invocations \cite{REMC:81}.

More than 20 years later, \citeauthor{N:05} \cite{N:05} considered undesirable
blocking effects on multiprocessors due to ADA 95's \emph{protected
actions}. Specifically, \citeauthor{N:05} identified that, if
low-priority tasks spread across several processors issue a continuous
stream of requests for a certain type of ADA operations (namely, \emph{entries} protected by \emph{barriers})
to be carried out on a \emph{protected
object} currently locked by a higher-priority task, then, according to the
language standard, these operations could potentially all be serviced by
the higher-priority task in its exit path (\ie, when trying to release
the protected object's lock) \cite{N:05}, which theoretically can lead to
unbounded delays. As an aside, \citeauthor{TS:97}'s SPEPP approach~\cite{TS:97} offers an
elegant solution to this particular problem since it is starvation-free. Similarly, the
MrsP~\cite{BW:13a} could be applied in this context, as suggested by \citeauthor{BW:13}~\cite{BW:13} in their
investigation of protected objects in ADA 2012.

Even today, predictable multiprocessor synchronization in ADA
remains a point of discussion. \citeauthor{LWB:13a}~\cite{LWB:13a,LWB:13,L:13} revisited the
support for analytically sound multiprocessor real-time synchronization
in ADA 2012 and still found it to be wanting. At the
same time, they  also found the multiprocessor real-time locking protocols
available in the literature unsatisfactory, in the sense that there is
no clear ``best'' protocol that could be included in the standard to the
exclusion of all others. To resolve this mismatch in needs and capabilities,
\citeauthor{LWB:13a} argued
in favor of letting programmers provide their own locking protocols,
so that each application may be equipped with a
protocol most suitable for its needs, and presented a flexible framework
for this purpose as well as a number of reference implementations of
well-known protocols on top of the proposed framework \cite{LWB:13a,LWB:13,L:13}.

Most recently, \citeauthor{GZAJ:17} \cite{GZAJ:17} investigated the question of
predictable multiprocessor real-time locking within the constraints of
the ADA Ravenscar profile \cite{B:99} for safety-critical hard real-time systems.
In particular, they compared implementations of the MSRP \cite{GLD:01}
(based on non-preemptive sections) and the MrsP \cite{BW:13a} (based on
allocation inheritance), and found that the simpler MSRP is preferable in
the restricted Ravenscar context, whereas the MrsP is suitable for use in a
general, full-scope ADA system.

In work on other programming environments and languages, \citeauthor{ZC:04} \cite{ZC:04,ZC:06}
investigated a range of multiprocessor real-time locking protocols in
the context of a CORBA middleware, and \citeauthor{SPS:17} \cite{SS:15,SPS:17} proposed
and evaluated hardware implementations of real-time synchronization
primitives in a native Java processor for embedded safety-critical
systems. Also targeting Java, \citeauthor{WLB:11} \cite{WLB:11} studied the
multiprocessor real-time locking problem from the point of view of the
needs and requirements of the \emph{Real-Time Java} (RTSJ) and
\emph{Safety-Critical Java} (SCJ) specifications, and found a
considerable gap between the (restrictive) assumptions underlying the
(at the time) state-of-the-art real-time locking protocols and the broad
flexibility afforded by the RTSJ and, to a lesser degree, SCJ
specifications. As a step towards closing this gap,
\citeauthor{WLB:11} \cite{WLB:11} suggested changes to the
RTSJ and SCJ specifications that would ease a future integration of
analytically sound multiprocessor real-time locking protocols.

The first to discuss in detail the implementation of a multiprocessor
real-time locking protocol in an actual RTOS were \citeauthor{SZ:92}
\cite{Z:92,SZ:92}, who proposed a real-time threads package, including support
for predictable synchronization as discussed in \secref{saw:part}, for use on top of the Mach microkernel,
which has since been superseded by later generations of microkernels (\eg, the L4 family).

\citeauthor{TS:96} considered the design of a multiprocessor RTOS in light
of the scalability of worst-case behavior \cite{TS:96,T:96}.
Among other techniques, they proposed a scheme called \emph{local
preference locks} \cite{TS:96}, where resources local to a particular processor
are protected with priority-ordered spin locks, but request priorities
do \emph{not} depend on task priorities. Instead, the local processor
accesses a local preference lock with higher priority than remote
processors, which ensures that processors quickly gain access to local
resources (\ie, with $O(1)$ s-blocking) even if they are shared with
multiple remote processors.

In work throughout the past decade, many locking
protocols have been implemented and evaluated in \litmus \cite{B:11,E:15}.
Already in 2008, \citeauthor{BA:08a} \cite{BA:08a} provided a
detailed discussion of the implementations of several locking protocols in \litmus,
including the FMLP and the MPCP.

Several authors have reported on MPCP implementations. 
In work targeting Linux with the PREEMPT\_RT patch, \citeauthor{CDF:14} \cite{CDF:14}
provided details on in-kernel implementations of the FMLP \cite{BLBA:07} and a
non-preemptive MPCP variant \cite{CO:12a}. A particularly low-overhead
implementation of the MPCP for micro-controllers that avoids the need
for expensive wait-queue manipulations was proposed by \citeauthor{MDPL:14}
\cite{MDPL:14}. Targeting a very different system architecture, \citeauthor{IYY:17}
implemented the MPCP on top of a CAN bus to realize mutual exclusion in
a distributed shared memory (DSM) \cite{IYY:17}.

In recent overhead-oriented work, \citeauthor{NAA:17} \cite{NAA:17,NAA:19}
added a \emph{fastpath} to the RNLP \cite{WA:12} to optimize for the common case
of non-nested lock acquisitions, and \citeauthor{AVGB:16} \cite{AVGB:16} presented an
implementation of spin locks with flexible spin priorities \cite{ABBN:14}.
Motivated by the fact that the
contention-sensitive C-RNLP~\cite{JWA:15} exhibits relatively high acquisition
and release overheads due to its complex request-sequencing rules, \citeauthor{NAA:18}~\cite{NAA:18} went a significant step further and introduced a novel and rather unconventional
approach to implementing locking protocols. Specifically,
\citeauthor{NAA:18} centralized the acquisition and release logic in dedicated \emph{lock servers} that run the locking protocol in a cache-hot manner. Importantly, in contrast to the centralized semaphore protocols discussed in \secref{rpc}, a lock server does not actually centralize the execution of critical sections; rather, it centralizes only the execution of the locking protocol itself, and still lets tasks execute their actual critical sections in-place. In other words, in \citeauthor{NAA:18}'s design, a lock server decides in which order critical sections are executed, but does not execute any critical sections of tasks.  As a result of the cache-hot, highly optimized implementation of the contention-sensitive RNLP, \citeauthor{NAA:18} were able to demonstrate a reduction in acquisition and release overheads by over 80\%~\cite{NAA:18}.

To date, multiprocessor real-time locking protocols have received scant
attention from a WCET analysis perspective, with \citeauthor{G:13}'s thesis
\cite{G:13} and recent proposals for synchronization support in time-predictable multicore platforms~\cite{SPS:17,SS:18,MHKS:19} being notable exceptions. 

Last but not least, \citeauthor{GAB:16} \cite{GAB:16}
recently reported on a verified implementation of priority inheritance
with support for nested critical sections in the RTEMS operating system.

\section{Conclusion, Further Directions, and Open Issues}\label{sec:conc}

Predictable synchronization is one of the central needs in a
multiprocessor real-time system, and it is thus not surprising that
multiprocessor real-time locking protocols, despite having received considerable
attention already in the past,  are still a subject of ongoing research.
In fact, the
field
has seen renewed and growing interest in recent years due to the emergence and
proliferation of multicore processors as the \emph{de facto} standard
computing platform. 
Looking back at its history over the course of the past three
decades---starting with \citeauthor{RSL:88}'s pioneering results
\cite{RSL:88,R:90,R:91}---it is fair to say that the community has gained
a deeper understanding of the multiprocessor real-time locking problem and 
amassed a substantial body of relevant
knowledge. In this survey, we have attempted to systematically document and 
structure a current snapshot of this knowledge and the relevant literature, in hopes of making it more easily accessible to researchers and practitioners alike.

\subsection{Further Research Directions}\label{sec:misc}

In addition to the topics discussed in the preceding sections,
there are many further research directions related to multiprocessor
real-time locking protocols that have been explored in the past. While
these topics are beyond the scope of this already long survey, we do mention a few
representative publications to provide interested readers with starting points for further exploration.

One important resource in multiprocessor real-time systems that we
have excluded from consideration herein is energy. Unsurprisingly, energy management policies, in
particular schemes that change processor speeds, can have a significant
impact on synchronization \cite{CCK:08,HWZJ:12,FTCS:13,TFCY:16,W:17}.
Another architectural aspect that can interact negatively with multiprocessor real-time locking protocols 
is \emph{simultaneous multithreading} (SMT)~\cite{L:06}.

Targeting high-integrity systems subject to both real-time
and security requirements, \citeauthor{VEHH:13} \cite{VEHH:13} studied a number of real-time
locking protocols (including the MPCP \cite{R:90} and the clustered OMLP~\cite{BA:11}) from a timing-channel perspective  and identified 
\textit{confidentiality-preserving} progress mechanisms and locking protocols that prevent shared resources from being repurposed as covert timing channels.

A number of authors have studied multiprocessor real-time
synchronization problems from an optimality perspective and have
obtained speed-up and resource augmentation results for a number of
protocols and heuristics \cite{AE:10,HYC:16,RAB:11,RNA:12,R:13,AR:14,HCR:16,BCHY:17,CBSU:18}. These results are largely based on rather limiting
assumptions (\eg, only a single critical section per job), and in
several instances pertain to protocols purposefully designed to obtain a
speed-up or resource augmentation result, which has limited practical
relevance~\cite{CBHD:17}.

Taking a look at synchronization in real-time systems from a foundational perspective,
\citeauthor{LS:92}~\cite{LS:92} established lower and upper bounds on the number of atomic registers required 
to realize mutual exclusion with deadlock avoidance. 

Targeting periodic workloads in which each task has at most one critical section, \citeauthor{SUBC:19}~\cite{SUBC:19}
explored an unconventional synchronization approach in which an explicit dependency graph of all jobs and critical sections in a hyperperiod is built \apriori and all critical sections are sequenced offline with a list scheduling heuristic. 

All of the protocols discussed in this survey assume sequential tasks.
Going beyond this standard assumption, 
\citeauthor{HBL:12} \cite{HBL:12} studied the multiprocessor real-time locking problem in the context of parallel
real-time tasks \cite{H:12,HBL:12}. Exploring a similar direction, \citeauthor{DLAG:17}~\cite{DLAG:17} proposed an analysis of parallel tasks
using spin locks in the context of federated scheduling. Most recently, \citeauthor{JGLY:19}~\cite{JGLY:19} presented an analysis of parallel tasks using semaphores under the same scheduling assumptions.

Over the years, many results pertaining to the task and resource mapping
problems as well as related optimization problems have appeared \cite{TL:94,SBL:94,LDG:04,NBN:09,NNB:10,FM:10,RAB:11,FM:11,HLK:11,N:12,HWZJ:12,WB:13,HZWY:14,SRM:14,BMV:14,ASZD:15,HYC:16,HIS:17,BCHY:17,HTZY:17,DLBC:18,CLYL:19}.
Particularly well-known is \citeauthor{LNR:09}'s task-set partitioning
heuristic for use with the MPCP \cite{LNR:09}. Alternative heuristics and
strategies have been proposed by (among others) \citeauthor{NNB:10} \cite{NNB:10}, \citeauthor{WB:13}
\cite{WB:13}, and \citeauthor{ASZD:15} \cite{ASZD:15}.  Most recently, \citeauthor{DLBC:18}~\cite{DLBC:18} presented 
a detailed study of several task- and resource-assignment heuristics following a resource-centric schedulability analysis approach~\cite{HYC:16,HCR:16} in the context of a semaphore protocol with centralized critical sections, and \citeauthor{CLYL:19}~\cite{CLYL:19} presented an ILP-based partitioning solution for the same workload and platform model.
Techniques for scheduling task graphs with precedence
constraints and end-to-end deadlines in distributed systems~\cite{TL:94,SBL:94} are particularly relevant 
in the context of the DPCP and its related 
task and resource mapping problems.

Another system configuration aspect that has received considerable
attention is the policy selection problem, where the goal is to choose
an appropriate synchronization method for a given set of tasks, a set of shared resources, and the tasks' resource needs
 \cite{BNNG:11,HLLW:12,HZDL:14,AKNN:15,ASZD:15,BB:16,BCBL:08}. \citeauthor{BCBL:08}~\cite{BCBL:08} compared spin- and
suspension-based locking protocols (namely, the FMLP variants for short
and long resources \cite{BLBA:07}) in \litmus under consideration of
overheads with each other, and also against non-blocking synchronization
protocols. The choice between spin-based locking protocols and
non-blocking alternatives has also recently been considered by
\citeauthor{ASZD:15} \cite{ASZD:15} and
\citeauthor{BB:16} \cite{BB:16}. \citeauthor{S:03} proposed to reduce the impact of remote blocking
by combining locks with a versioning mechanism to reduce critical section lengths~\cite{S:03}.

In work on the consolidation and integration of legacy systems on
multicore platforms, \citeauthor{ABN:13}~\cite{ABN:12,ABN:13}, \citeauthor{NN:13}~\cite{NN:13}, and \citeauthor{NBN:11}~\cite{NBN:11} explored
the use of abstract interfaces that allow to represent a component's
resource needs and locking behavior without revealing detailed
information about the component's internals.

Finally, several authors have considered synchronization needs in
\emph{hierarchical} multiprocessor real-time systems (\ie, systems in which there is a hierarchy of schedulers)~\cite{NBNB:09,NBN:09,KWR:14,BBB:15,AKBB:16,ABBN:15a,ABBN:15}. In such systems, tasks are typically
encapsulated in processor reservations, resource servers, or, in the case of
virtualization, virtual machines, and thus prone to preemptions in
the middle of critical sections. As first studied by \citeauthor{HA:02}~\cite{HA:02}
in the context of Pfair-scheduled systems, this poses considerable
challenges from a locking point of view \cite{HA:02,HA:02a,H:04,HA:06}, and requires special rules to
either prevent lock acquisitions shortly before a job or reservation's
(current) budget allocation is exhausted \cite{HA:02}, or acceptance of (and
appropriate accounting for) the fact that jobs or reservations may
overrun their allocated budget by the length of one critical section \cite{BBB:15}.
\label{sec:placement}

\subsection{Open Problems}

As already mentioned in \secref{intro}, the ``last word'' on
multiprocessor real-time resource sharing has not yet been spoken, and
will likely not be spoken for a long time to come.
Without seeking to detract from other interesting
directions, we briefly highlight three largely unexplored opportunities
for future work.

First, there is a need for more flexible blocking analyses that can
handle multiple lock types simultaneously. Practical systems typically use
multiple types of locks for different purposes (\eg, both spin locks
and semaphores), and while many lock types and real-time locking
protocols have been investigated and analyzed in isolation, few results
in the literature explicitly account for effects that arise from the
combination of different lock types (\eg, blocking bounds for
semaphores in the presence of non-preemptive sections due to spin
locks \cite{BLBA:07,WA:13}). Worse, few (if any) of the existing analyses focused on individual lock types
and protocols  compose soundly without modification. To
better support the needs of real-world systems, clearly further advances
will be required in this direction.

Second, all major existing blocking analyses pertain to the worst case. While
this is clearly required for true hard real-time systems, given that
average-case contention in well-designed systems is usually low, the
resulting bounds can seem extremely pessimistic relative to observable
blocking delays. For firm real-time systems that do not require hard
guarantees, or for systems where there are strong economic incentives to
not provision based on an absolute worst-case basis (which is arguably
the majority of systems in practice), there is unfortunately little support
in the existing literature. Thus, practitioners today must choose
between pessimistic hard-real-time  blocking bounds that tend to result in
over-provisioning, or no analysis at all (\ie, rely purely on
measurements instead). To extend the range of systems to which
analytically sound blocking bounds are
applicable, we will need means for reasoning about
anticipated blocking delays that are both more rigorous than
average-case observations and less taxing than hard-real-time analyses based on 
worst-case assumptions at every step.

Last but not least, we would like to highlight the need for a rigorous foundation and
formal proofs of correctness for blocking analyses. In particular for
\emph{worst-case} blocking analyses, which by their very nature are
intended to be used in critical systems, it is essential to have utmost
confidence in the soundness of the derived bounds. However, as
blocking bounds become more accurate, task models more detailed, and
synchronization techniques more advanced, the required blocking analyses
also become more tedious to derive, more challenging to validate, and
ultimately more error-prone. If the goal is to support safety-critical
systems in practice, and to use multiprocessor real-time locking
protocols and their analyses as evidence of system safety in
certification processes, then this is a very dangerous trend, in particular
in the light of prior missteps that have only recently come to light
\cite{YCH:17,CB:17,CNHY:19,GZBW:17}. As a first step towards a higher degree of
confidence in the correctness of advanced blocking analyses, recent
LP- and MILP-based blocking analyses \cite{B:13,WB:13a,BB:16,YWB:15} offer
the advantage that each constraint can be checked and proven correct
individually (rather than having to reason about the entire analysis as
a whole), which  simplifies the problem considerably. However, while this
is a much needed improvement, it is clearly not yet enough. In the longterm, it
will be desirable (if not outright required at some point) for analyses and protocols
intended for use in safety-critical systems---such as blocking bounds
for multiprocessor real-time locking protocols---to come with a
machine-checked proof of soundness, or other equivalent soundness
guarantees backed by formal verification methods. 
Much interesting and challenging work remains to be done before a full formal verification
of a multiprocessor real-time locking protocol and its timing properties can become reality. 

\clearpage
\bibliographystyle{plainnat} %
\bibliography{multiprocessor-real-time-locking.bib}

\end{document}